\documentclass[11pt]{article}

\usepackage{amsmath,amssymb}
\usepackage{graphicx}
\usepackage{xcolor}
\usepackage[nosort]{cite}
\usepackage{sidecap}

\sidecaptionvpos{figure}{c}

\begin{document}

\vspace*{-1.5cm}
\begin{flushright}
  {\small
  LMU-ASC 05/20\\
  MPP-2020-14
  }
\end{flushright}

\vspace{1.75cm}

\begin{center}
{\LARGE
Connecting Dualities and Machine Learning}
\end{center}

\vspace{0.4cm}

\begin{center}
  Philip Betzler$^{1,2}$, Sven Krippendorf$^1$
\end{center}

\vspace{0.3cm}

\begin{center} 
\textit{$^{1}$\hspace{1pt} Arnold Sommerfeld Center for Theoretical Physics\\[1pt]
Ludwig-Maximilians-Universit\"at \\[1pt]
Theresienstra\ss e 37 \\[1pt]
80333 M\"unchen, Germany}
\\[1em]
\textit{$^{2}$\hspace{1pt} Max-Planck-Institut f\"ur Physik\\[1pt]
F\"ohringer Ring 6 \\[1pt]
80805   M\"unchen, Germany}
\end{center} 

\vspace{0.8cm}

%%%%%%%%%%%%%%%%%%%%%%%%%%%%%%%%%%%%%%%%%%%%%%%
%%%%%%%%%%%%%%%%%%%%%%%%%%%%%%%%%%%%%%%%%%%%%%%
%%%%%%%%%%%%%%%%%%%%%%%%%%%%%%%%%%%%%%%%%%%%%%%
%%%%%%%%%%%%%%%%%%%%%%%%%%%%%%%%%%%%%%%%%%%%%%%

\begin{abstract}
\noindent
Dualities are widely used in quantum field theories and string theory to obtain correlation functions at high accuracy. Here we present examples where dual data representations are useful in supervised classification, linking machine learning and typical tasks in theoretical physics. We then discuss how such beneficial representations can be enforced in the latent dimension of neural networks. We find that additional contributions to the loss based on feature separation, feature matching with respect to desired representations, and a good performance on a `simple' correlation function can lead to known and unknown dual representations. This is the first proof of concept that computers can find dualities. We discuss how our examples, based on discrete Fourier transformation and Ising models, connect to other dualities in theoretical physics, for instance Seiberg duality.
\end{abstract}

%%%%%%%%%%%%%%%%%%%%%%%%%%%%%%%%%%%%%%%%%%%%%%%
%%%%%%%%%%%%%%%%%%%%%%%%%%%%%%%%%%%%%%%%%%%%%%%
%%%%%%%%%%%%%%%%%%%%%%%%%%%%%%%%%%%%%%%%%%%%%%%

\clearpage

\tableofcontents

%%%%%%%%%%%%%%%%%%%%%%%%%%%%%%%%%%%%%%%%%%%%%%%
%%%%%%%%%%%%%%%%%%%%%%%%%%%%%%%%%%%%%%%%%%%%%%%
%%%%%%%%%%%%%%%%%%%%%%%%%%%%%%%%%%%%%%%%%%%%%%%
%%%%%%%%%%%%%%%%%%%%%%%%%%%%%%%%%%%%%%%%%%%%%%%
%%%%%%%%%%%%%%%%%%%%%%%%%%%%%%%%%%%%%%%%%%%%%%%
%%%%%%%%%%%%%%%%%%%%%%%%%%%%%%%%%%%%%%%%%%%%%%%
%%%%%%%%%%%%%%%%%%%%%%%%%%%%%%%%%%%%%%%%%%%%%%%
%%%%%%%%%%%%%%%%%%%%%%%%%%%%%%%%%%%%%%%%%%%%%%%

\section{Introduction}
In many cases, when we want to describe a dynamical system in physics we identify the effective field theory governing its dynamics. However, in some cases there are multiple effective field theories describing the same system. This phenomenon is referred to as duality. Dualities are a very powerful tool in fundamental physics, ubiquitously used in dynamical systems involving gauge theories, and are extremely explored and utilised in the context of string theory (cf.~\cite{Quevedo:1997jb,Polchinski:2014mva} for an overview). Such dualities provide two descriptions -- often two Lagrangians with distinct sets of fields and associated couplings -- of the same dynamical system. The difference between these effective field theories is that they describe certain properties of the system, i.e.~correlation functions, in a more efficient way.

The efficient calculation of correlation functions or estimates of them based on sample data is also relevant in typical data science applications such as classification. Here, we present examples of data questions where dualities prove to be useful (cf.~Section~\ref{sec:examples}). For simplicity we restrict ourselves at this stage to data questions in physical systems where we know a useful dual description. This has the added benefit that the results can be compared to interpretable solutions. We show that the classification with `simple' standard network architectures works much better for data in the dual representation. Better accuracy is achieved in the dual frame with less training effort.

We then show that finding a similar level of classification is not easily possible, i.e.~by examining several standard changes to the architectures such as wider and deeper networks. In particular, this includes architectures which in principle have the capability to perform the duality transformation. We find that the network generically does not find this beneficial configuration. As a next step, we then explore opportunities how to enforce such dual representations, beyond a `trivial' enforcing of dual variables when the duality transformation is known (cf.~Section~\ref{sec:enforcing}). In particular, we find positive results when we demand feature separation in the latent space. We also identify good representations with a modified autoencoder structure where we put an additional constraint (good performance on simple classification tasks) on the latent dimension. Finally we provide and exemplify a method how to enforce certain distributional properties of the dual representation. These representations found by the networks are the first examples where dual representations are obtained without the network ``knowing" them a priori.

Before concluding, we comment on the connection to other dualities in physics (cf.~Section~\ref{sec:connections}).

\section{Benefits of Dual Representations} 
\label{sec:examples}
Here we present several examples where dualities prove useful to address supervised classification tasks.

\subsection{Discrete Fourier Transformation}\label{DiscreteFourierTransformation}
\begin{figure}
\includegraphics[width=0.245\textwidth]{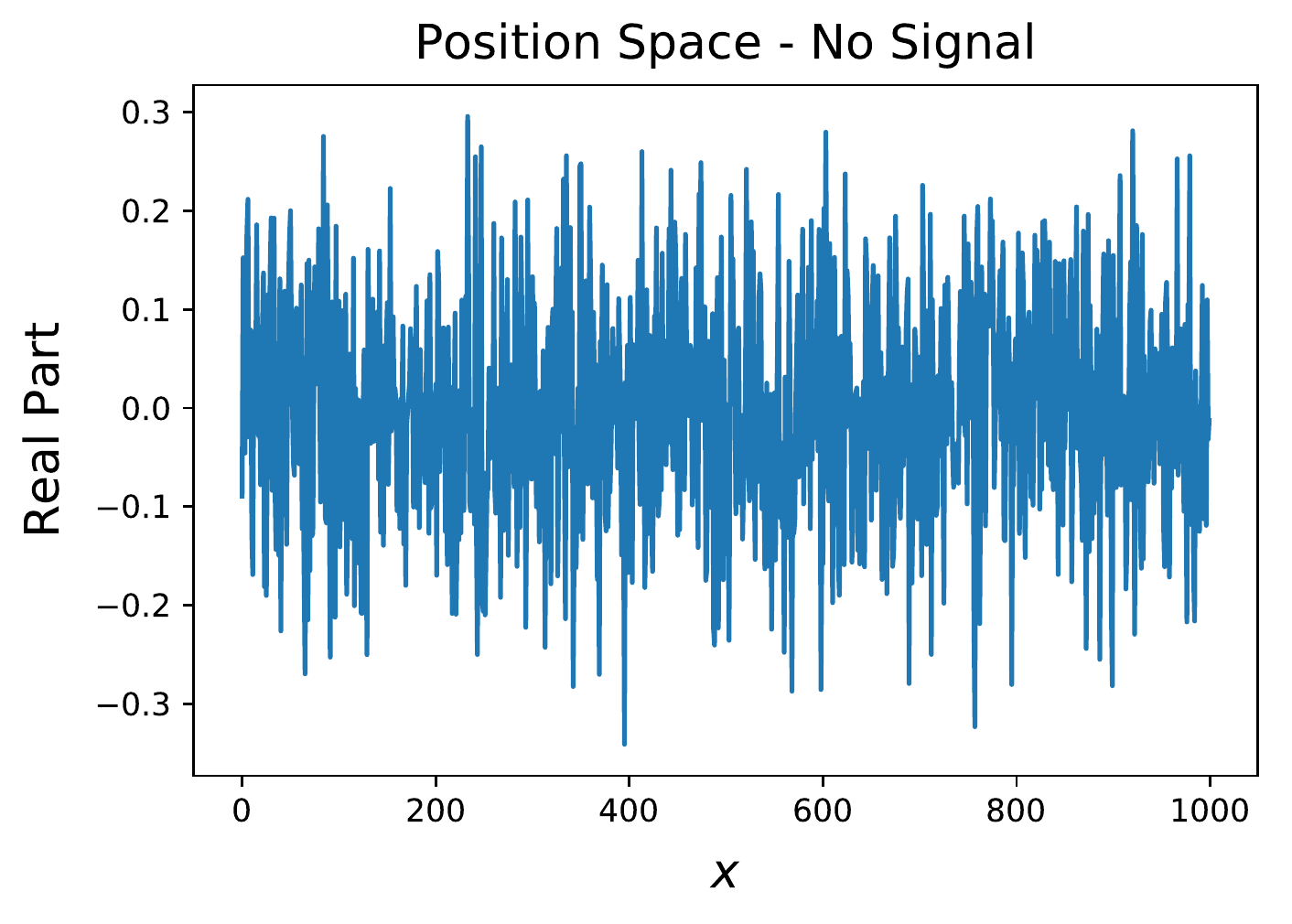}
\includegraphics[width=0.245\textwidth]{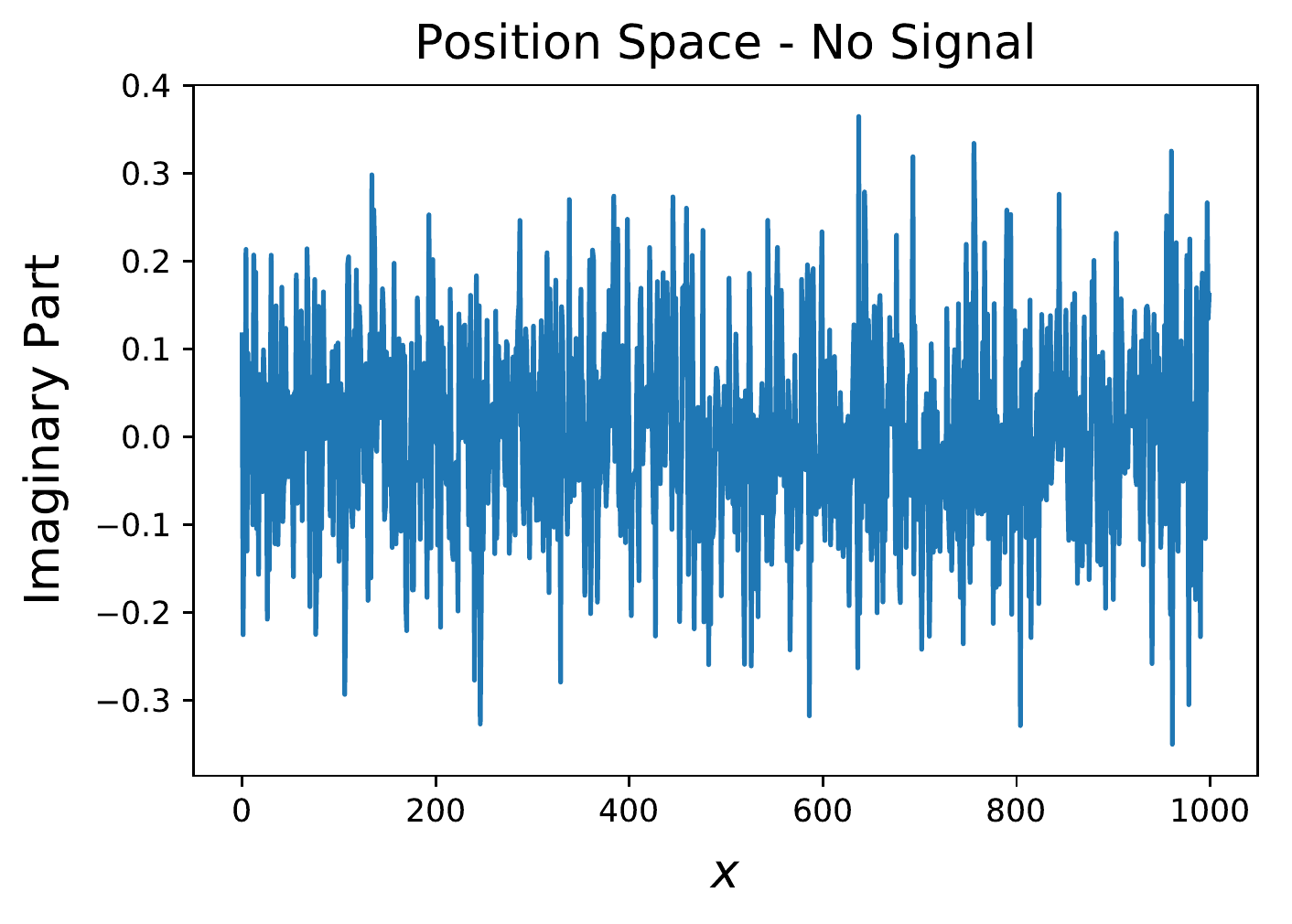}
\includegraphics[width=0.245\textwidth]{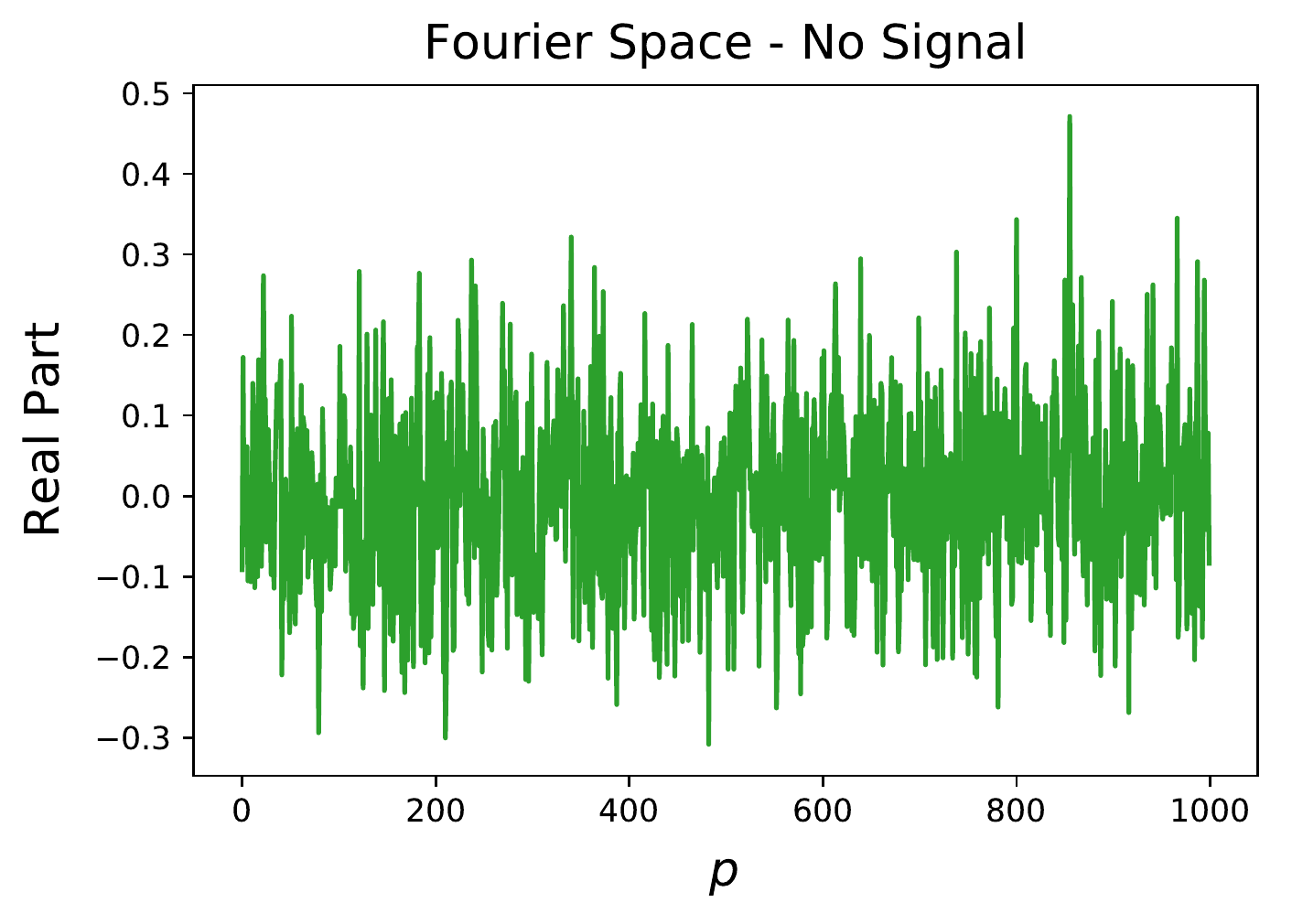}
\includegraphics[width=0.245\textwidth]{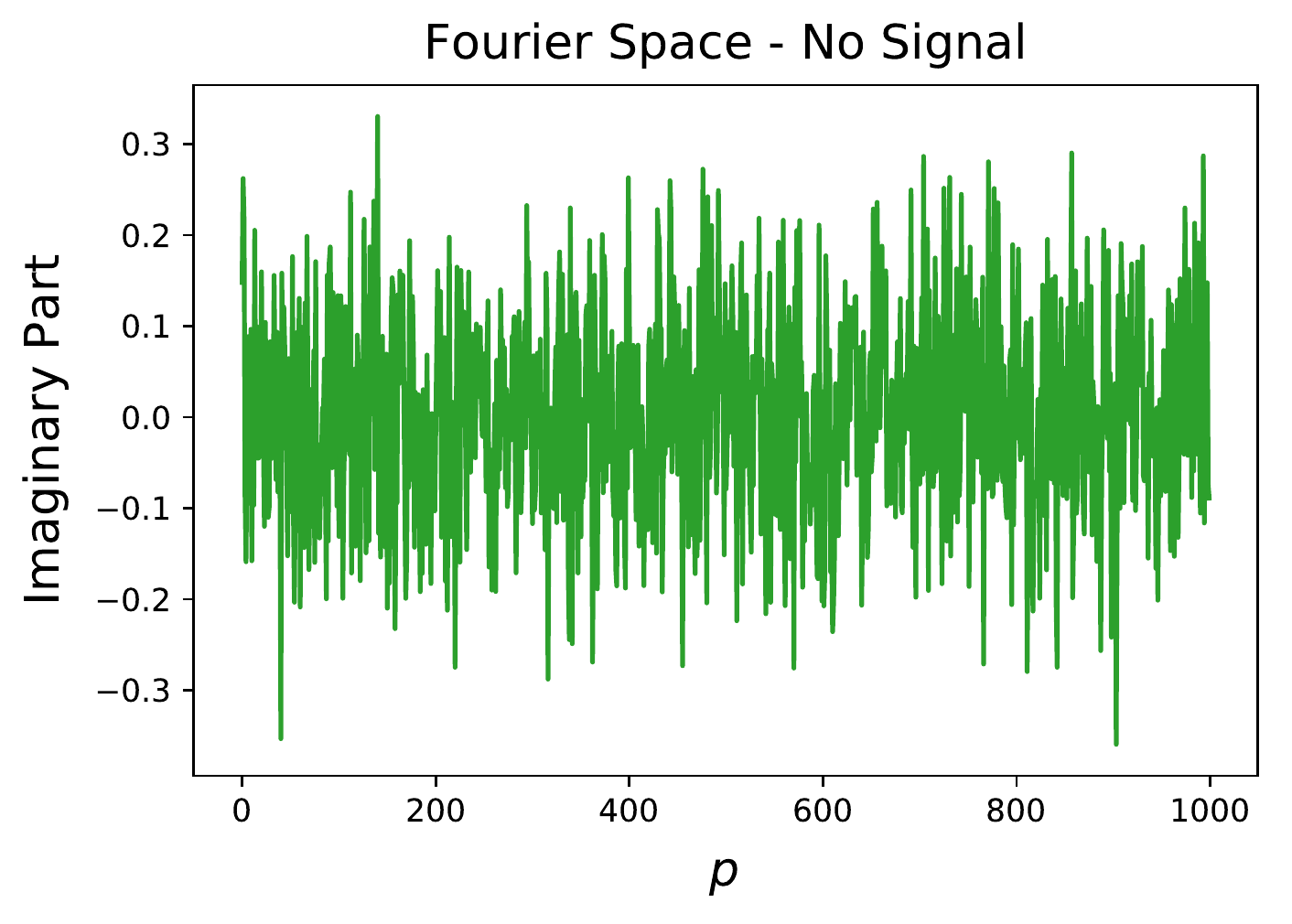}\\
\includegraphics[width=0.245\textwidth]{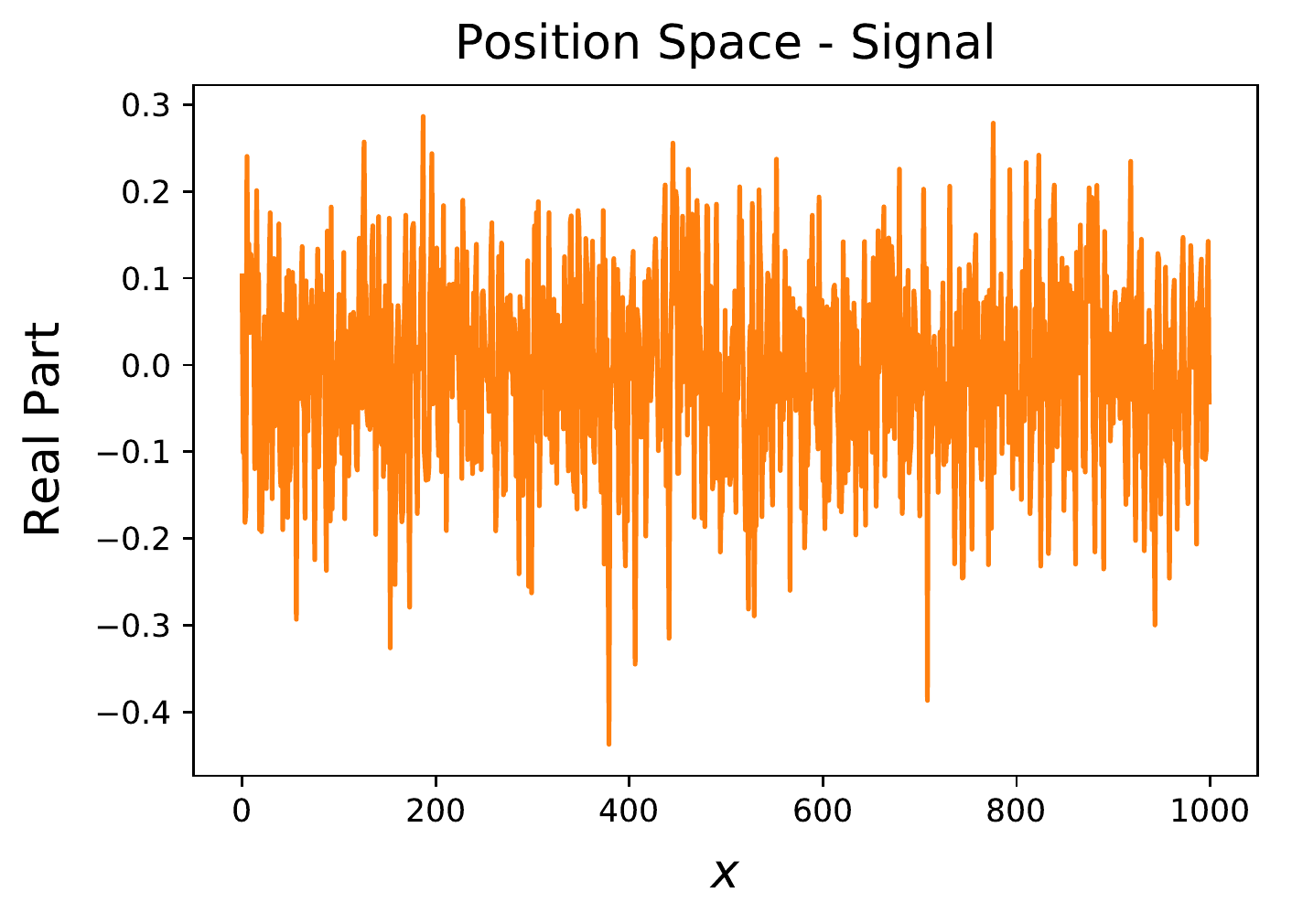}
\includegraphics[width=0.245\textwidth]{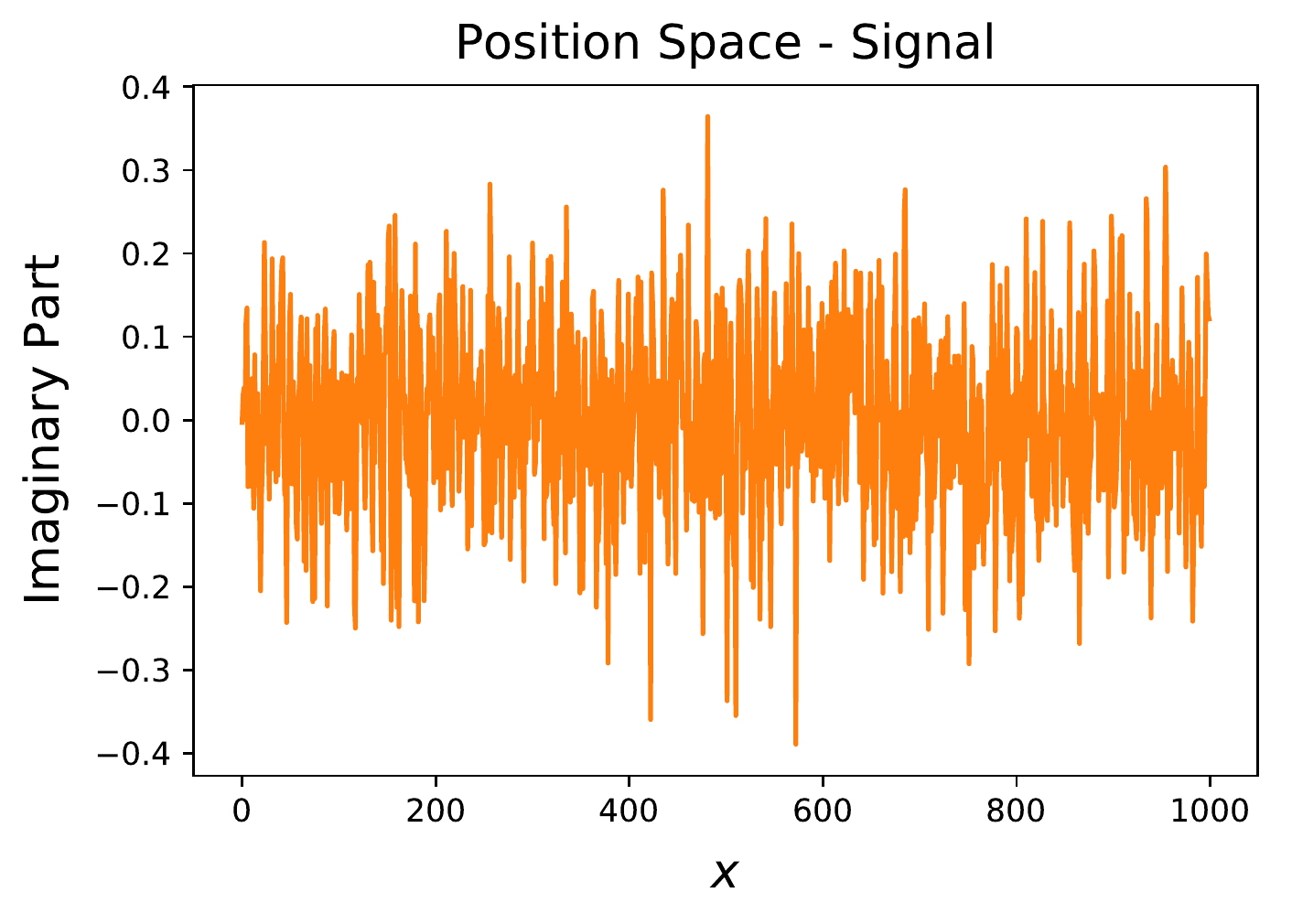}
\includegraphics[width=0.245\textwidth]{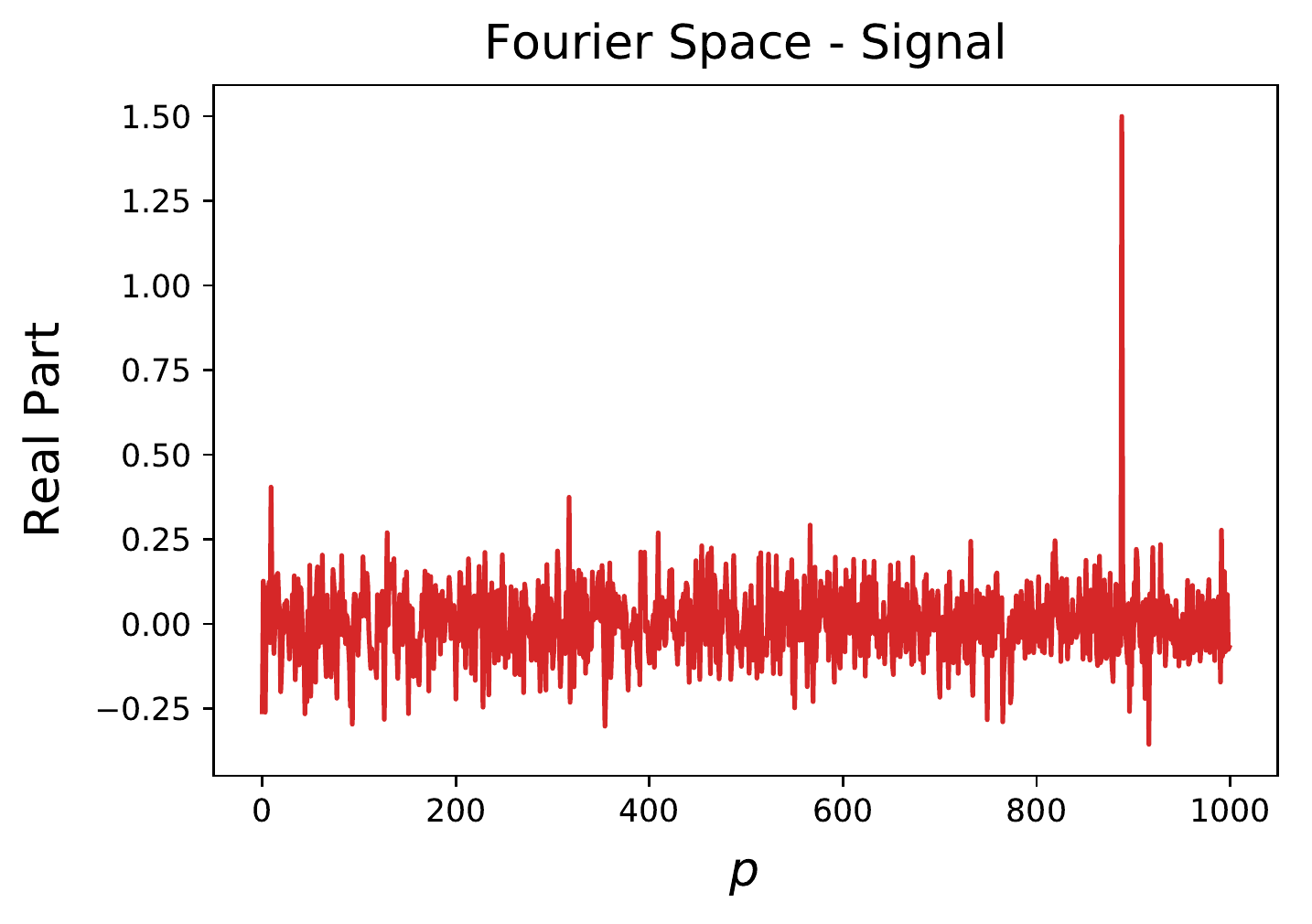}
\includegraphics[width=0.245\textwidth]{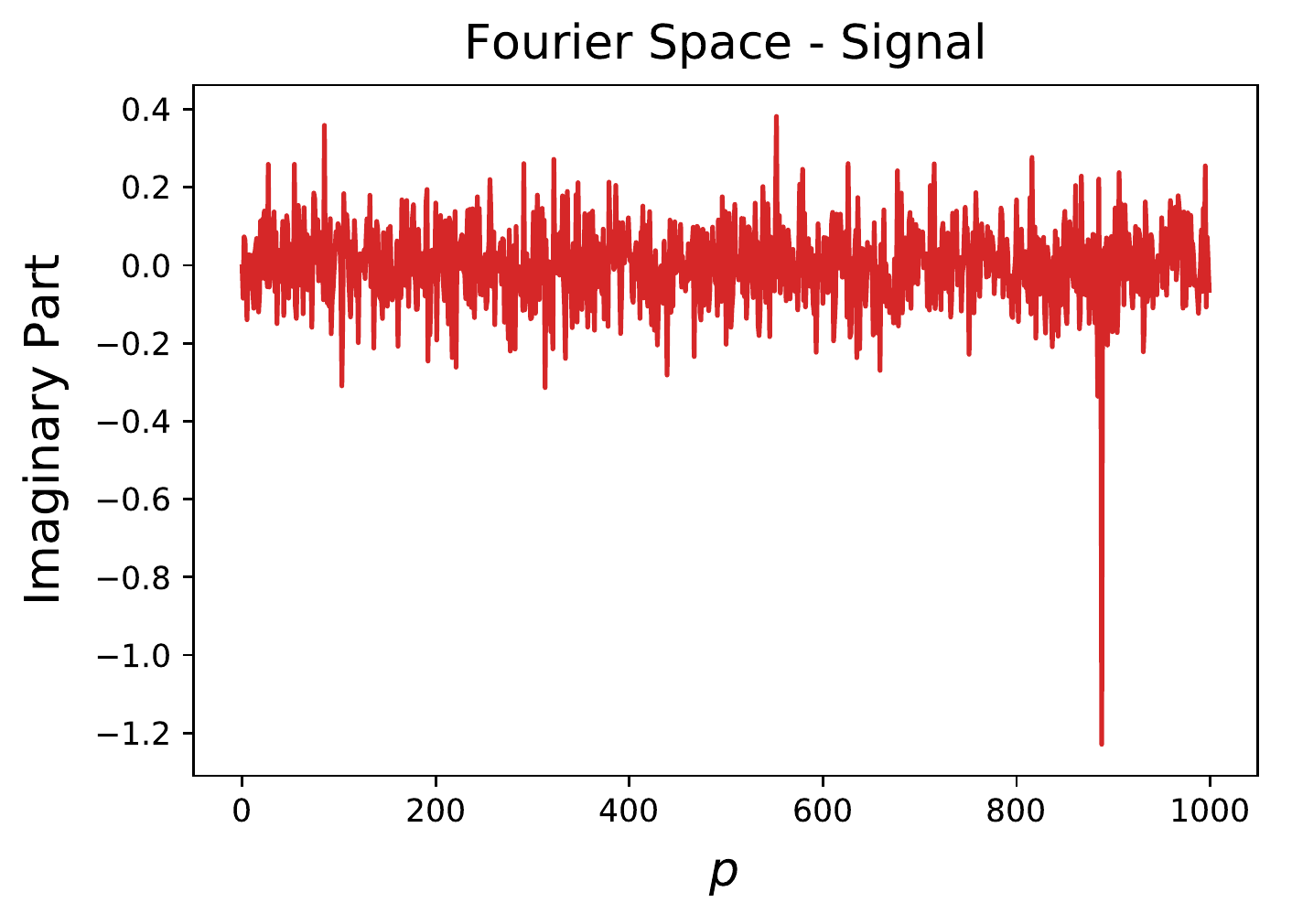}
\vspace{-1.cm}\caption{Comparison of noisy signals and pure noise in position and Fourier space.}\label{fig:Fourier_example}
\end{figure}

The Fourier transformation captures the essence of many dualities relating strongly-coupled and weakly-coupled field theories (cf. also Section \ref{sec:connections}). Strongly coupled theories feature non-vanishing correlations over large distances whereas weakly coupled theories only feature seizable correlations at short distances. This is resembled in Fourier transformation, where a delta-peak in momentum space is spread out over all of position space. {\it When is it useful to use position or momentum space representations?} A simple example is given by identifying whether there is a signal hiding under Gaussian noise. For concreteness we consider a signal which is a single peak in momentum space. An example of the data for each class in this binary classification problem is shown in Figure~\ref{fig:Fourier_example} and the details of the construction and our neural networks and numerical experiments can be found in Appendix~\ref{app:fourier}.

When performing classification with a simple neural network\footnote{Here we perform a classification with a single Conv1D layer with 4 filters and ReLU activation followed by a Dense layer with a single neuron and sigmoid activation. Details on the experiment can be found in Appendix~\ref{app:fourier}.}, we find that a classification is possible for the data in the momentum representation (test accuracy $0.9835$) but not for the position representation (test accuracy at pure guessing $\sim 0.5$). 

When adding a single or several hidden dense layers to the position space network, we find only a marginal improvement (again details can be found in Appendix~\ref{app:fourier}). As the reached performance does not even come close to the perfect score in the momentum space representation, it is clear that our deeper neural networks are not adapting the position space representation.

\subsection{2D Ising Model}
\label{sec:2DIsing}
A very well-known example of duality in physics is that of the high-low temperature duality in the 2D Ising model~\cite{Kramers:1941kn,Kramers:1941zz,Onsager:1943jn} (cf. also ~\cite{Savit:1979ny} for a review).
\begin{figure}[t]
\includegraphics[width=0.49\textwidth]{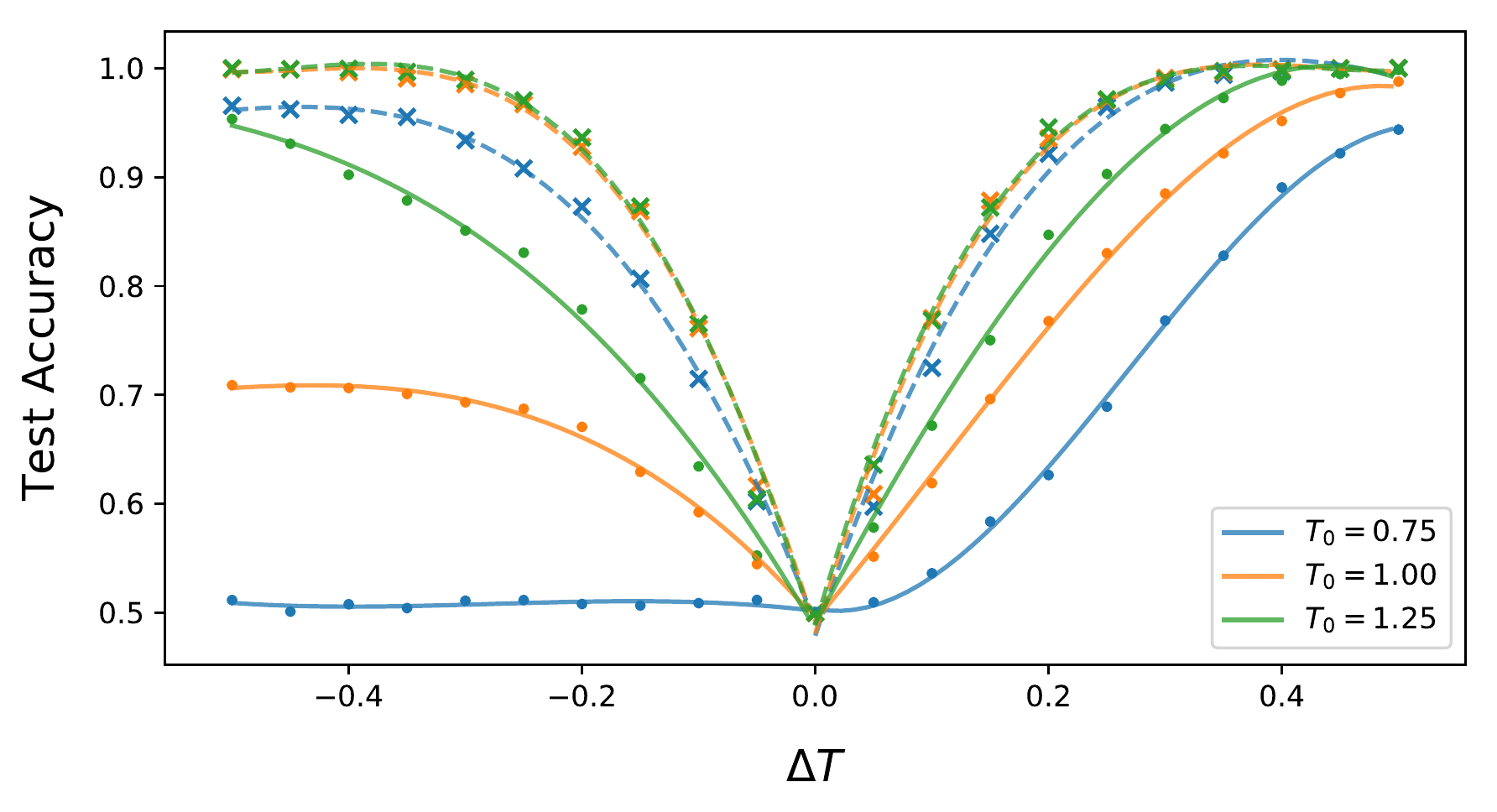}
\includegraphics[width=0.49\textwidth]{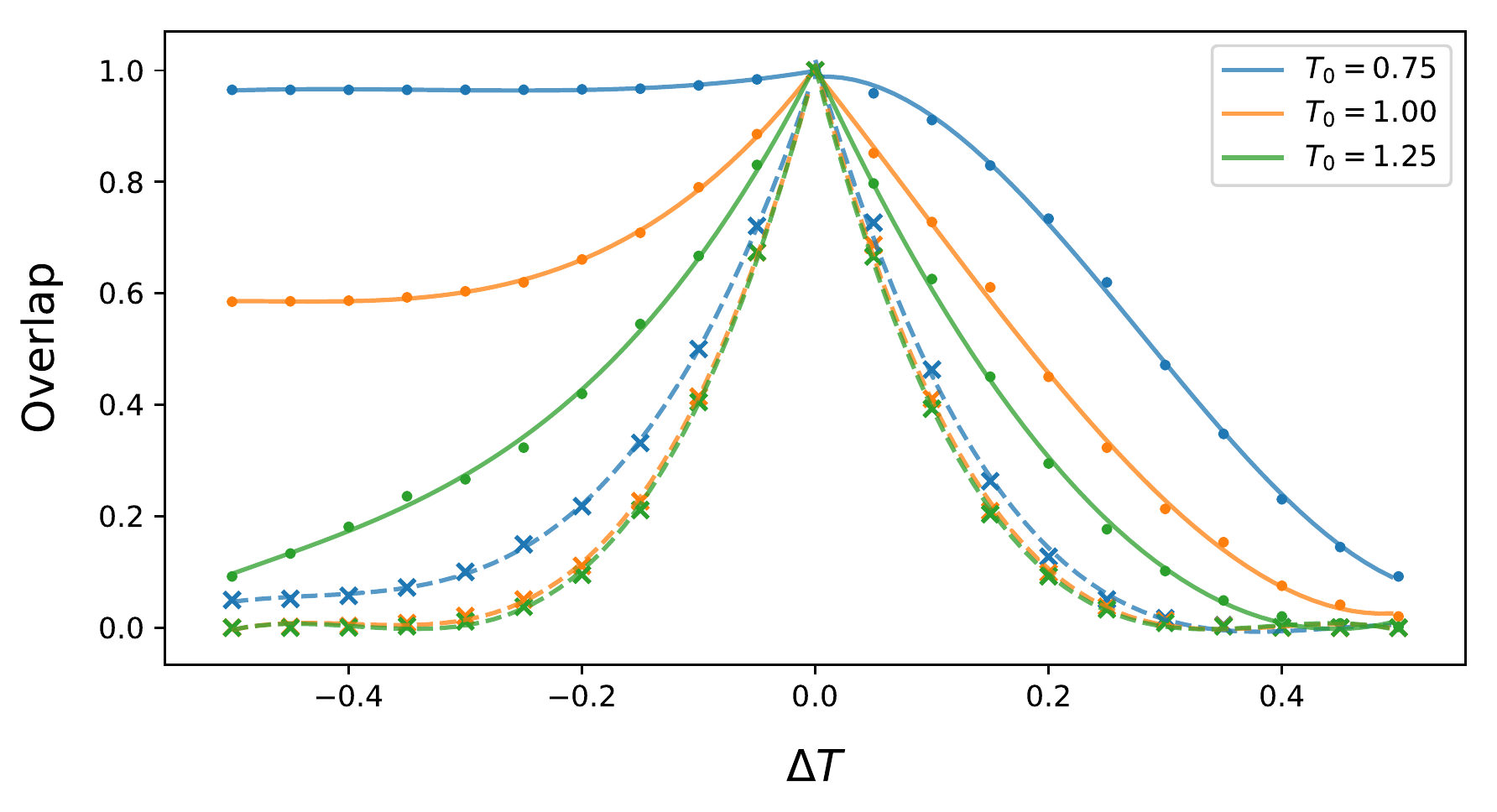}
\vspace{-0.5cm}\caption{Classification of states according to their temperature in the square-lattice Ising model. Solid lines and dots indicate data for the original temperatures, dashed lines and crosses for the dual temperatures. Pairs of temperatures $T_{0},T=T_{0}+\Delta T$ were chosen by fixing a reference point $T_{0}$ and gradually increasing $\Delta T$ by increments of $0.05$.
\label{fig:Ising2D_OriginalDual_Performance}}
\end{figure}

This Ising model lives on a $N\times N$ square lattice with periodic boundary conditions.
On each lattice site there is a spin degree of freedom $s_i,$ which can take values $\pm 1$. The Hamiltonian of a given state $s$ in the original description is given by
\begin{equation}
 H\left( s \right)=-J \sum_{\langle i,j\rangle}s_i s_j~,
\end{equation}
where we take the interaction to be ferromagnetic $J>0$ and from now set $J=1,$ $k_B=1.$ The partition function of this system at finite temperature $T$ is given by
\begin{equation}\label{2D-SL-Ising_PartitionFunction}
Z\left( \beta \right) = \sum_{s}e^{-\beta H\left( s \right)}\,,
\end{equation}
where $\beta=1/T.$ The duality in this Ising model is as follows. The partition function $Z\left( \beta \right)$  of the above system is related to that of another system at a dual temperature $\tilde{\beta}=-\frac{1}{2}\ln \tanh \beta$ by the dependency
\begin{equation}
Z\left( \beta \right) = \frac{1}{2}\left( \sinh (2 \tilde{\beta}) \right)^{-N}\sum_{\sigma}e^{-\beta H\left( \sigma \right)}\,,
\end{equation}
where the dual spins $\sigma_i$ also take values $\pm 1$ on a lattice with the same geometry, and the dual system shows the same coupling strength $J$. This is known as the Kramers-Wannier duality~\cite{Kramers:1941kn,Kramers:1941zz} which relates a description at low temperature with long-range correlations (strong coupling) and high temperature with short range correlations (weak coupling).\footnote{The fact that both partition functions describe the same 
type of Ising model implies the existence of a critical temperature $\beta_{\mathrm{crit}}\approx 0.4407$ at which a transition between ordered and disordered phases occurs.}

\begin{figure}[t]
\includegraphics[width=0.49\textwidth]{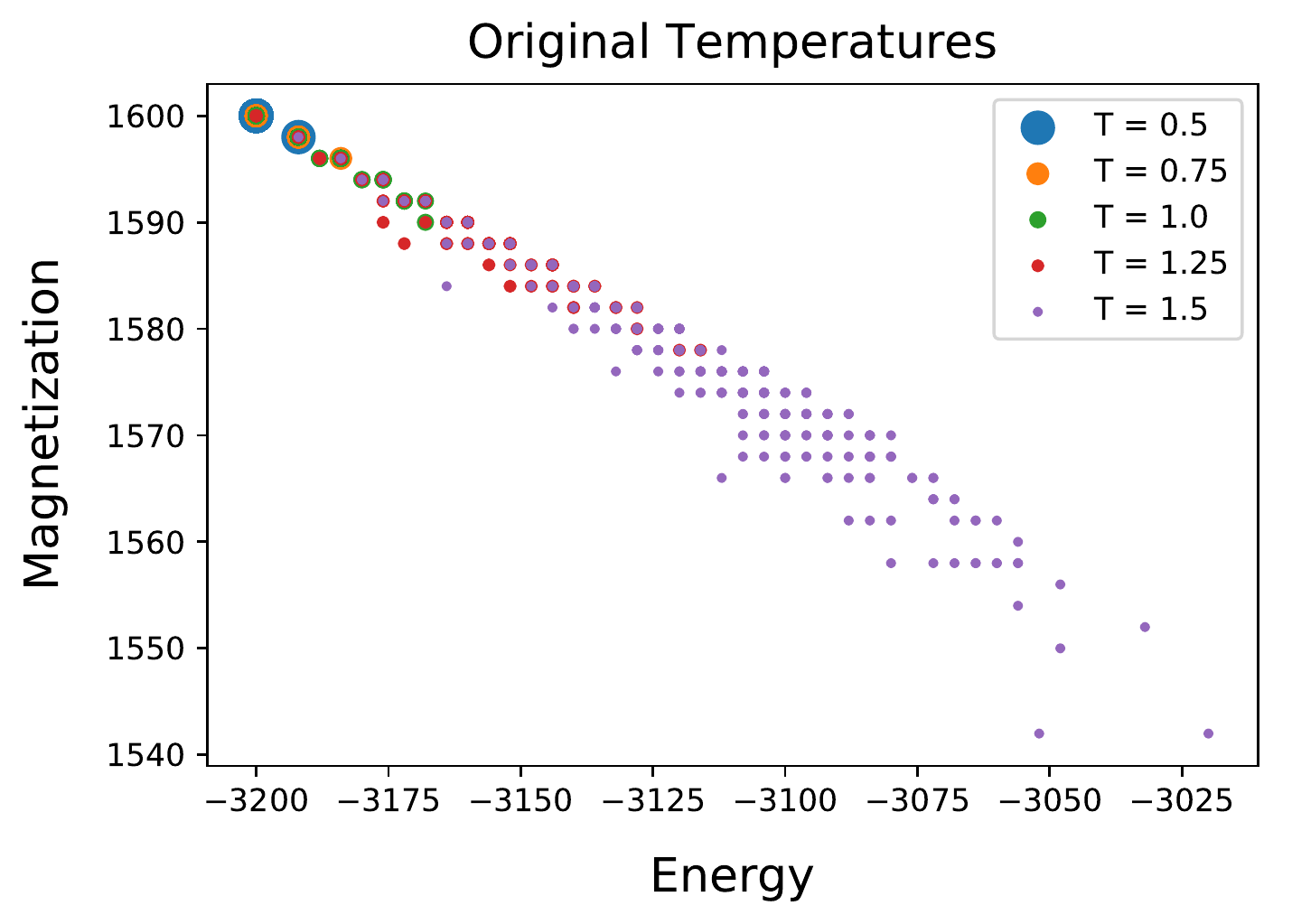}
\includegraphics[width=0.49\textwidth]{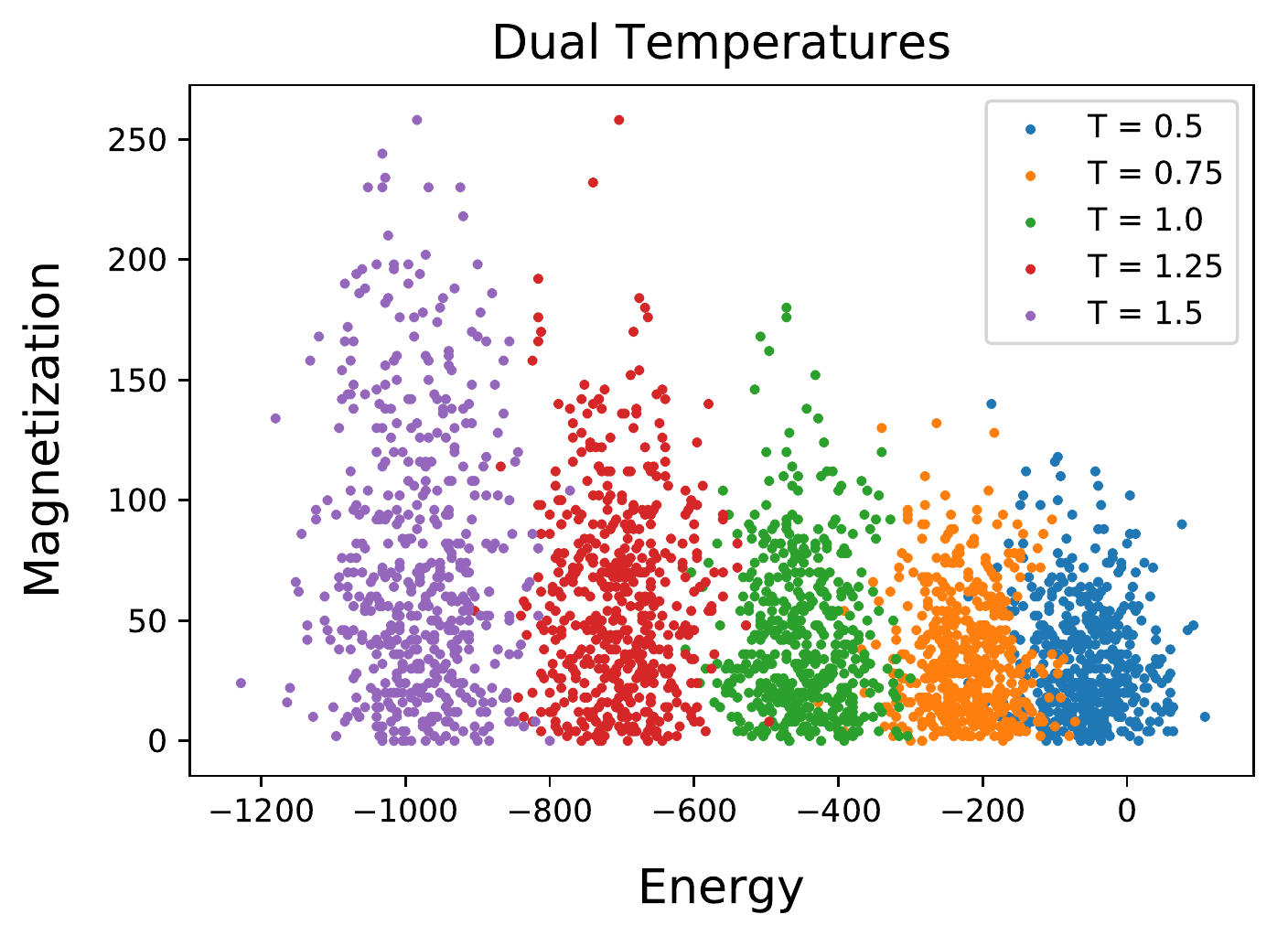}
\vspace{-0.5cm}\caption{Distribution of energies and magnetizations of a square-lattice Ising model for various temperatures and their duals. \label{fig:overlap1}}
\end{figure}

\begin{figure}[t]
\includegraphics[width=0.49\textwidth]{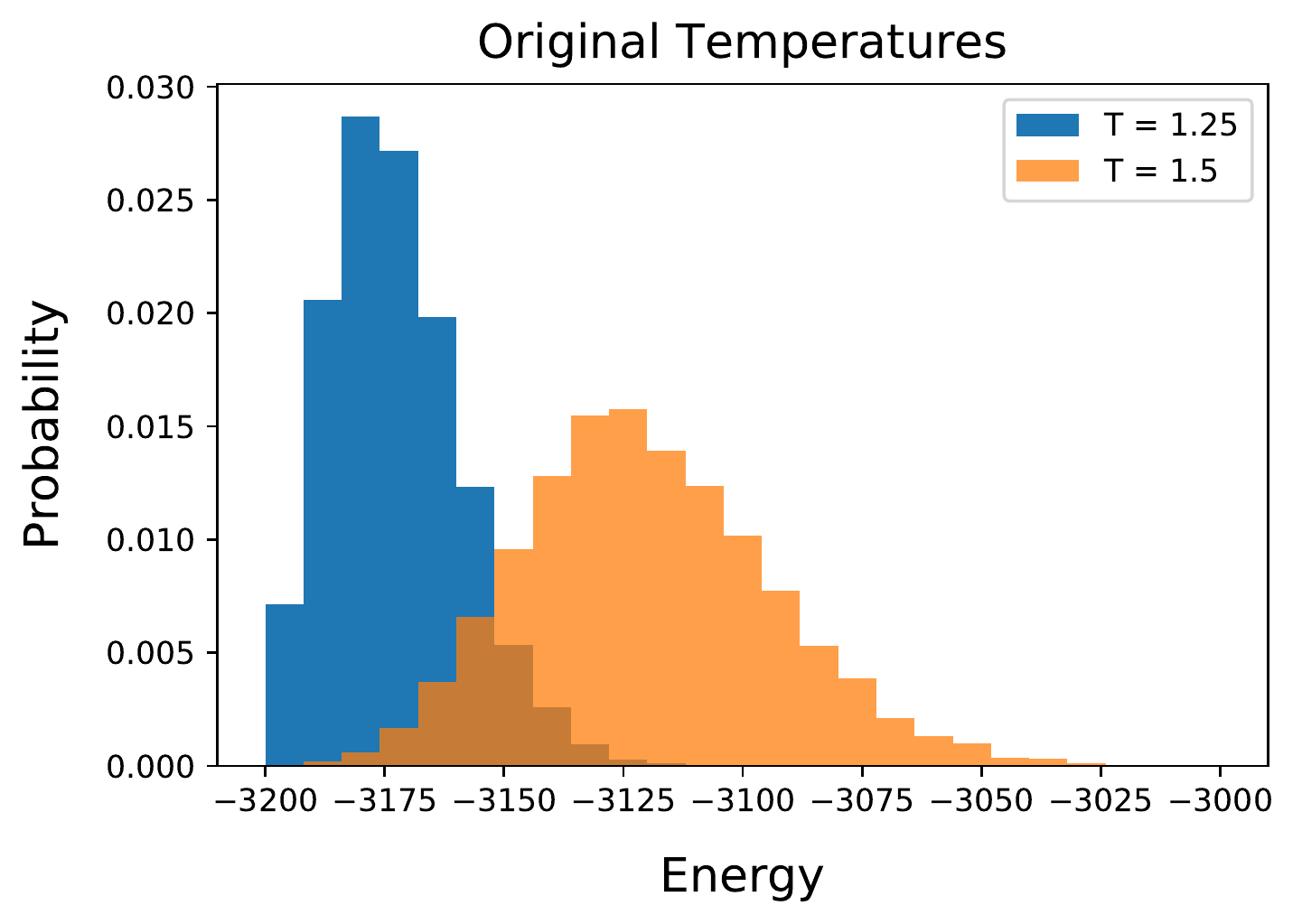}
\includegraphics[width=0.49\textwidth]{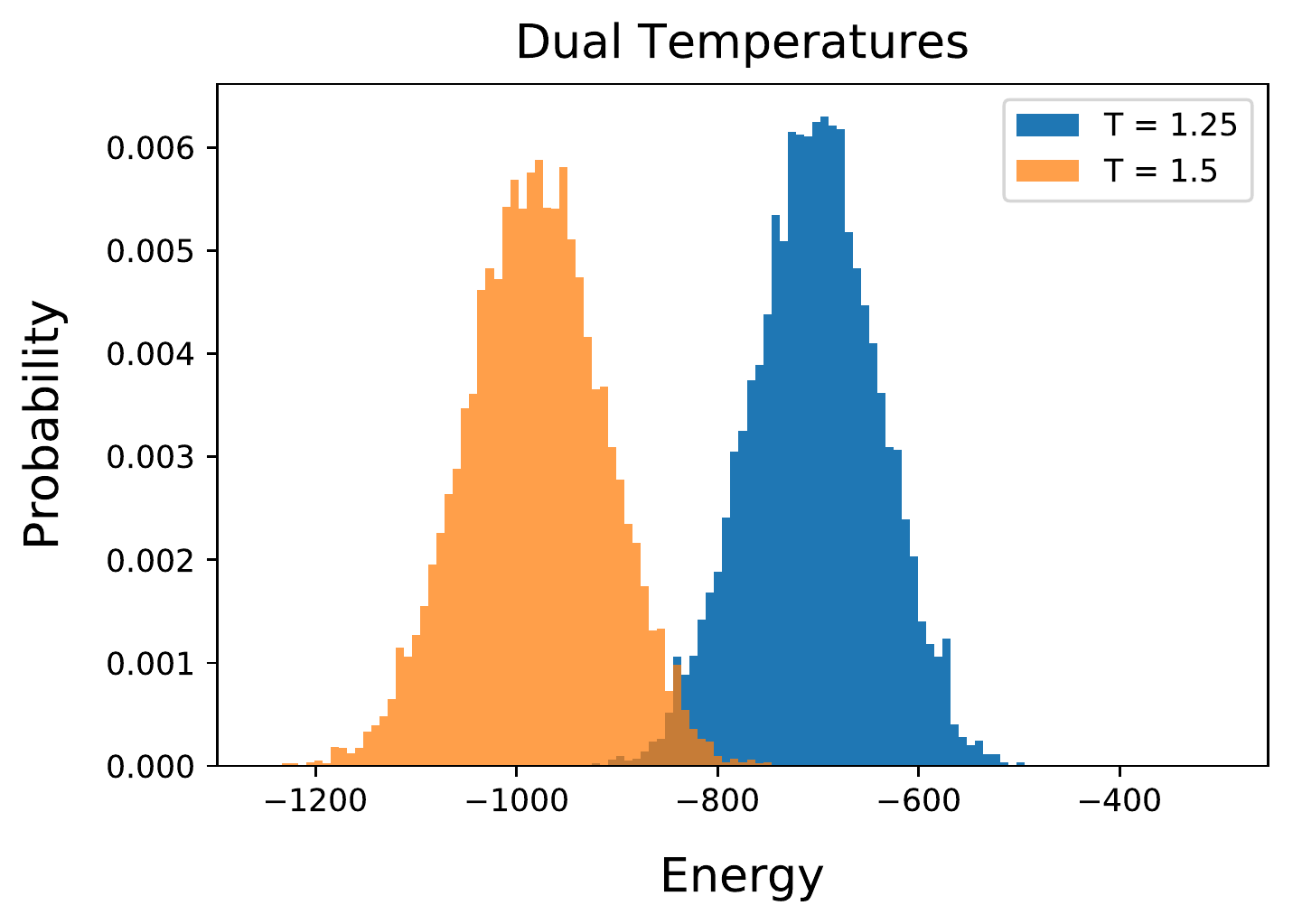}
\vspace{-0.4cm}\caption{Energy distributions of the square-lattice Ising model for $T=1.25, 1.5$ and their
respective dual temperatures. The energies of the original representation concentrate on a very small
region and show a significant overlap. Both diagrams use bins of the same width respectively. \label{fig:overlap2}}
\end{figure}
\subsubsection*{Classification of Temperature}
{\it When is it useful to use the high temperature and when is it useful to use the low temperature phase?}
Similar in spirit to the Fourier case, we start with a classification task. In particular we are interested in predicting which temperature a sample is drawn from. Our experimental setup is as follows: We considered a square-lattice Ising model on a $40\times 40$ lattice at temperatures $T=0.25, 0.5, \dots , 2.25$ and their corresponding dual temperatures. The dataset for each temperature was split into 16000 training samples and 4000 test samples. Networks  were then trained to classify states drawn from two datasets according to the respective temperature of the set they were drawn from (binary classification). We chose as architecture a simple convolutional neural network consisting of one $2\times 2$-convolutional layer with 8 filters and ReLU activation followed by a linear layer with sigmoid activation. The overall performance did not change significantly when increasing the number of layers to up to five and varying the umber of filters between $8,~12,~16,$ and $32.$ Weights were initialised randomly; training was performed using standard Nesterov Adam optimiser with initial learning rate $0.002$ and learning rate decay. No significant changes in performance were observed after a maximum of $200$ training epochs.

Dataset generation and training was performed for ten different seeds to prevent outliers in performance from distorting the results. The best test set accuracies reached after 200 epochs were then averaged over the ten test-runs. The average best test set accuracies for various pairs of temperatures are shown in Figure~\ref{fig:Ising2D_OriginalDual_Performance}.

As can be seen, the classification performance improves substantially when performed for the dual temperatures. This can be seen when visualising the energy and magnetisation for both representations, cf.~Figure~\ref{fig:overlap1}. An example of the overlap in the energy distributions for temperatures $T_1=1.0$ and $T_2=1.25$ is shown in Figure~\ref{fig:overlap2}. The correlation with the classification performance and the overlap of the energy distributions is shown in Figure~\ref{fig:Ising2D_OriginalDual_Performance}.

Further uses could be looked for in determining other correlation functions. In particular, we investigated several disorder correlation functions, e.g.~correlators of the type $\langle \sigma_i \sigma_j\rangle.$ However, as the performance difference between the two representations are not as dramatic as in the temperature classification we leave a detailed discussion of these correlators to the future.

\subsection{1D Ising Models}
\label{sec:1DIsing}
Other lattice systems offer different types of dualities, and here we present an example where the dual representation features a different Hamiltonian, i.e.~there is no self-duality of the same system. Simple examples of this type of duality are given in the context of one-dimensional Ising models on a finite spin-chain with $N$ spins, $n$-spin interactions and free boundary conditions. A discussion of such systems can be found for instance in~\cite{2016JPhA...49I5002T}, and we summarise here the important system properties for our sub-sequent analysis.

For $n$-spin interaction models, the Hamiltonian $H(s_1,\ldots,s_N)$ takes the form
\begin{equation}
H\left( s \right) =- J\sum_{k=1}^{N-n+1}\prod_{l=0}^{n-1}s_{k+l} -  B\sum_{k=1}^{N}s_k\,.
\end{equation}
The free boundary conditions are to be understood in the sense that one considers only interactions of $n$-spin chains which can be fully embedded into the system $(s_1, s_2,\dots s_N)$, and there are no identifications or interactions connecting both ends of the chain. Furthermore, there do not exist any relations which fix the values of boundary (or other) spins to specific values.

\begin{figure}[t]
\begin{center}
\includegraphics[width=0.8\textwidth]{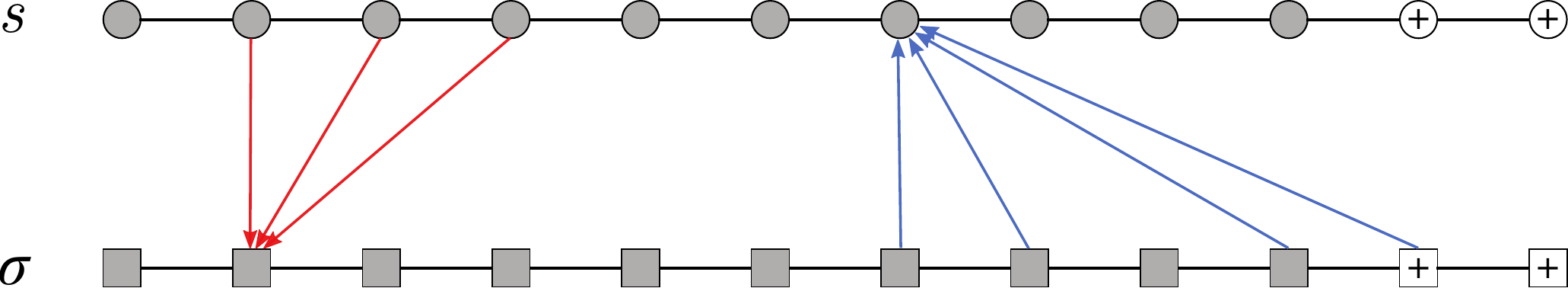}
\end{center}
\vspace*{-12pt}
\caption{Structure of the duality mappings \eqref{1DIsing_DualityTransformation} (red) 
 and the inverse duality mappings \eqref{1DIsing_DualityTransformation_inverse} (blue)
 for $N=10$ and $n=3$. White spins with  ``$+$"-sign inside indicate ghost spins
with fixed value 1.} 
\label{fig:Ising1D_Dualities_n3}
\end{figure}

Let us now consider the special case of a purely interacting theory with $B=0$. The Hamiltonian 
then reduces to 
\begin{equation}
\label{Ising1D_Hamiltonian}
H\left( s \right) =- J\sum_{k=1}^{N-n+1}\prod_{l=0}^{n-1}s_{k+l}\,,
\end{equation}
This can be bijectively mapped to a non-interacting theory with external field $J$ and Hamiltonian 
\begin{equation}
\label{Ising1D_DualHamiltonian}
H\left( \sigma \right) =  -J\sum_{k=1}^{N-n+1}\sigma_k\,.
\end{equation}
The corresponding duality transformation exchanges the roles of the spins and their interaction terms,
\begin{equation}
\label{1DIsing_DualityTransformation}
\sigma_k = \prod_{l=0}^{n-1}s_{k+l}, \hspace{50pt} k=1,\dots N\,,
\end{equation}
where spins $s_{l}$ with $l>N$ are to be understood as ghost spins taking the fixed value $1$. The inverse
transformation is given by 
\begin{equation}
\label{1DIsing_DualityTransformation_inverse}
s_{k}= \prod_{r=0}^{q}\sigma_{k+rn}\sigma_{k+rn+1}\,,
\end{equation}
where $q$ is to be chosen as the maximum value such that $k+qn \leq N$ and one again introduces a ghost
spin $\sigma_{N+1}=1$ (further ghost spins can be introduced to generate representations of the same dimension, but they do not play any role in the inverse transformation). 
\begin{figure}[t]
\begin{center}
\hspace{-10pt}
\includegraphics[width=0.39\textwidth]{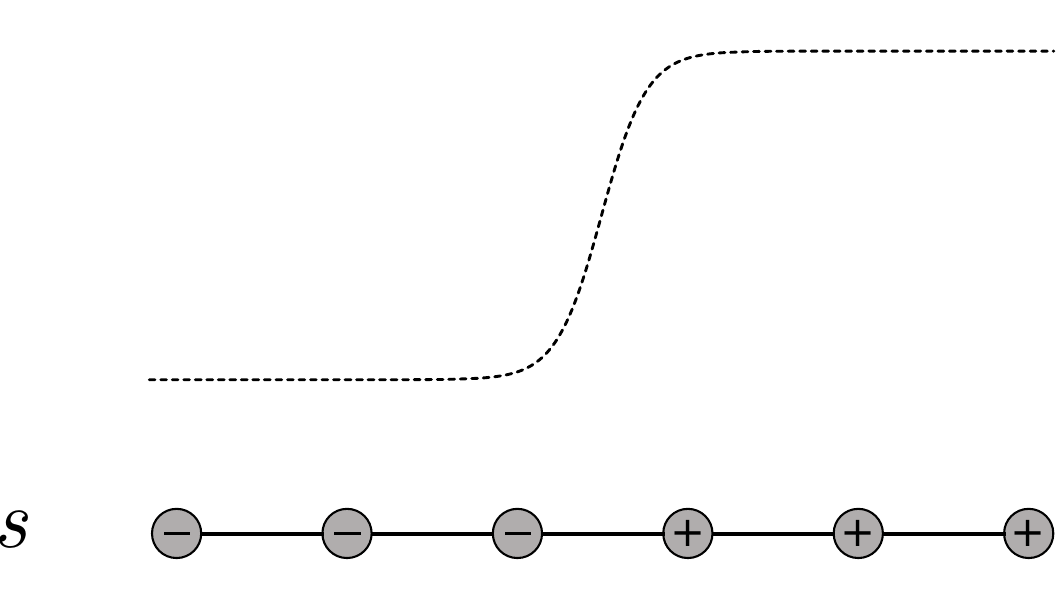}
\hspace{40pt}
\includegraphics[width=0.39\textwidth]{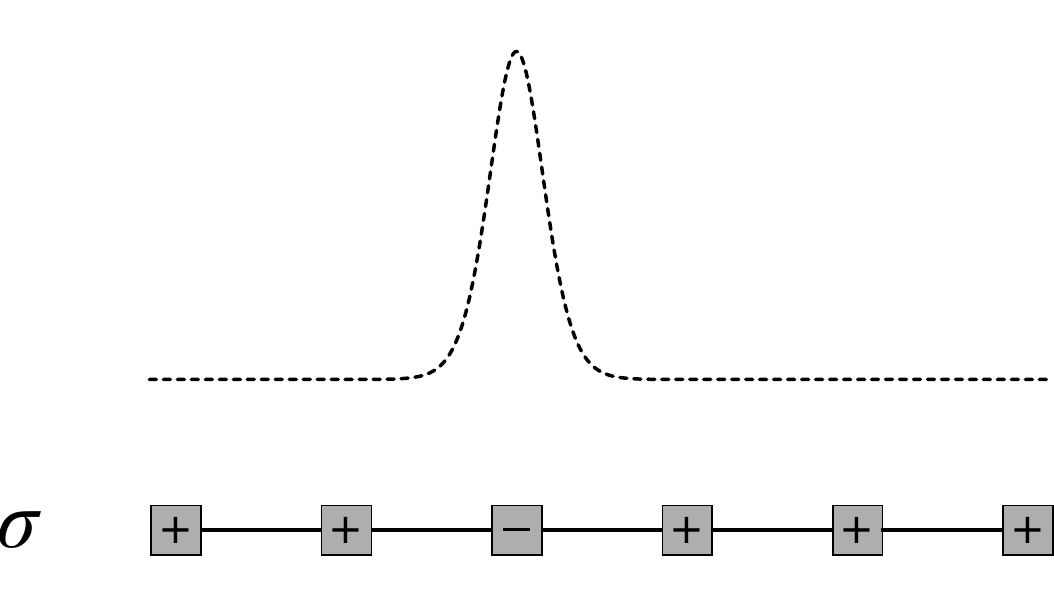}
\end{center}
\vspace{-0.8cm}\caption{Comparison of spin configurations in a two-spin interaction model and a scalar field kink. Dual spins
located on the interaction links represent the energy distribution of a ``kink" in the spin model.}
\label{fig:1DIsing_KinkVariables}
\end{figure}
For $n$-spin interactions, the product runs over pairs of adjacent spins, starting from the 
position $k$ and skipping $n-2$ spins between the individual pairs. The involvement of spins in the duality
transformation~\eqref{1DIsing_DualityTransformation} and its inverse~\eqref{1DIsing_DualityTransformation_inverse}
is exemplified in Figure~\ref{fig:Ising1D_Dualities_n3} for the case $N=10$ and $n=3$. Notice that this can be considered a direct generalisation of the special case $n=2$, for which the the duality transformation
corresponds to an exchange of roles between the original spins and their kink variables (cf.~Figure~\ref{fig:1DIsing_KinkVariables}).

\subsubsection*{Identifying (Meta-)Stable States}

{\it Which task is more easily addressed in the dual representation?} 
A simple example for this would be to compute the total energy of a given spin configuration $(s_1,\ldots,s_N)$, which can involve high-order
products in the original frame and simplifies to summing over the first $N-n+1$ spins in the dual frame. Of course, 
this is more of an ad-hoc example since the duality transformation by construction computes the local energy
contributions.

Generally speaking, there exist more sophisticated tasks where no such hand-crafted frame can be constructed. These tasks also can be drastically simplified by applying duality transformations known from or learned in a different context.

One such instance is the detection of states $s$ which are
(meta-)stable with respect to single-flip spin dynamics. Such single-flip stable states are defined as
configurations for which flipping any of the spins causes the energy of the system to increase.\footnote{Such metastable states can cause standard MCMC-algorithms to be trapped in a local minimum as the temperature approaches zero and is a major reason why the performance of common simulation algorithms tends to deteriorate at low temperatures.} 

\subsubsection*{Effect on Simple Networks}
In order to get an idea whether the duality \eqref{1DIsing_DualityTransformation} is a viable tool to improve the classification of metastable states, we choose as a first benchmark how ``simple" architectures of neural networks 
can handle this classification problem and whether transforming our variables to the dual frame can
improve their performance. While, in practice, any improvement from utilising the dual frame might also be achieved by using more sophisticated
architectures, this setting nevertheless serves as an important first step. A positive result justifies a further scrutinising whether the same principles also hold for tasks which state-of-the-art models fail to solve.

Since the duality transformations \eqref{1DIsing_DualityTransformation} are themselves highly nontrivial from 
the perspective of computational complexity, some caution is needed here to prevent distorting our results by
limitations arising from a mere lack of capacity. Taking, for instance, our toy-example of energy regression, 
it is clear that the task cannot be solved by a linear network in the normal frame, while even a simple perceptron
with sufficiently high number of neurons can do so at ease. In this case, the only benefit coming from using the
dual frame thus lies in a lower network complexity, which is, however,  in parts nullified  by the computational
complexity of the duality transformation itself.

Taking this into account, we chose a suitable benchmark for our tests a single-layer perceptron with 128 hidden neurons,
ReLu activation for the hidden layer and sigmoid activation for the output layer. This architecture shows a sufficiently-high
capacity to easily learn the transformation \eqref{1DIsing_DualityTransformation} directly, while at the same time keeping 
a relatively simple structure. 

We generated all $2^{18}$ states for the 1D Ising chain
with $N=18$ spins and tested different networks for varying $n$. We 
split the data into states labeled as ``not (meta-)stable" (0) or ``(meta-)stable"~(1) and normalised the training and test
sets to contain an equal number of samples for each class. We furthermore checked the performance for varying amounts
of training data in order to properly analyse effects on generalisation errors and data efficiency.

The average best test accuracies and losses achieved in 10 training runs of 500 epochs are listed in Table~\ref{table:val_acc_1DSimpleNets_original}.
Average training curves for the case $n=8$ and varying amounts of training data can be found in 
Figure~\ref{fig:Ising1DN18n8_TrainingCurves}. Further details on the training and
testing modalities are discussed in Appendix \ref{app:1DIsing}.

\begin{figure}[t]
\includegraphics[width=0.33\textwidth]{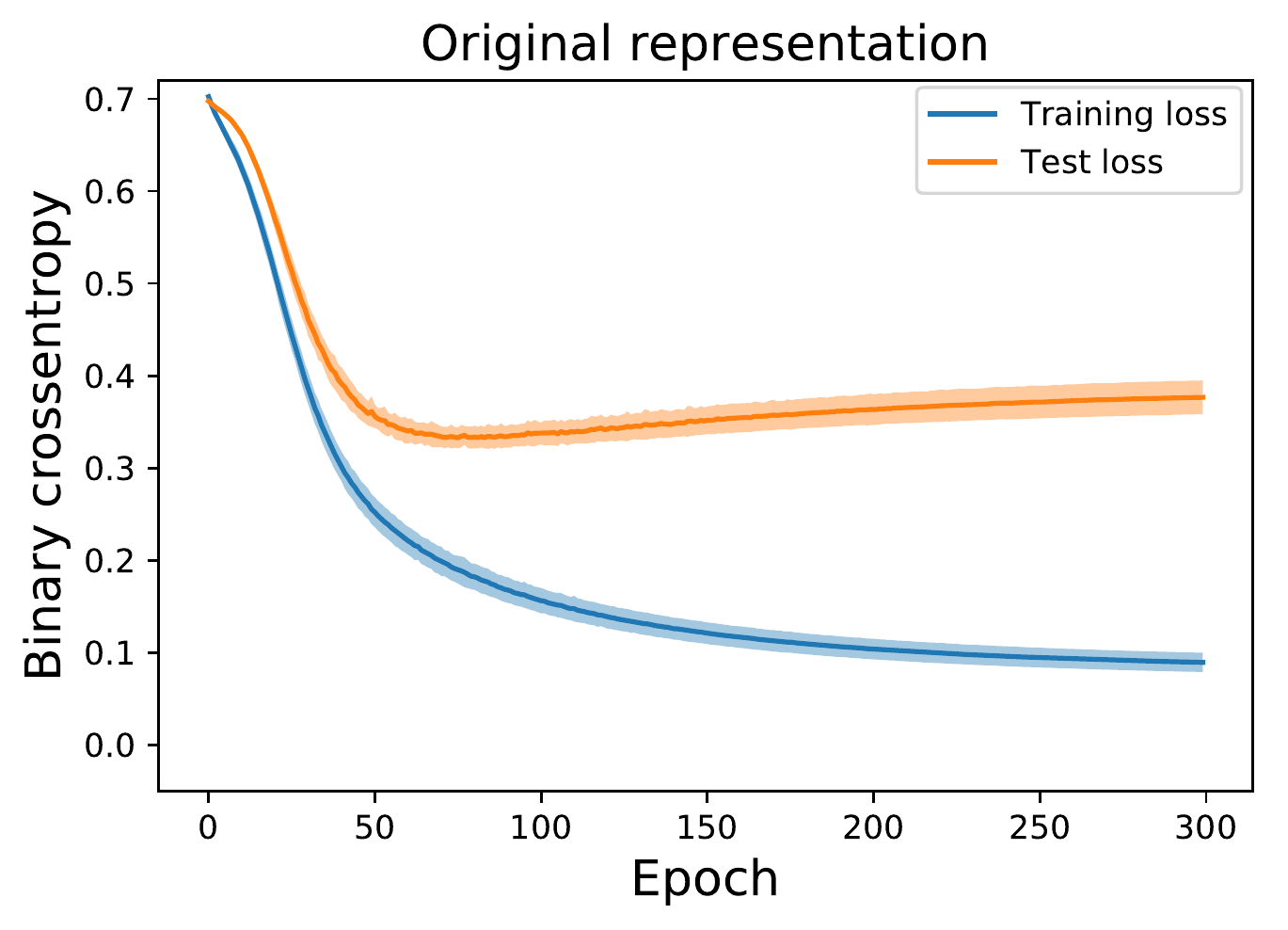}
\includegraphics[width=0.33\textwidth]{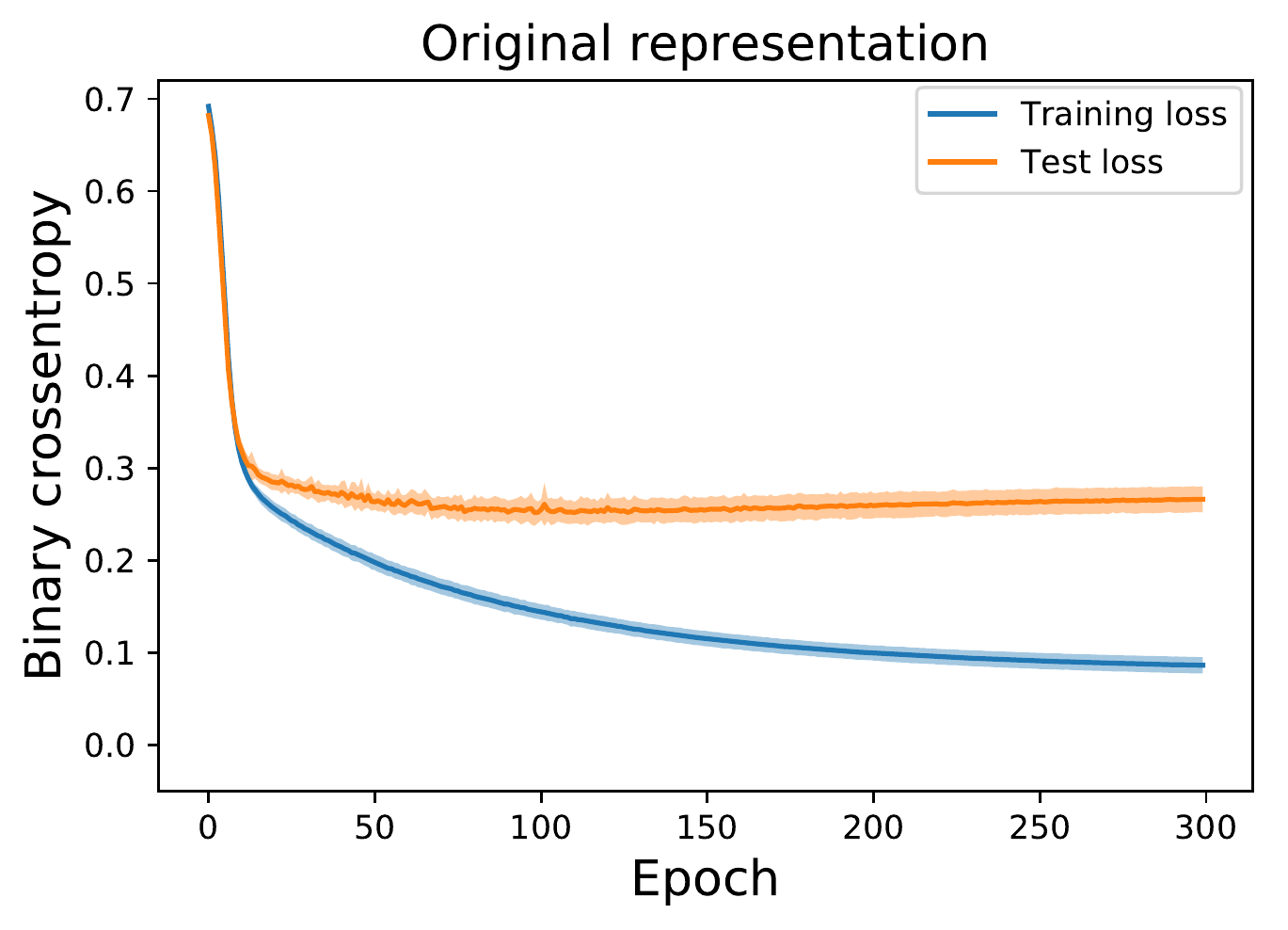}
\includegraphics[width=0.33\textwidth]{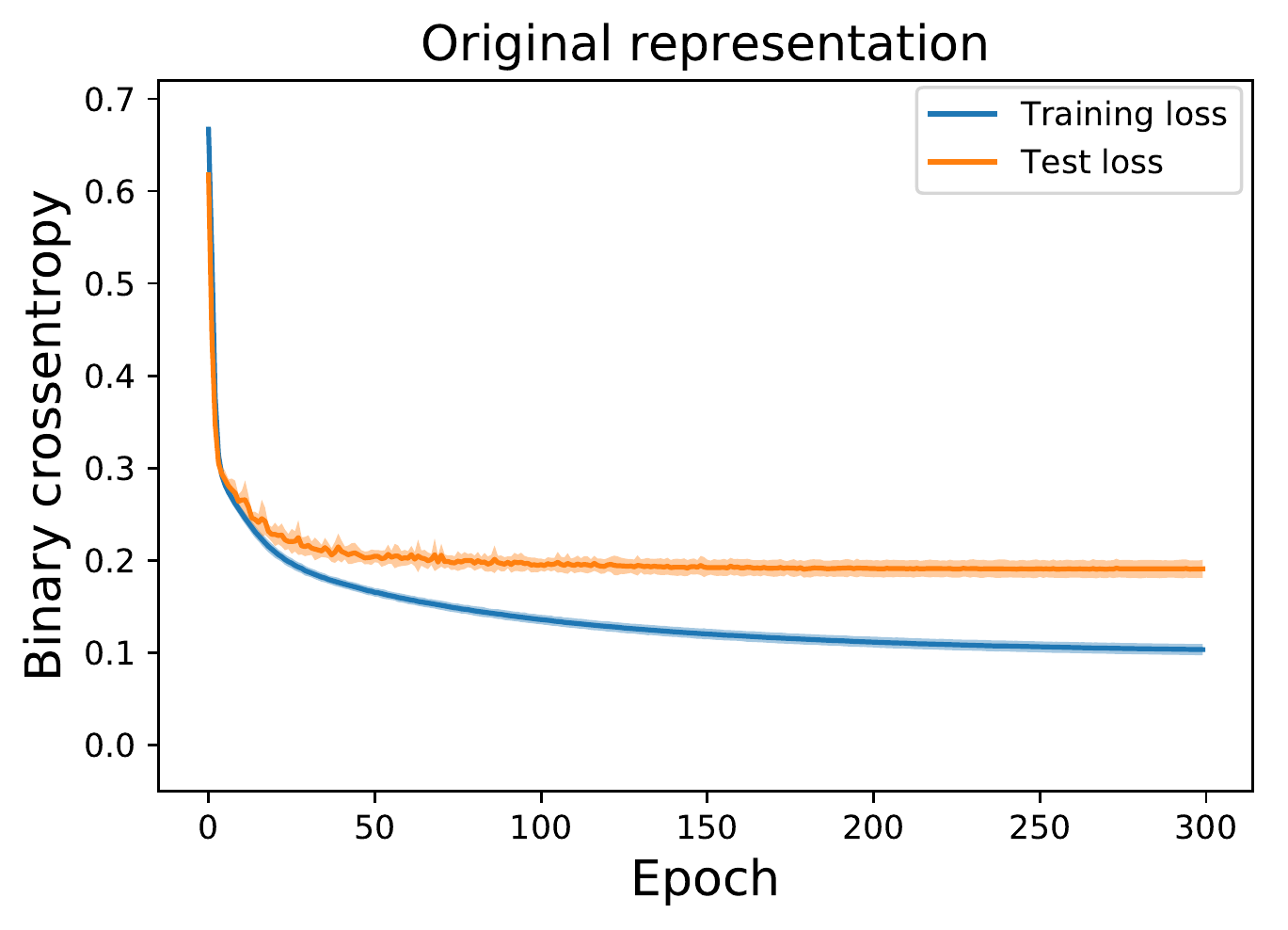}\\
\includegraphics[width=0.33\textwidth]{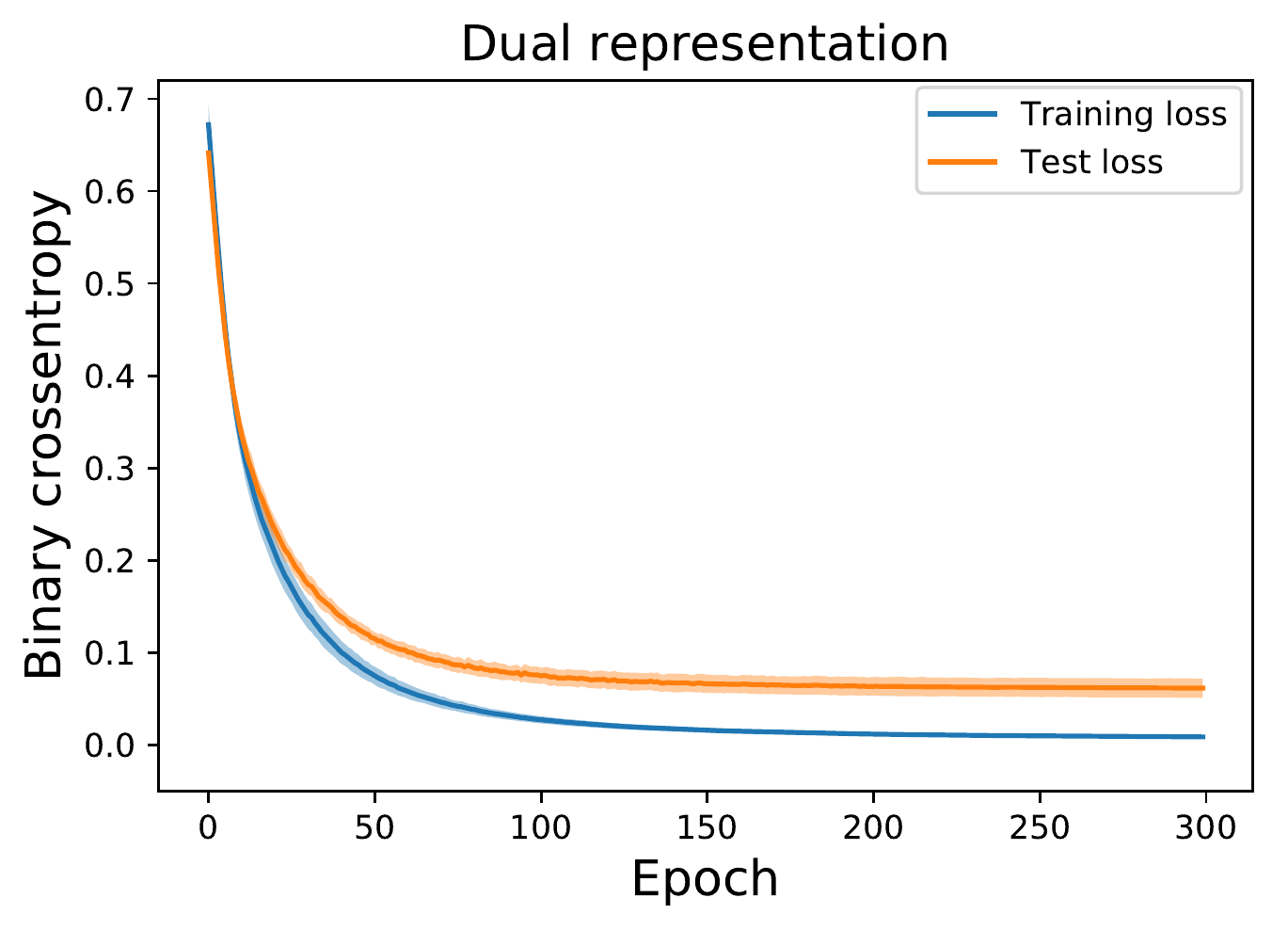}
\includegraphics[width=0.33\textwidth]{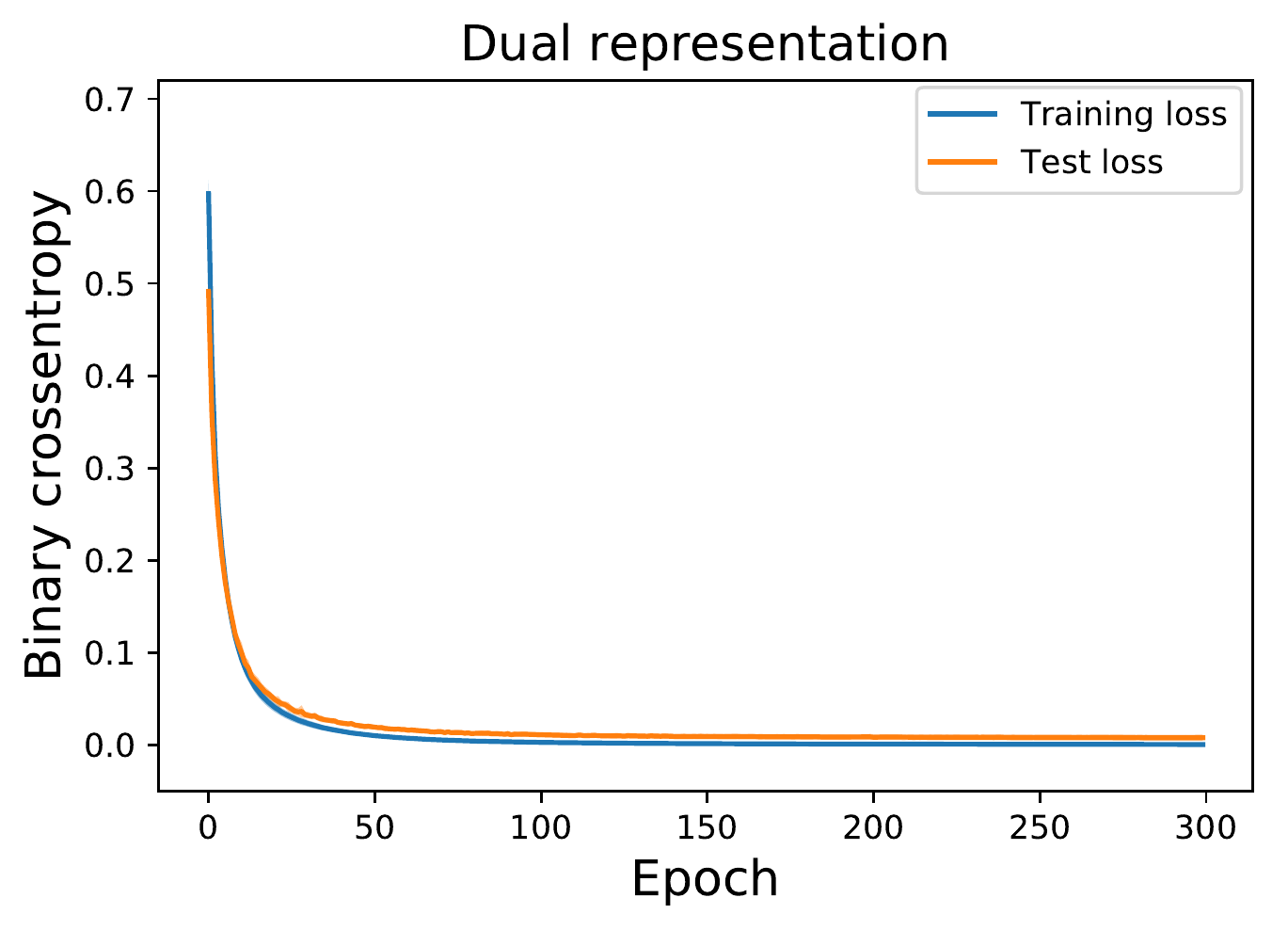}
\includegraphics[width=0.33\textwidth]{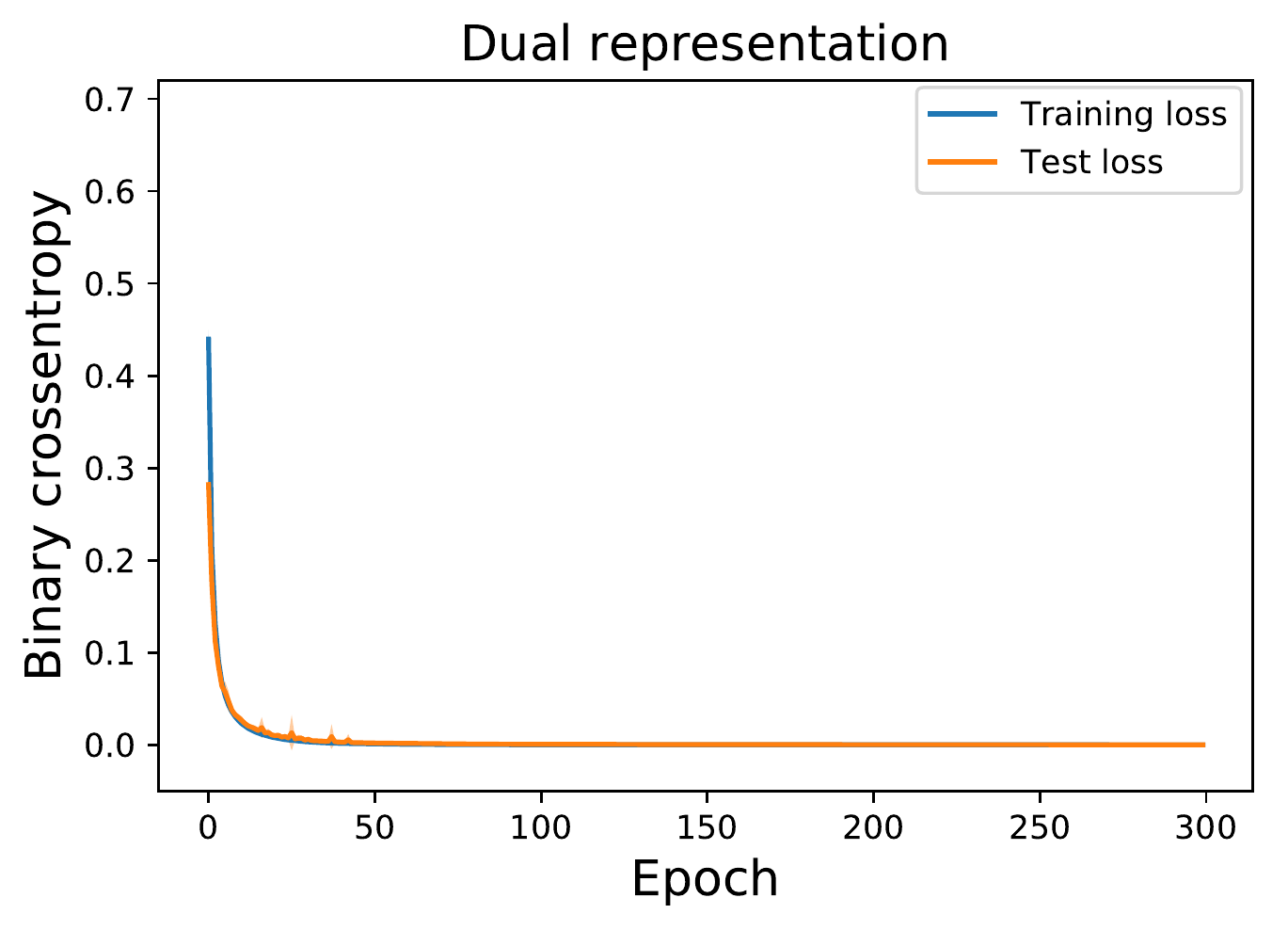}
\vspace*{-20pt}
\caption{
Example histories of training loss (blue) and test loss (orange) over the course of 300 epochs for $n=8$ and various
numbers of training samples. The plots show averaged curves computed over ten test-runs; standard deviations are indicated
with shaded colours.
}
\label{fig:Ising1DN18n8_TrainingCurves}
\end{figure}

\begin{table}[t]
\begin{footnotesize}
\begin{center}
 \begin{tabular}{c||ccccc}
 normal &  n=4 &  n=5 &  n=8  & n=9  & n=12  \\
 \hline\hline
 $6\cdot 10^2$ & 0.9113 &  0.8688  &  0.8788  &  0.8813 & 0.8803 
 \\
$3\cdot 10^3$ & - &   0.9243  &   0.9215  &  0.9223  & 0.9295  
 \\
$9.5\cdot 10^3$ & - &   -  &   0.9424  &  0.9475  & 0.9739 
 \end{tabular}\qquad 
 \begin{tabular}{c||cccccc}
 dual &  n=4 &  n=5 &  n=8  & n=9  & n=12  \\
 \hline\hline
 $6\cdot 10^2$ & 0.9911 & 0.9783  &  0.9819  &  0.9855 &  0.9909 
 \\
$3\cdot 10^3$ & - &   0.9958  &   0.9977  &  0.9994  & 1.0000  
 \\
$9.5\cdot 10^3$ & - &   -  &   1.0000  &  1.0000  & 1.0000  
 \end{tabular}
\end{center}
\end{footnotesize}
\vspace*{-10pt}
\caption{Detection of (meta-)stable states in the 1D Ising chain for different interactions and amounts of training data.
The listed numbers describe the average best test accuracy over 10 training runs of 500 epochs each.
Missing values indicate that the number of required samples exceeds the total number of metastable states for the considered setting. On the left are the results for the normal variables, and the right side shows the results for the dual variables.}
\label{table:val_acc_1DSimpleNets_original}
\end{table}

\subsubsection*{Results}

The results show that there is indeed a major improvement of performance in the dual representation. While all networks
are able to detect at least some patterns in either frame, we find several advantages from using the dual representations:
\begin{itemize}
\item{The best performance achieved for low numbers of training samples is notably higher in the dual representation,
implying that the duality transformation \eqref{1DIsing_DualityTransformation} can be useful to prevent overfitting
and improve data efficiency.}
\item While increasing the amount of training data gradually tightens the performance gap between the original and dual
representations, the learning curves in the latter remain much steeper in all cases, leading to shorter and more stable training. 
\item Even in cases for which the best test accuracies are high in both representations, there remains a significant 
difference in the actual binary cross-entropy,
\begin{equation}
\begin{split}
\mathcal{L}=-[y_{\textrm{true}}\textrm{log}(y_{\textrm{pred}})+(1-y_{\textrm{true}})\textrm{log}(1-y_{\textrm{pred}})]\,,
\end{split}
\end{equation}
implying that networks trained on the dual representation perform classifications  
with a considerably higher degree of certainty. This is also reflected in the model outputs, which are commonly closer
to 0 or 1 in the dual representation than in the original variables, even in settings with high test accuracies in both representations (cf. Figure~\ref{fig:Ising1D_NetworkOutput}).

\begin{SCfigure}
%\begin{center}
\includegraphics[width=0.55\textwidth]{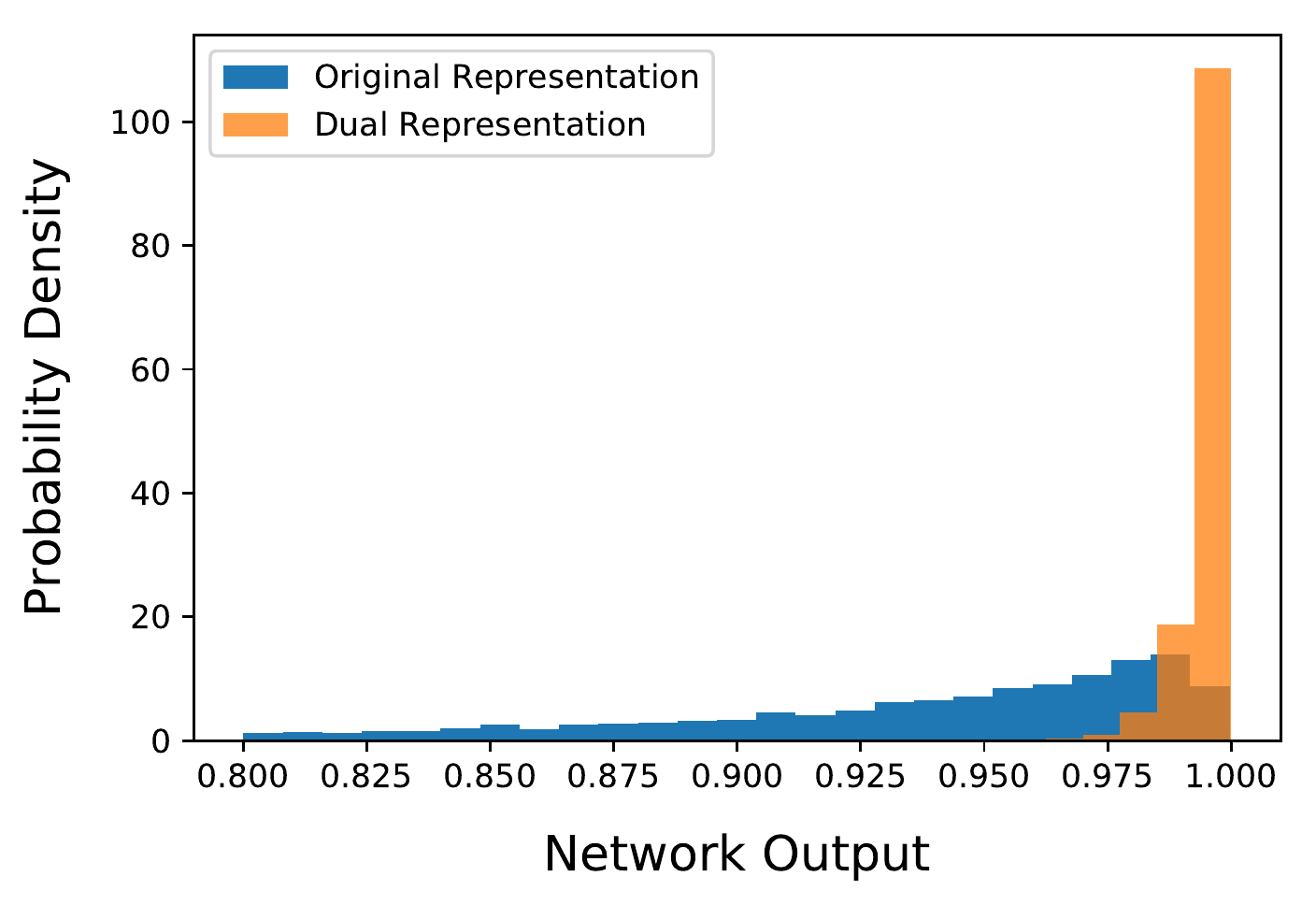}
%\end{center}
\caption{Output distribution of simple neural networks for states classified as \mbox{(meta-)stable} for $N=18$ and $n=8$. Both networks were trained on 3000 samples. Only values for the dual representation accumulate very close to one, implying a higher degree of certainty in this frame. \label{fig:Ising1D_NetworkOutput}} 
\end{SCfigure}
\item{While overfitting is prevalent in the original representation, the loss curves additionally show signs of underfitting.
This can be remedied by increasing the capacity of the network, which, however, leads to even stronger overfitting. We found 
that regularization techniques can slightly improve performance in this case, however, there remained a significant difference
between both representations for all tested methods. Details on this are discussed in Appendix~\ref{app:1DIsing}.}
\end{itemize}

\subsubsection*{Interpretation}

\begin{figure}[t]
\begin{center}
\includegraphics[width=0.9\textwidth]{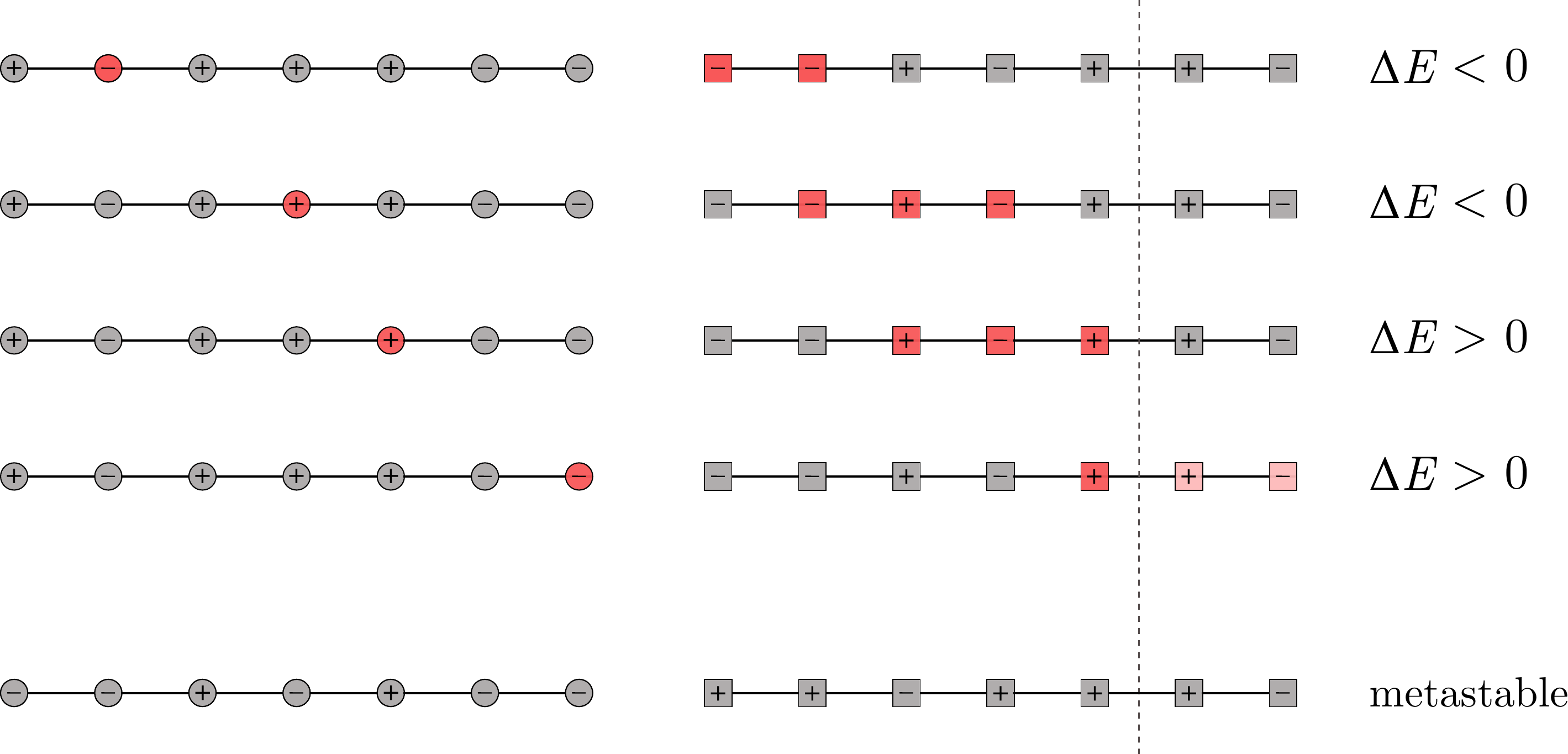}
\end{center}
\vspace{-0.8cm}
\caption{
Single-spin flip dynamics and metastability in the normal and dual representation.
{\bf Top:}~Flipping a single spin $s_i$ in the normal representation (left) causes $n$ dual spins $\sigma _j$ with $j=i-n+1,\dots, i$
to change sign (right), as indicated by red color for the case $n=3$. The overall energy increases iff more than half of the involved
dual spins  have positive sign (counting only values $j$ with $1\leq j \leq N-n+1$). {\bf Bottom:} Example of a metastable state for $n=3$ in the normal (left) and dual (right) representation. Flipping any of the spins in the original representation causes the overall energy
of the state to increase.}
\label{fig:1DIsing_Dualities_n3_Metastability}
\end{figure}

Some sense can be made out of this result when addressing the problem from a naive analytical viewpoint. In the original representation,
checking whether flipping a particular spin $s_{i}$ increases the total energy of the system requires taking into account
$n$ interaction terms containing  $s_{i}$, some of whose contributions might cancel each other. On the other hand,
these $n$ interaction terms are represented by a cluster of $n$ spins $\sigma_{j}, \, j=i-n+1,\dots ,  i$ in the dual frame,
and flipping $s_{i}$ causes all of those $n$ dual spins to change sign. Since the total energy of the system can
be computed by simply adding up the first $N-n+1$ dual spins of the complete system,
an overall increase in energy then occurs precisely iff more than half of the flipped  dual spins take the value $1$ (not counting those spins
$\sigma_{j}$ with $j\geq N-n$) . 
In other words,  the transformation~\eqref{1DIsing_DualityTransformation} maps the single-flip dynamics of the original
system to $n$-spin-cluster dynamics in the system governed by the Hamiltonian~\eqref{Ising1D_DualHamiltonian}, thus creating 
a ``dual task" which is considerably easier to learn for neural networks. An illustrative example for the case 
$N=10$ and $n=3$ is given in Figure~\ref{fig:1DIsing_Dualities_n3_Metastability}.

\subsubsection*{Discussion and Limitations}

There are several important aspects as well as limitations of the considered experimental settings, which shall be briefly commented on in this subsection.

\begin{SCfigure}
%\centering
\resizebox{0.55\textwidth}{!}{%
\includegraphics[width=300pt]{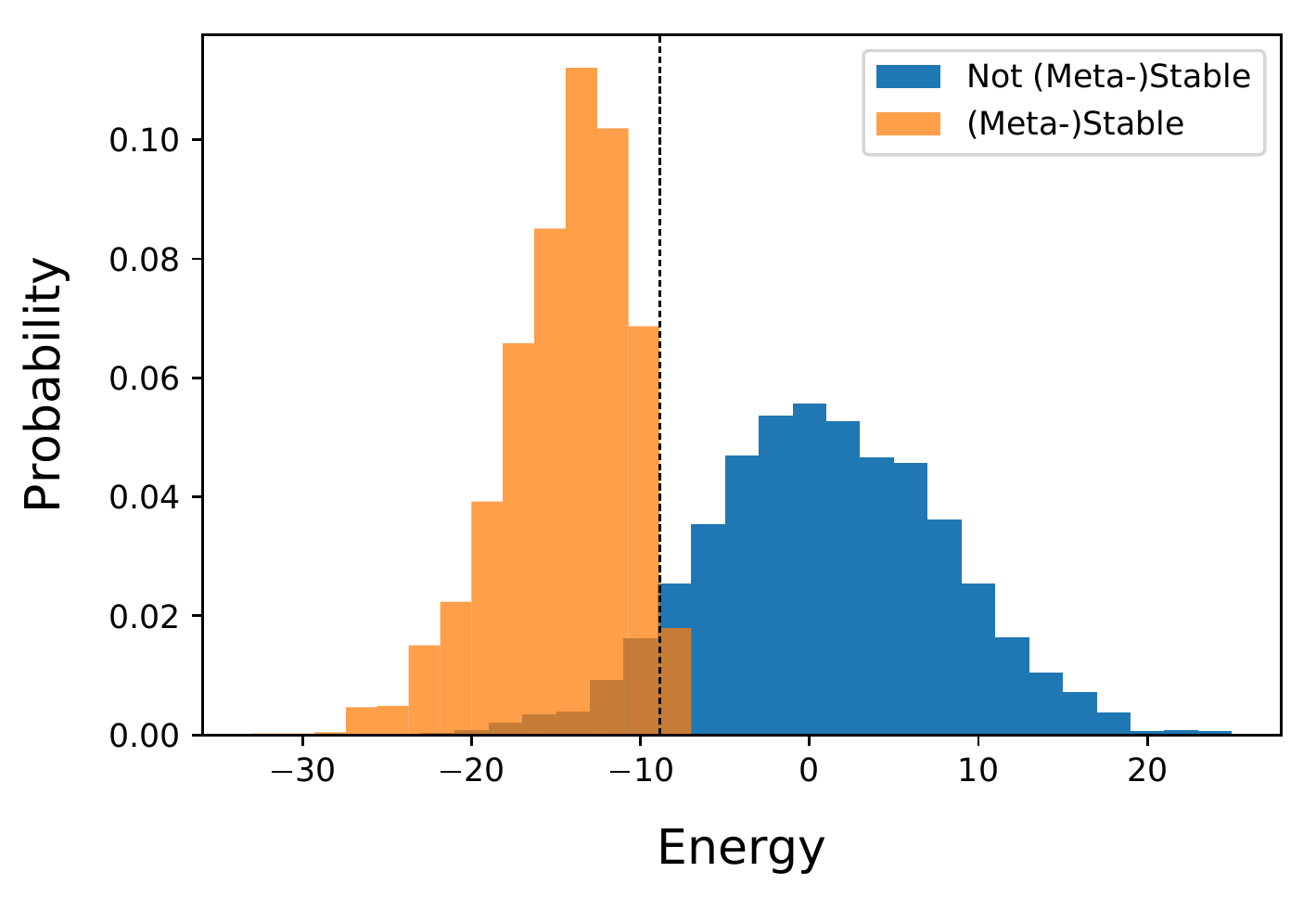}
}
\caption{Energy distributions of normal and (meta-)stable states for $N=100$ and $n=50$ (choice for illustrative reasons). Relatively
high accuracies can be obtained by choosing a fixed energy cutoff for classification (dashed line).} 
\label{fig:Metastability_Energy_schematic2}
\end{SCfigure}

\begin{itemize}

{\item{\bf Modifications to Setup}

A first important point to remark is that the discussed setting describes a very low number of spins and is therefore to be understood as a toy model. While a large-scale simulation of realistic systems is beyond the scope of this work, it is worth mentioning that we found
a more drastic difference in performance as more complex settings such as $N=100$ and $n=50$ were considered.
This commonly led to pure guessing on the original data, whereas accuracies higher than
0.95 could be reached with as few as 1500 training samples in the dual representation. The benefits of dualities might thus extend beyond simple
toy-settings, however, further testing is required to confirm this.}

{\item{\bf Sensitivity to Architecture}

While the above tests were performed for a rather large number of different systems and training set sizes, defining a clear benchmark naturally required the utilization of a fixed model to test performance. In light of this, a natural question is to which degree the improvement is owed to the choice of architecture, and whether the results remain valid if a wider class of architectures is considered. We therefore checked the effect of various modifications on our results, as described in more detail Appendix \ref{app:1DIsing}.

We found that, except for strong results of convolutional neural networks on very simple systems with $n\leq 4$,
none of the above modifications led to a significant change in the overall results. It cannot be excluded that a similar improvement in performance
can alternatively be obtained by more sophisticated network architectures. However, our tests clearly demonstrate that the benefits of the dual representation is not isolated to our experimental setting, but does extended to a wider class of architectures.}

{\item{\bf Avoiding Shortcutting Predictors}

Since the dual representation by definition describes the spin system in terms of its local energy contributions, 
there is one particular pitfall here which has to be treated with caution: (Meta-)stable states commonly accumulate at low energies, and relatively high accuracies in our classification task can be obtained by simply choosing a fixed energy cutoff to label states
as ``(meta-)stable" (see Table~\ref{table:val_acc_1D_energetic} and Figure~\ref{fig:Metastability_Energy_schematic2}). In such situations, a neural network can be prone to adopting shallow heuristics which perform well in many cases (in this case the total energy) instead of learning the actual task it is supposed to solve.

We found, however, that 
the networks trained for the settings listed in Table~\ref{table:val_acc_1DSimpleNets_original} do
not rely purely on the lower energy of metastable states, and the difference in performance remains the same if tested
in low-energy regions where the ratio of each class is roughly the same. Going one step further, the element of energy can be eliminated completely
by additionally training only on states with fixed energies. While this drastically tightens the performance gap in simple settings,
a similar difference as before remains at more complex settings like  $N=100$ and $n=50$. \label{Discussion:Mestability-Energy}}

\end{itemize}

%%%%%%%%%%%%%%%%
%%%%%%%%%%%%%%%%

\begin{table}[b]
\centering
\begin{equation*}
\begin{split}
 \renewcommand{\arraystretch}{1.1}
 \arraycolsep12pt
 \begin{array}{c||cccccc}
  &  n=4 &  n=5 &  n=8  & n=9  & n=12 & \\
 \hline\hline
 \textrm{Energy cutoff} & 0.9925 & 0.9605  &  0.9535  &  0.9269&  0.8985
 \end{array}
\end{split}
\end{equation*}
\vspace{-0.8cm}\caption{Classification accuracy for (meta-)stable states using only a fixed energy cutoff (cf. Figure~\ref{fig:Metastability_Energy_schematic2}).}
\label{table:val_acc_1D_energetic}
\end{table}
%%%%%%%%%%%%%%%%
%%%%%%%%%%%%%%%%
%%%%%%%%%%%%%%%%
%%%%%%%%%%%%%%%%

\section{Enforcing Good Representations}
\label{sec:enforcing}
Having established that dual representations can be `beneficial' for classification tasks, we now turn to the question how such representations can be adapted by the network dynamically. When a duality map is known explicitly, it could easily be learned by a neural network with appropriate regression. Although this can be of interest in principle, we here focus on unsupervised learning techniques for adapting dual representations.

To do this, we discuss three different training strategies which we find to lead to `dual-like' representations:
\begin{enumerate}
\item Feature separation in the latent space. 
\item An autoencoder setup with an additional latent loss. In this case, the output of the encoder is the dual-like representation.
\item Demanding properties of the dual representation, for instance that it resembles the correct energy distribution.
\end{enumerate}

\subsection{Feature Separation}\label{FeatureSeparation}
For the discrete Fourier transform described in Section~\ref{DiscreteFourierTransformation} and Appendix~\ref{app:fourier}, the momentum space defines a valuable data representation in which the previously infeasible task of detecting
signals in noisy data becomes easy to solve. Based on our finding that deeper networks do not adapt this representation (cf.~Appendix~\ref{app:fourier}), we now pursue the question how one can assist the neural network to find such a beneficial representation without knowledge  about its explicit form. 

\subsubsection*{Basic Idea and Motivation}

Heuristically, the benefit is likely to come from the information of a non-localised signal in the space domain being collected in one single (complex) bin of the momentum space domain. This causes the signal in the momentum space domain being clearly separated from the background noise, which takes the same non-local form in both frames (cf. again Figure~\ref{fig:Fourier_example}). 
%%%%%%%%%%%%%%%%
%%%%%%%%%%%%%%%%
\begin{figure}[t]
\centering
\vspace*{20pt}
\centering
\resizebox{0.95\textwidth}{!}{%
  \begin{minipage}[b]{.95\linewidth} 
\includegraphics[width=\linewidth]{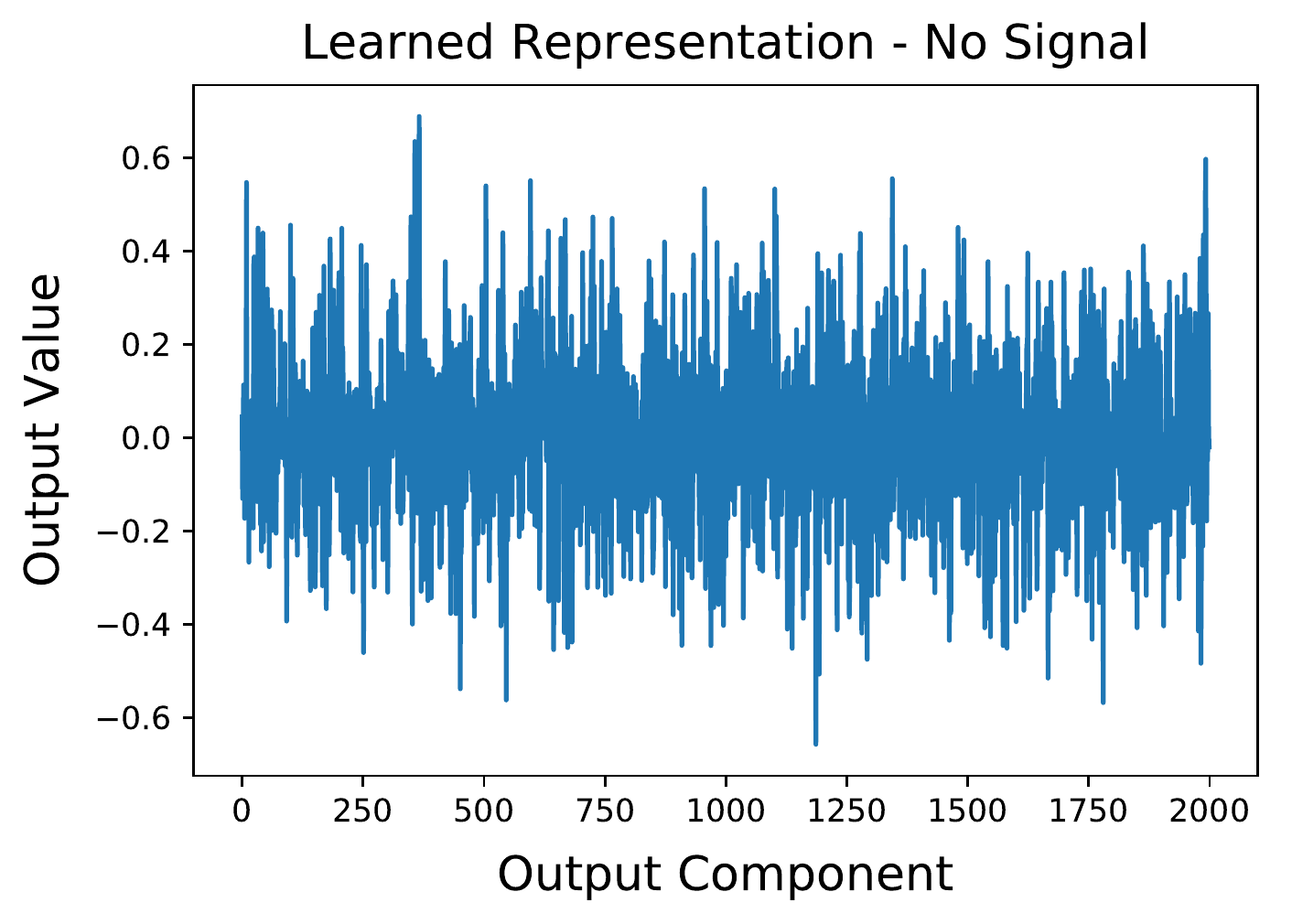}
   \end{minipage}
\hspace{.04\linewidth}
  \begin{minipage}[b]{.95\linewidth} 
\includegraphics[width=\linewidth]{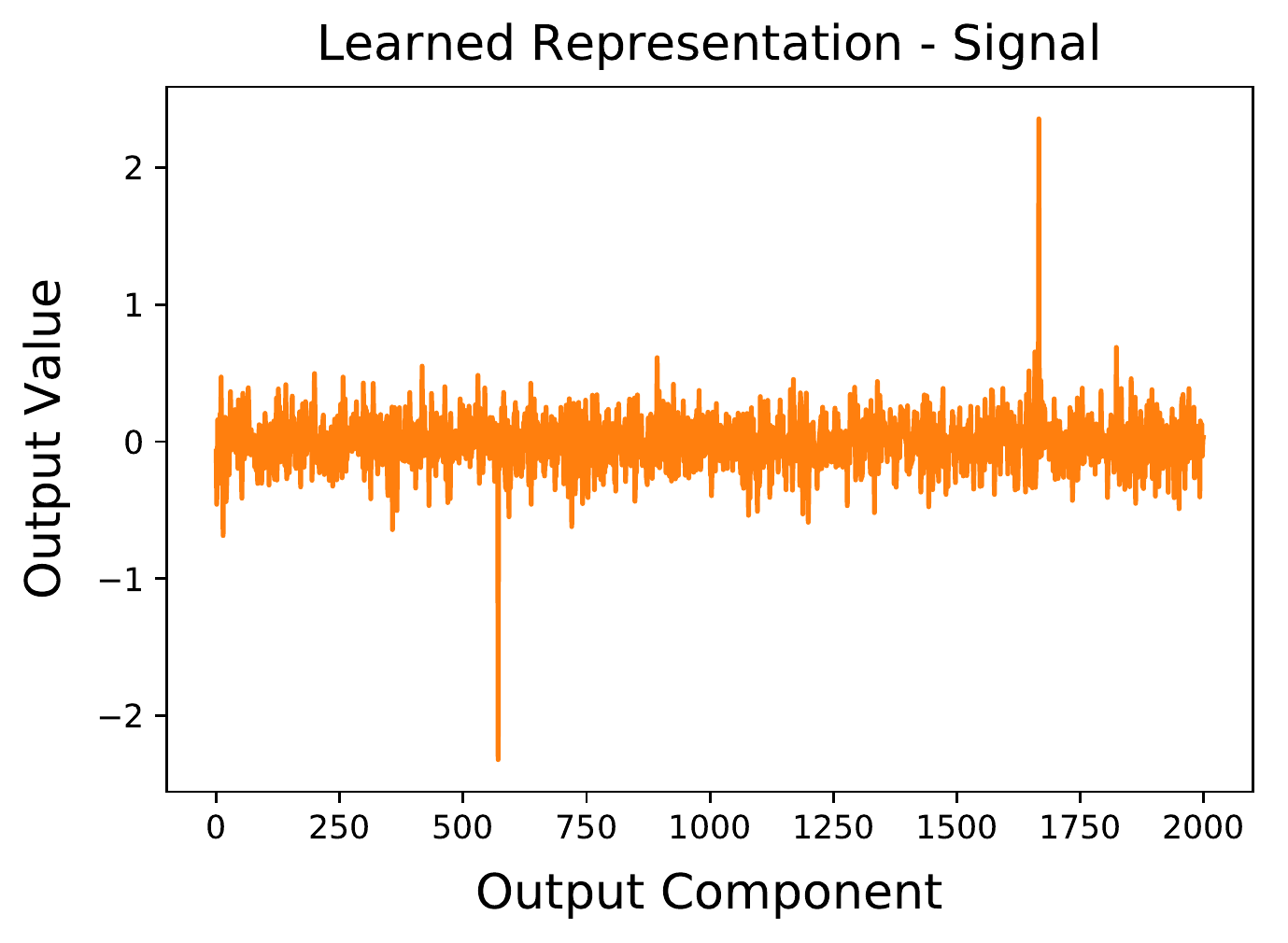}
   \end{minipage}
}
\vspace{-0.4cm}
\caption{
Output of the feature-separation network for pure noise and noisy signal.
\label{fig:LearnedRep_Output}
}
\end{figure}
%%%%%%%%%%%%%%%%
%%%%%%%%%%%%%%%%

{\it Can this ``feature separation" be exploited to automatically learn such favourable representations without analytic knowledge about the structure of the signal?} Assuming for the moment that there exists only one non-vanishing frequency, we would like to train a neural network to find a representation in which the outputs for pure signals and pure noise satisfy
\begin{equation}
|y_{\rm signal}|^2-|y_{\rm noise}|^2\geq\alpha\,.
\label{eq:loss1}
\end{equation}
Here, $\alpha >0$ denotes a margin where we want to push the latent representation. Formulated as a loss function, at values larger than $\alpha,$ this function shall take the value $0,$ which avoids a runaway of the signal (vanishing gradients). 
Notice that this task resembles the minimisation of a triplet loss~\cite{Chechik2009LargeSO,2015arXiv150303832S}, with the location of the noise fixed at zero. To apply this strategy to a setup with $N=1000$ different frequencies, two aspects have to be taken into account:
\begin{enumerate} 
\item{The relation \eqref{eq:loss1} should be satisfied for any frequency. }
\item{The information of different frequencies should be collected at different locations. Otherwise, the mapping might not be able to distinguish between clear signals and ``noisy" inputs with small components in many different frequencies (as is the case for the background noise in our setting).}
\end{enumerate}
A viable ansatz to achieve this is by defining a loss function
\begin{equation}
\label{eq:loss2}
\mathcal{L} = \textrm{max}(0,\alpha - (\xi _1 ^2 + \xi _2 ^2 ))\,,
\end{equation}
where $\xi _1 ^2$ and $\xi _2 ^2$ are defined as the two largest squared values of the  $2N$ outputs for a given input sample. When using pure single-frequency signals as training data, this loss effectively urges the sum of only the two output components with largest absolute value to be pushed away from zero until the margin $\alpha$ is reached. The aim of this is to enforce a data representation similar to the actual Fourier transform, in which all information of the single-frequency signals is concentrated in the real and imaginary parts of the $p_k$. 

At the same time, we keep the complexity of the network as low as possible (in this case linear). This is necessary because the loss~\eqref{eq:loss2} alone does not prevent  the occurrence of representations in which an arbitrarily large number of bins is maximised for any frequency. As a consequence, enforcement of sparse and local representations of signals would not  take place. In practice, such cases of ``overfitting" are possible for any network architecture, however, we observe that they commonly occur at higher degrees of complexity, whereas the constrained parameter space of low-capacity networks seems to act as an efficient preventive measure. Somewhat remarkably, this heuristic approach clearly outperformed more elaborate methods such as forcing sparse outputs via L1 penalty or penalising for correlation of latent variables.

Note that the network has no further knowledge on the structure of Fourier transformation or the structure of background noise.

\subsubsection*{Performance and Structure of Representation}
%%%%%%%%%%%%%%%%
%%%%%%%%%%%%%%%%
\begin{figure}[t]
\centering
\vspace*{10pt}
\centering
\resizebox{0.99\textwidth}{!}{%
  \begin{minipage}[b]{.95\linewidth}
\includegraphics[width=\linewidth]{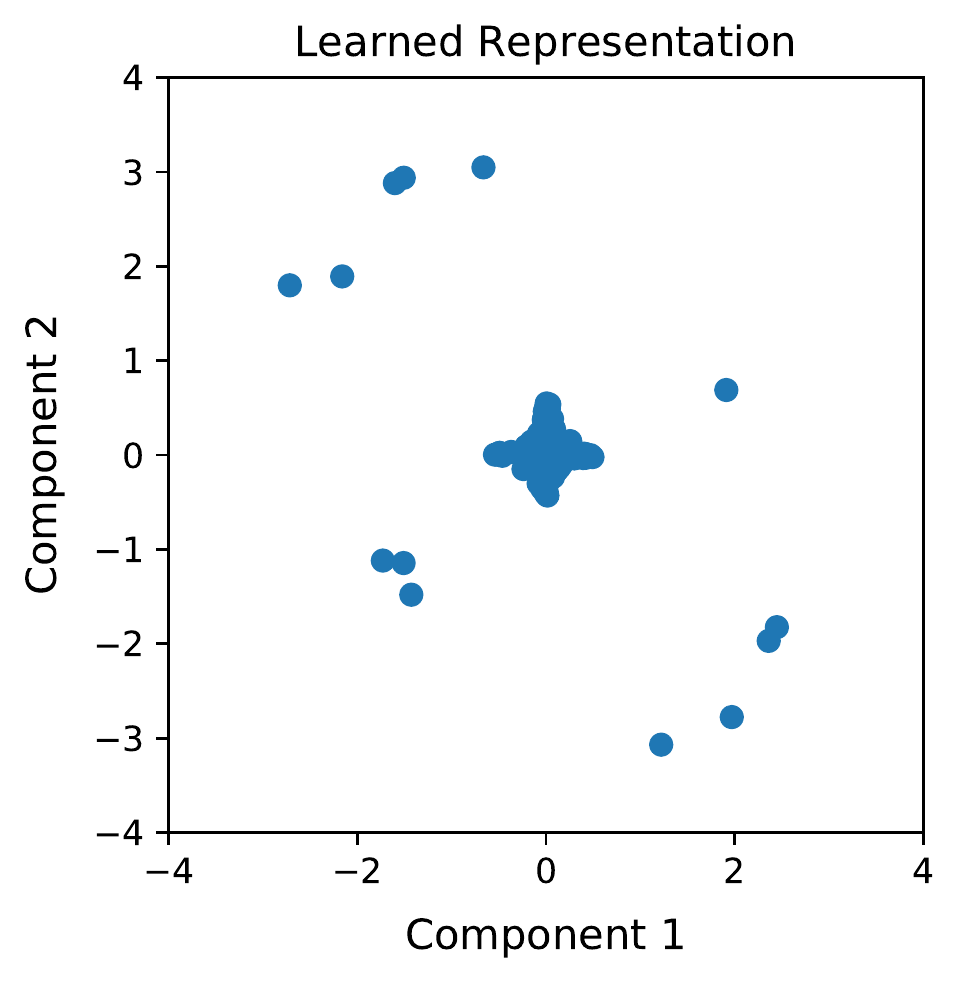}
   \end{minipage}
\hspace{.04\linewidth}
  \begin{minipage}[b]{.95\linewidth} 
\includegraphics[width=\linewidth]{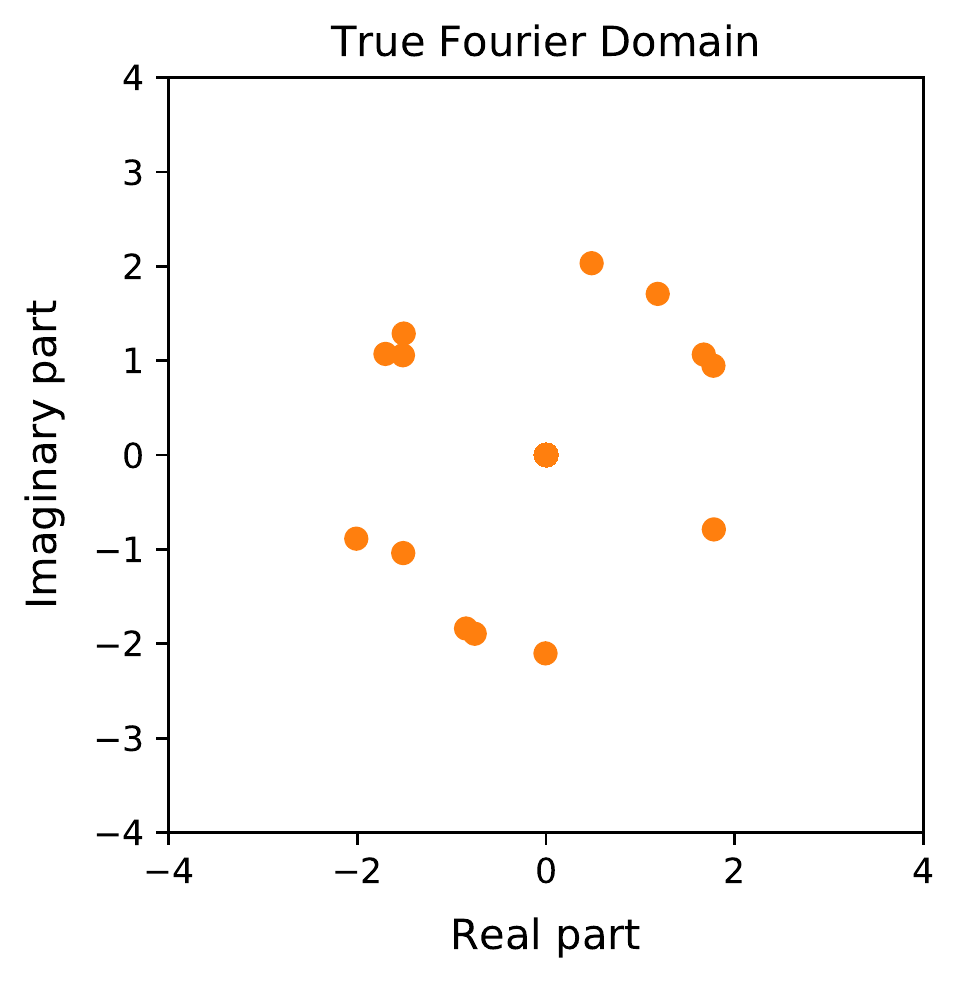}
   \end{minipage}
%}
\hspace*{35pt}
%\vspace*{10pt}
%
%\centering
%\resizebox{0.65\textwidth}{!}{%
  \begin{minipage}[b]{.95\linewidth} 
\includegraphics[width=\linewidth]{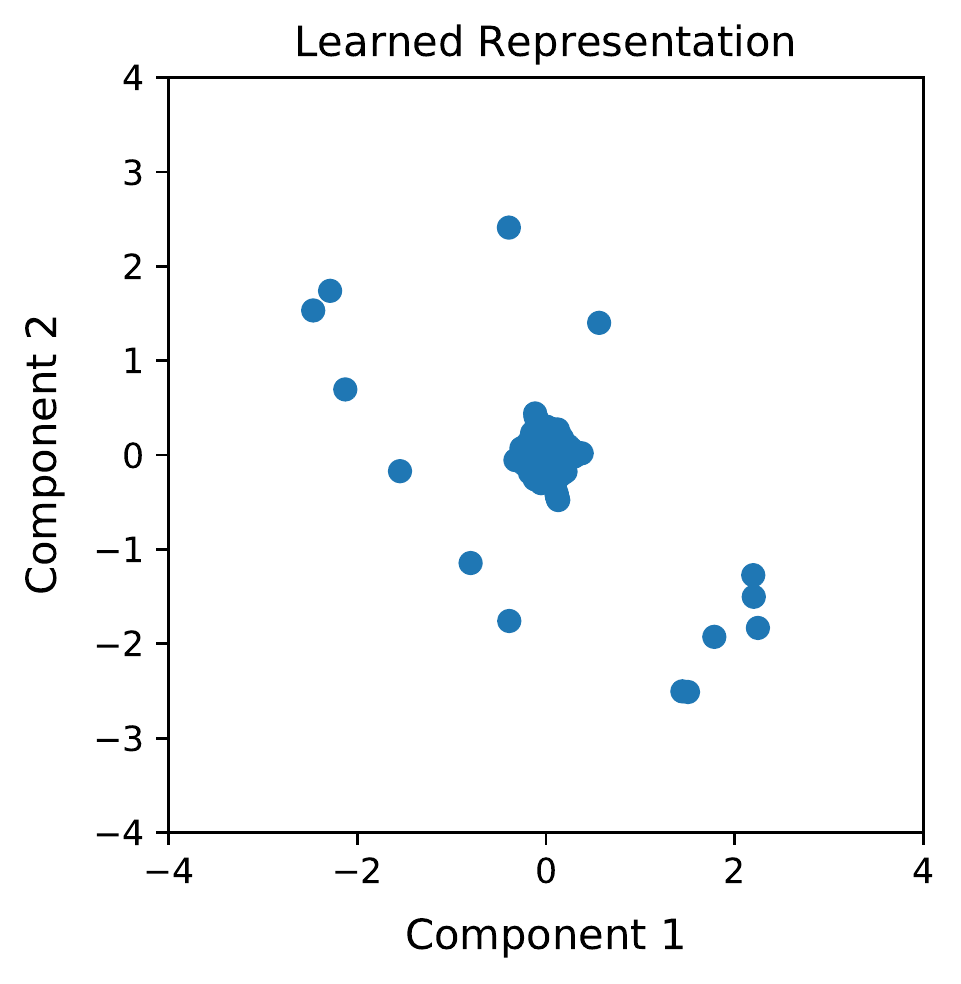}
   \end{minipage}
\hspace{.04\linewidth}
  \begin{minipage}[b]{.95\linewidth}
\includegraphics[width=\linewidth]{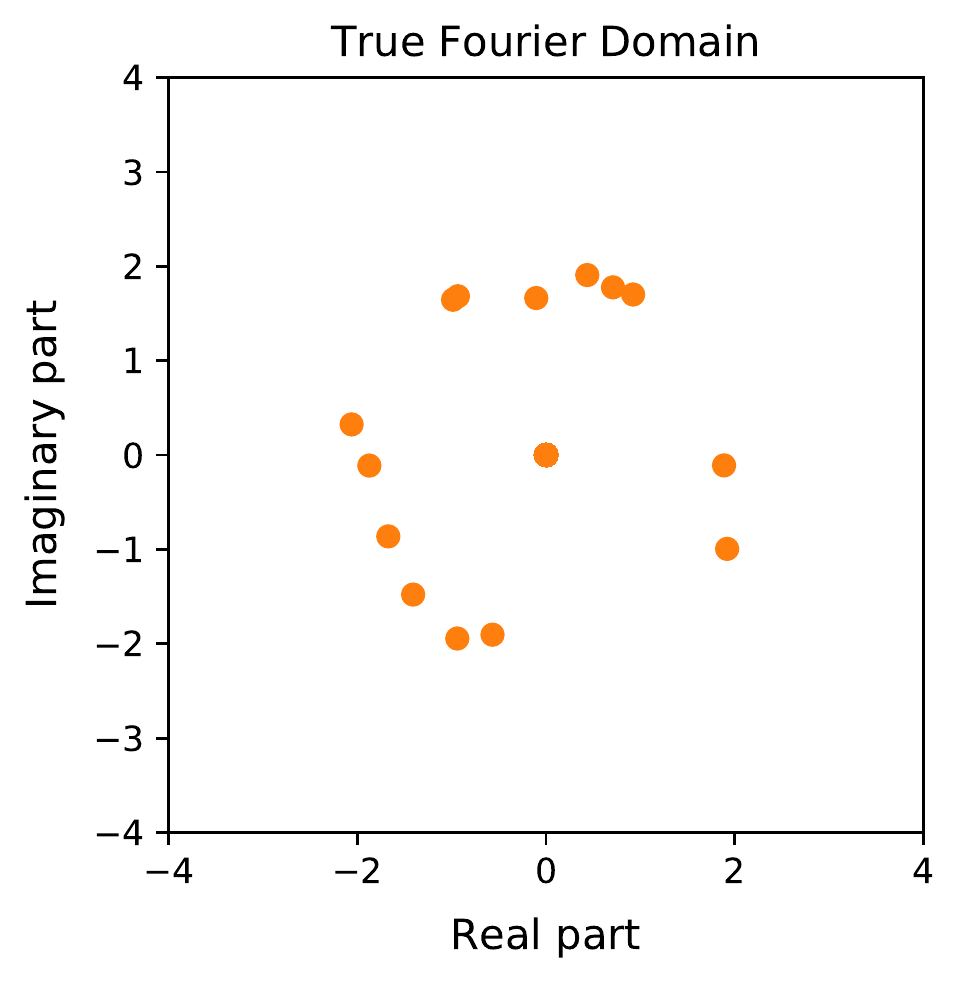}
   \end{minipage}
}

\caption{
Comparison of representations learned via feature separation and 
embedding into true momentum space domain. The above plots show examples of learned
representations and Fourier transforms of single-frequency signals 
at different frequencies without noise. Signals with non-vanishing component in
the respective frequency arrange in similar shapes, while the rest accumulates 
close to or at the origin.}
\label{fig:Comparison_Learned-Fourier}

\end{figure}
%%%%%%%%%%%%%%%%
%%%%%%%%%%%%%%%%
Training a linear network with $2N$ output nodes with Nesterov Adam optimiser, learning rate $1\cdot 10^{-3}$ and $\alpha = 5$ commonly led to close-to-zero losses after less than five epochs. As can be seen in  Figure~\ref{fig:LearnedRep_Output}, the learned representation  shows characteristic properties of the actual Fourier transform when we trained just with noisy signals as input. 
Using this representation for our previous task of signal detection in noisy data, the mean best test accuracy of the same simple one-layer convolutional neural network 
as described in section \ref{DiscreteFourierTransformation} (cf.~also Appendix~\ref{app:fourier} for more details) indeed improved to around 0.7717.

Interestingly, the 
learned data representations often take the form of transformations such as rescalings, reflections or rotations of the actual 
Fourier transform in the $2N$-dimensional space. Projecting the output of the network for a large number of samples onto
particular pairs of components, the distribution
of values then corresponds to that of the real and imaginary parts of a certain value $p_k$ in the Fourier domain.
This is exemplified for two instances in Figure~\ref{fig:Comparison_Learned-Fourier}.

\subsubsection*{Response to Single-Frequency Signals}

\begin{figure}[t]
\centering
\vspace*{20pt}
\centering
\resizebox{0.85\textwidth}{!}{%
  \begin{minipage}[b]{.95\linewidth}
\includegraphics[width=\linewidth]{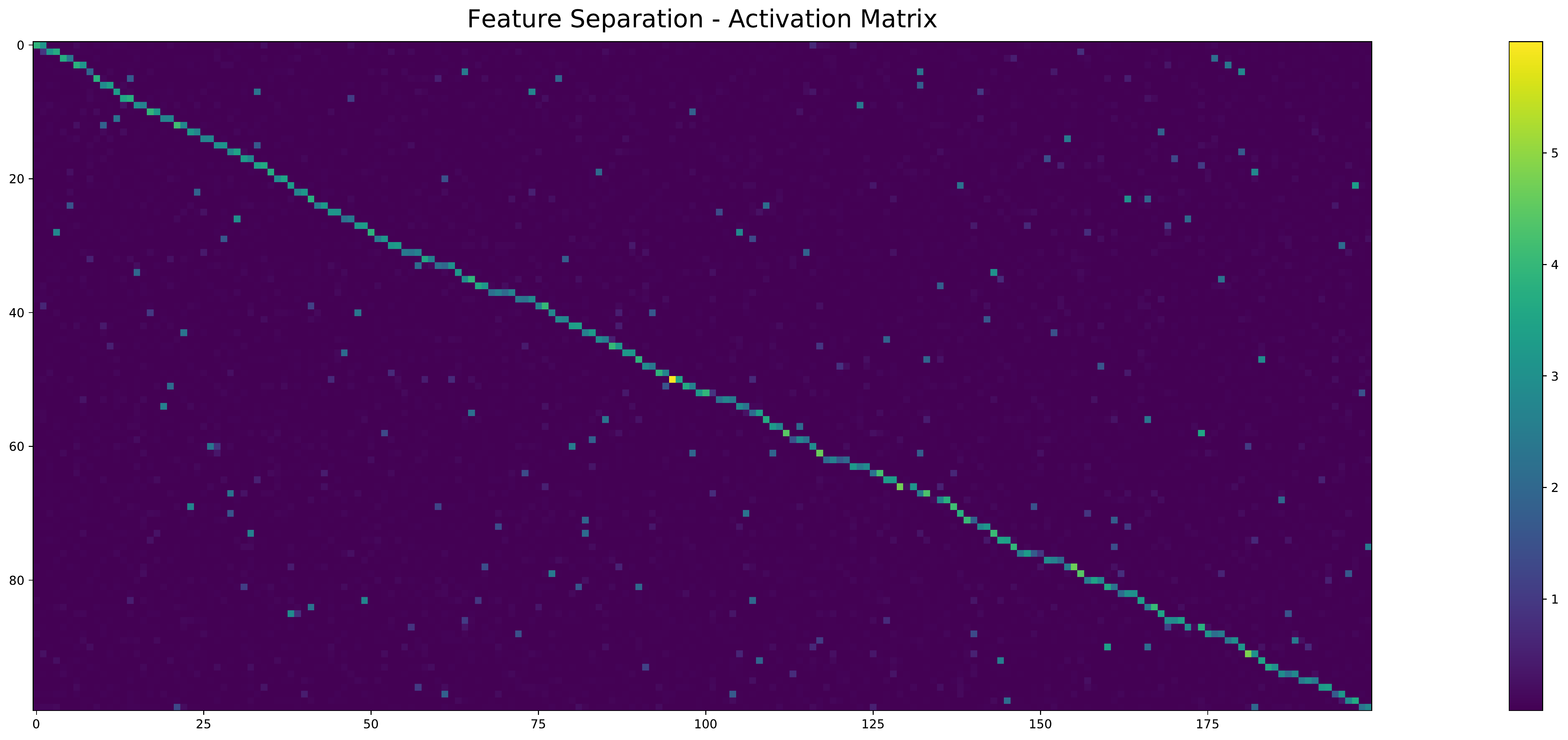}
   \end{minipage}
}
\vspace*{-7pt}
\caption{
Example plot of an activation matrix \eqref{Fourier-FeatureSep-ActivationMatrix} for the 
case $N=100$. The columns have been reordered according to the indices of their respective
largest entries. The number of non-vanishing values in a given row matches with the dimension of the subspace
of the $2N$-dimensional latent space into which the representation of signals with a corresponding
non-vanishing frequency $p_i$ is nontrivially embedded (cf. Figure~\ref{fig:FeatureSep_Activation2}).
\label{fig:FeatureSep_Activation}
}
\end{figure}

\begin{figure}[h]
\centering
%\vspace*{20pt}

\centering
\resizebox{0.95\textwidth}{!}{%
  \begin{minipage}[b]{1.00\linewidth} 
\includegraphics[width=\linewidth]{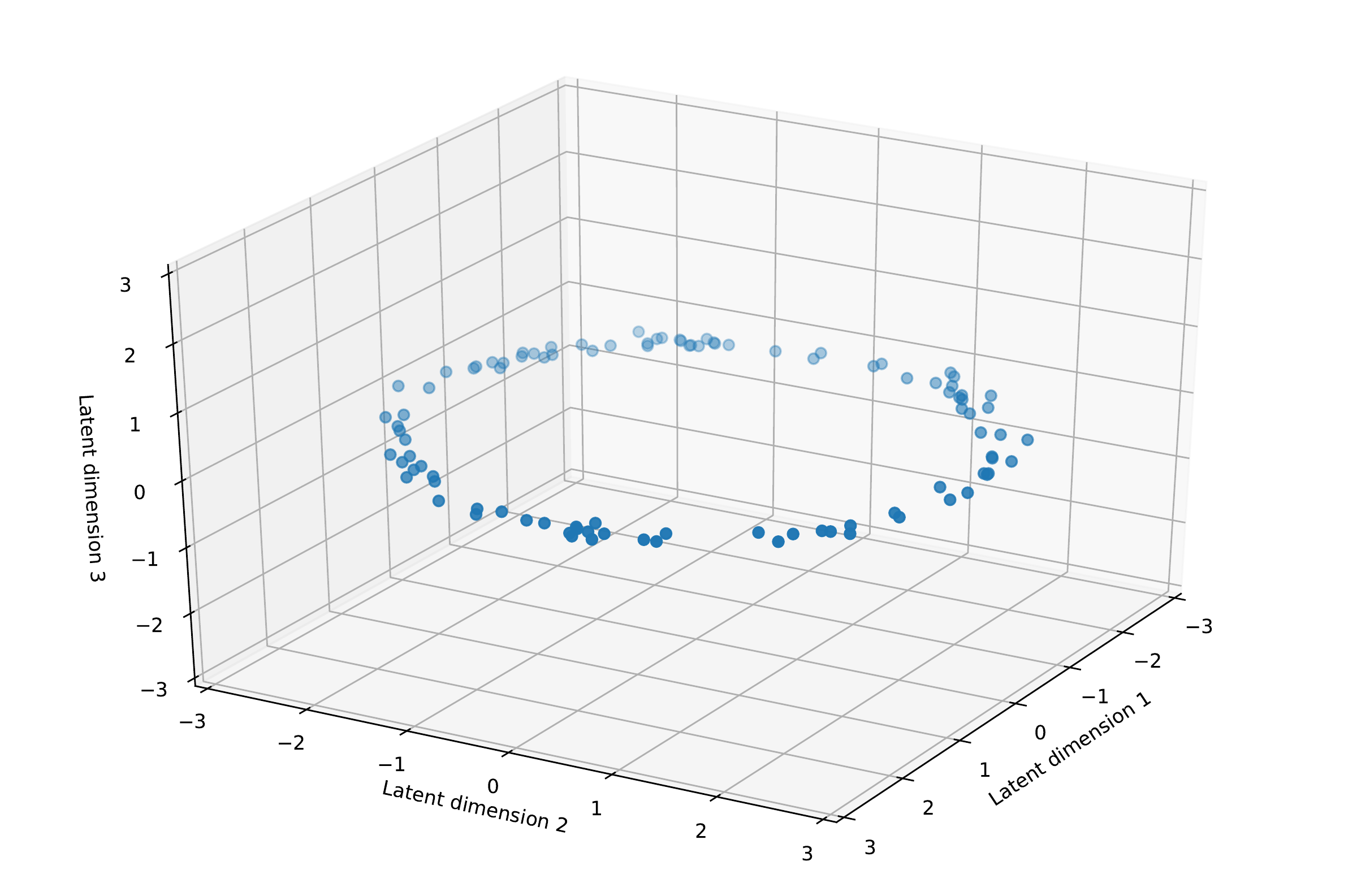}
   \end{minipage}
  \begin{minipage}[b]{1.00\linewidth} 
\includegraphics[width=\linewidth]{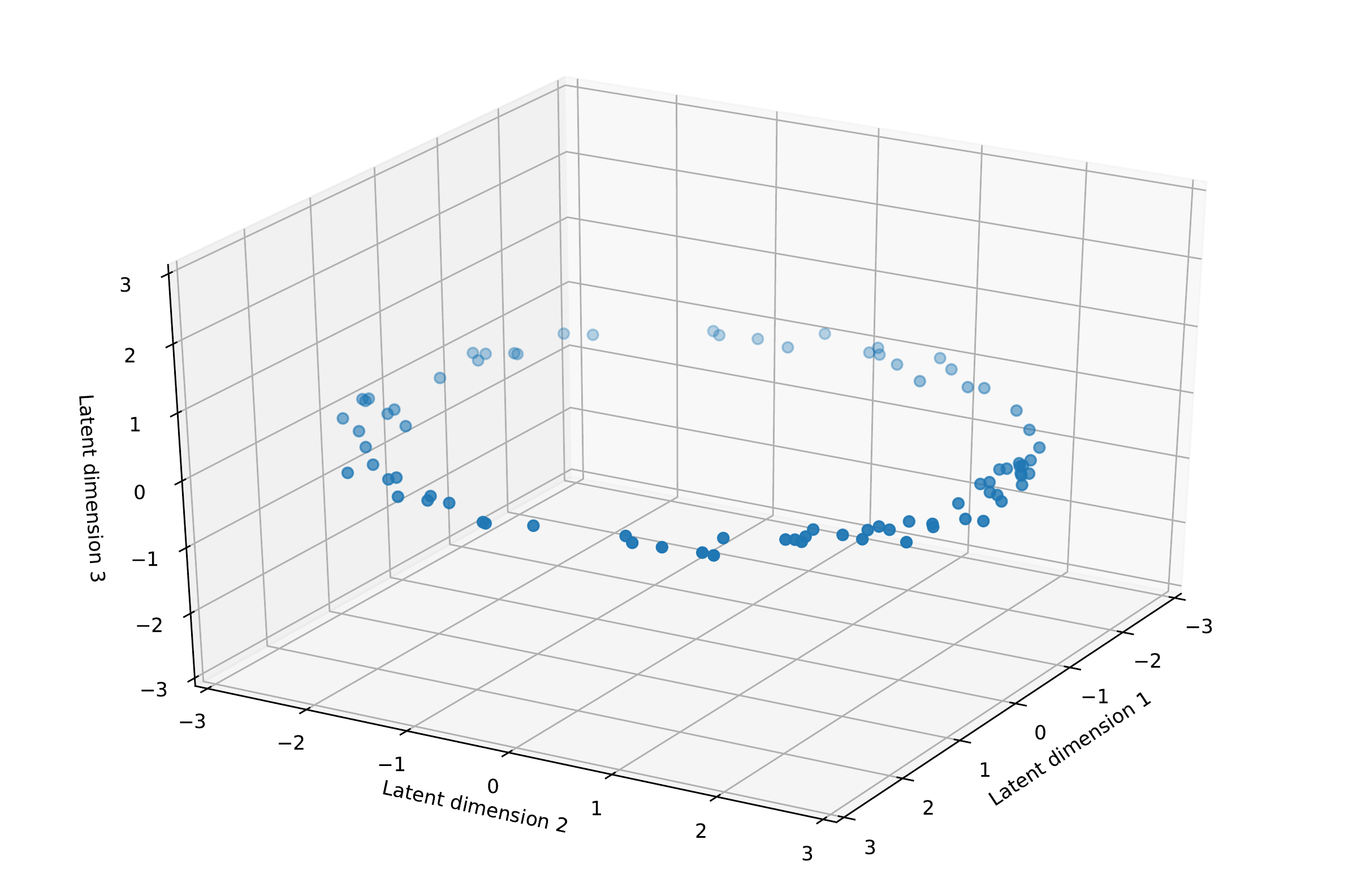}
   \end{minipage}
}

\vspace{6pt}

\centering
\resizebox{0.83\textwidth}{!}{%
\hspace*{10pt}
  \begin{minipage}[b]{1.00\linewidth}
\includegraphics[width=\linewidth]{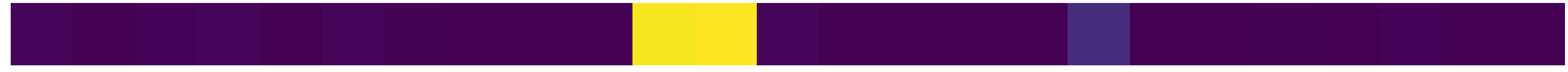}
   \end{minipage}
\hspace*{250pt}
  \begin{minipage}[b]{1.00\linewidth} 
\includegraphics[width=\linewidth]{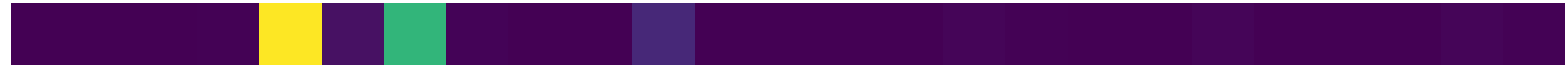}
   \end{minipage}
}

\vspace*{15pt}

\centering
\resizebox{0.95\textwidth}{!}{%
  \begin{minipage}[b]{1.00\linewidth} 
\includegraphics[width=\linewidth]{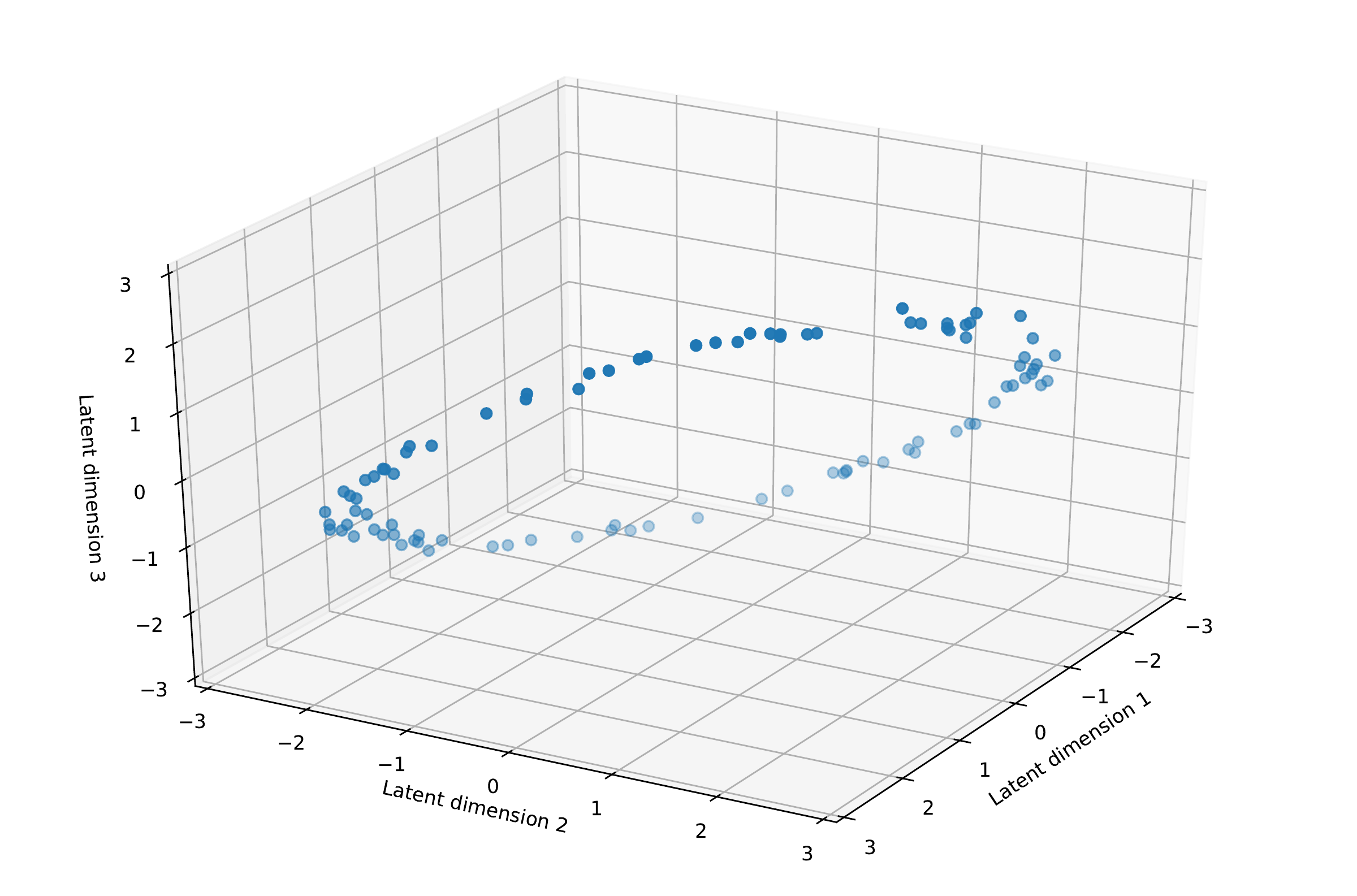}
   \end{minipage}
  \begin{minipage}[b]{1.00\linewidth} 
\includegraphics[width=\linewidth]{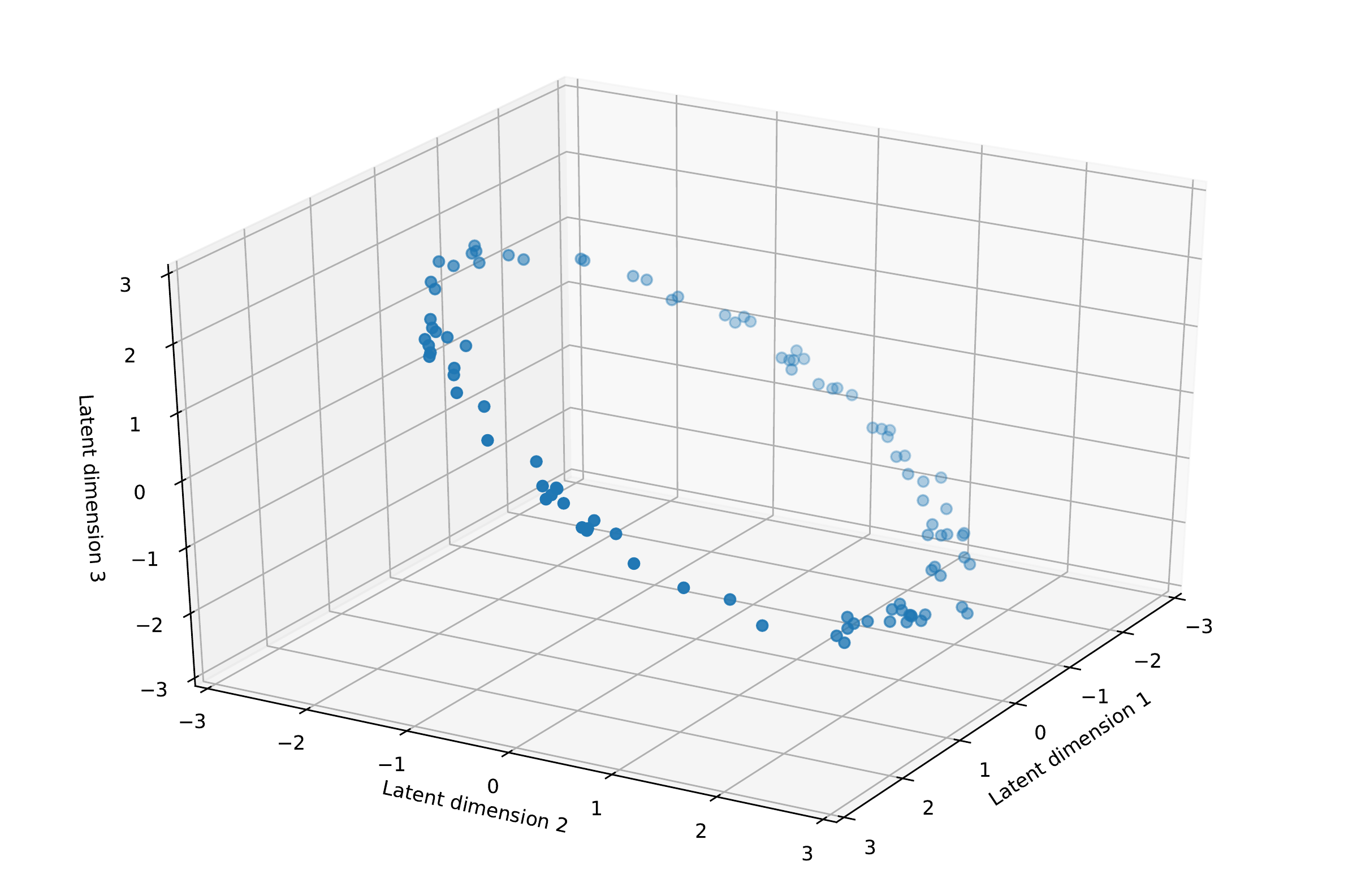}
   \end{minipage}
}

\vspace{6pt}

\centering
\resizebox{0.83\textwidth}{!}{%
\hspace*{10pt}
  \begin{minipage}[b]{1.00\linewidth} 
\includegraphics[width=\linewidth]{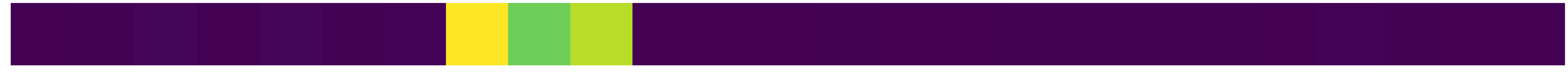}
   \end{minipage}
\hspace*{250pt}
  \begin{minipage}[b]{1.00\linewidth} 
\includegraphics[width=\linewidth]{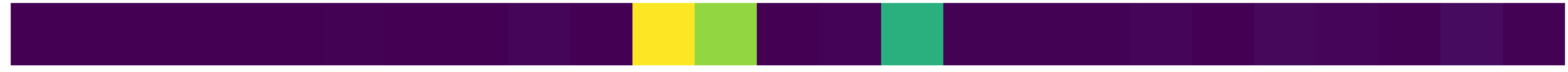}
   \end{minipage}
}

\vspace{-10pt}

\caption{
Interpretation of the activation matrix illustrated in Figure~\ref{fig:FeatureSep_Activation}. The plotted
latent dimensions correspond to the three  largest entries of a given row. {\bf (Top)} Two non-vanishing entries in one row.
The Fourier transform is completely embedded into two latent dimensions. {\bf (Bottom)} Three non-vanishing entries in one row. The Fourier transform is nontrivially embedded
into three latent dimensions.
\label{fig:FeatureSep_Activation2}
}
\end{figure}

Some more insights into the structure of the feature-separation network can be gained by analysing its outputs $f_j(x)$. Here,
we do this by analysing the $2N$ response values $f_j(x)|_{p_{i}\neq 0}$  when given pure signals with single non-vanishing frequency 
$p_i$.  These can be stored a $N\times 2N$ response matrix 
\begin{equation}\label{Fourier-FeatureSep-ActivationMatrix}
M_{ij} = \langle |\left. f_{j}(x) \vphantom{\frac{1}{2}}|^{2}\rangle\,\right|_{p_{i}\neq 0}  \,,
\end{equation}
where the mean is taken over all samples satisfying the condition $p_{i}\neq 0$. The matrix generally shows a high
degree of sparsity, and  we find that a fraction of higher than 0.8 of all 
rows contain at least one large value, implying that the network makes efficient use of the $2N$ dimensions to embed 
the signals into the latent space. An example plot of the matrix $M$ for the case $N=100$ can be found in Figure~\ref{fig:FeatureSep_Activation}. It can be observed that each row of the matrix commonly contains between 2 and 4 large
activations, with the remaining entries being close to zero. Visualising the corresponding latent dimensions, one finds that this
behaviour reflects precisely the way in which the Fourier-transform is embedded into the latent space. This is exemplified for various
cases in Figure~\ref{fig:FeatureSep_Activation2}.

%%%%%%%%%%%%%%%%
%%%%%%%%%%%%%%%%
%%%%%%%%%%%%%%%%
%%%%%%%%%%%%%%%%

%%%%%%%%%%%%%%%%
%%%%%%%%%%%%%%%%
%%%%%%%%%%%%%%%%
%%%%%%%%%%%%%%%%

\subsection{Autoencoder with Latent Loss}
\label{sec:CAE}
We now turn to the second example of adapting an appropriate latent dimension dynamically which is based on the 1D Ising setup already described in Section~\ref{sec:1DIsing}.

\subsubsection*{Motivation and Architecture}

We have seen that by exchanging the roles of individual spins and their interaction
terms, the task of detecting (meta-)stable states becomes more accessible due to the relevant information being easier to extract from a lower number of spins in the dual frame.
To find such a suitable representation, we here employ the following strategy: We use the fact that a simple task can be performed very efficiently in the dual representation. In this case this  is the (trivial) task of energy classification. 

By itself, this is not sufficient and we need to ensure that no information is lost in the latent representation.  
A viable method to achieve this goal is to use a autoencoder-like architecture whose `bottleneck' has (at least) the same 
dimension as the original input and is required to represent the data in a way that the total energy can be
extracted by a simple linear model. This way, the model is guaranteed to learn a representation which encodes
the energetic properties of a state in a manner similar to the dual frame (cf. Equation~\eqref{1DIsing_DualityTransformation}),
while at the same time the presence of an additional reconstruction loss forces the mapping to be information conserving.

In practice, this can be implemented by training a neural network to map an input state $s_1,\ldots,s_N$ to an intermediate
output of (at least) the same dimension, which in turn serves as input for a linear model extracting the total energy of 
the input state and another network reconstructing the initial input configuration. Figure~\ref{fig:TaskConstrainedAE} illustrates this architecture
schematically.

\subsubsection{Results and Discussion}

We tested the performance in classifying (meta-)stable states using the same setting as before,
with the duality transformation~\eqref{1DIsing_DualityTransformation} replaced by the intermediate output
of a constrained autoencoder with latent dimension 18 and 50. Details on the experimental conditions are provided in Appendix \ref{app:1DIsing}; results are shown in Table~\ref{table:val_acc_1DSimpleNets_learned}.

One again observes a significant improvement compared to the original representation (cf. Table~\ref{table:val_acc_1DSimpleNets_original}, left), albeit not as drastic as in the actual dual representation. Autoencoders with latent dimension 18 often suffered from underfitting problems, and further benefits were possible when increasing the latent dimension to 50.  Networks trained on the learned representation mostly outperformed accuracies reachable by pure energy cutoffs in particular at latent dimension 50, but showed a slight tendency to misclassify samples which are located in energy regions dominated by the respective other class.  While part of the improvement might therefore be attributed to the correlation between overall energy and (meta-)stability, the learned representation still allows to solve the classification task significantly better than by
training on the original representation directly, and the networks do not resort completely to superficial energetic arguments.

\subsubsection*{Further Applications}

Let us conclude this discussion by stressing that the main purpose of the above architecture is to realise transfer learning between different
physically related problems. This can be beneficial when training data is limited or expensive to 
generate for one task but can be efficiently acquired for a simpler task. In such cases, it might not be a reduction of required overall training data, but rather a change in the type of data that eventually leads to an improvement in overall performance.

In our considered setting,
we indeed found that benefits in performance are only possible when the constrained autoencoder is trained
on relatively large datasets. While this obviously nullifies the improvement in overall data efficiency of analytical dualities,
it can simplify the process of training due to the possibility to replace large datasets
of metastable states (which might not even exist for some settings) by corresponding pairs of random states and their energy.

Generally, finding such physically related tasks commonly requires domain knowledge or heuristic arguments,
but it nevertheless opens up a wide range of new possibilities going beyond known analytical dualities.

\begin{figure}[t]
\begin{center}
\includegraphics[width=0.75\textwidth]{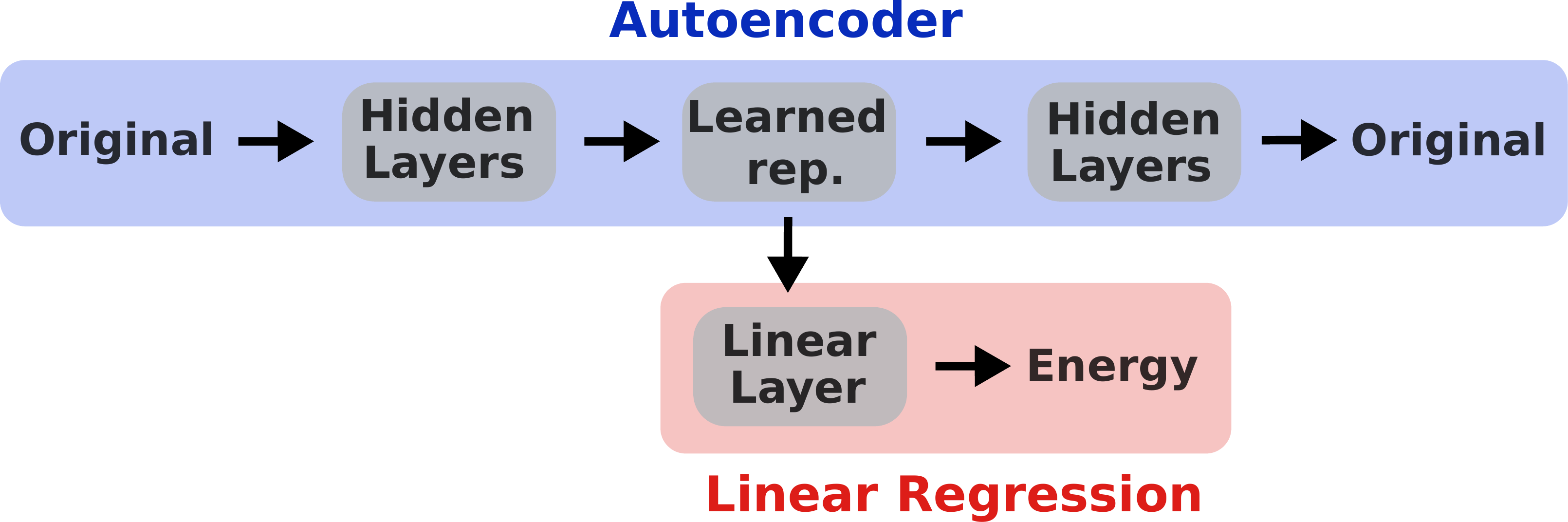}
\end{center}
\caption{Schematic illustration of a task-constrained autoencoder used to learn suitable representations
for difficult tasks. The intermediate output takes the role of the ``dual" representation.} 
\label{fig:TaskConstrainedAE}
\end{figure}

\hspace*{5pt} 

\subsubsection*{Interpretation of Intermediate Output}

Before we delve into the interpretation of the intermediate output, it is important to remark that we did not impose any further constraints regarding the structure
of the intermediate output as performance commonly suffered from reduced network capacity in such cases. As a consequence, the intermediate output has no obvious physical interpretation and relations to the true dual representation are a priori not obvious. 
\begin{table}[t]
\begin{footnotesize}
\begin{center}
 \begin{tabular}{c||ccccc}
 lat (18) &  n=4 &  n=5 &  n=8  & n=9  & n=12  \\
 \hline\hline
$6\cdot 10^2$ & 0.9880 &  0.9540 &  0.9180  &  0.9072 & 0.9228
 \\$
 3\cdot 10^3$ & - &   0.9677  &   0.9527  &  0.9353  & 0.9476
 \\
$ 9.5\cdot 10^3$ & - &   -  &   0.9607  &  0.9500  & 0.9597
 \end{tabular}\qquad 
 \begin{tabular}{c||cccccc}
 lat (50)  &  n=4 &  n=5 &  n=8  & n=9  & n=12  \\
 \hline\hline
$ 6\cdot 10^2$ & 0.9887 &  0.9526 &  0.9300  &  0.9304 & 0.9500 
 \\
$3\cdot 10^3 $& - &   0.9718  &   0.9787  &  0.9637  & 0.9829  
 \\
$9.5\cdot 10^3$ & - &   -  &   0.9910  &  0.9885  & 0.9968 
 \end{tabular}

\end{center}
\end{footnotesize}
\caption{Detection of (meta-)stable states in the 1D Ising chain for different interactions and amounts of training data.
The listed numbers describe the average best test accuracy over 10 training runs of 500 epochs each when trained
on the intermediate output of a constrained autoencoder with latent dimension 18 {\bf(Left)} and 50 {\bf(Right)}.
Missing values indicate that the number of required samples exceeds the total number of metastable states for the
considered setting.}
\label{table:val_acc_1DSimpleNets_learned}
\end{table}

\begin{figure}[t]
\begin{center}
\includegraphics[width=0.7\textwidth]{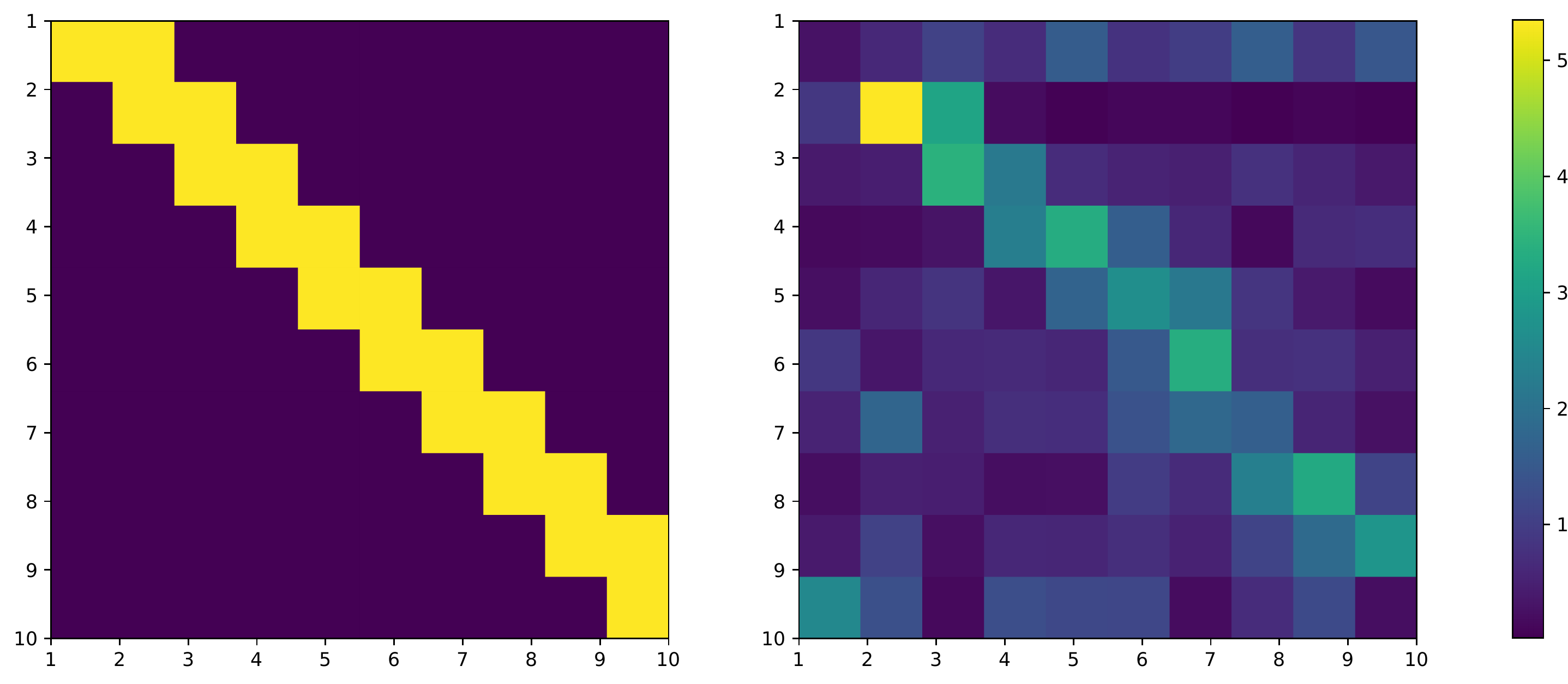}
\end{center}
\vspace*{-15pt}
\caption{Plots of the sensitivity matrix~\eqref{Ising1D_Sensitivity} for the actual duality transformation (left) and a learned
representation of a constrained autoencoder (right) for $N=10$ and $n=2$. Both matrices show characteristic nearest neighbour
interactions; the latter contains additional nonlocal components.} 
\label{fig:Ising1DN10n2_Sensitivity}
\end{figure}

An interesting question in this context is whether there is some way to make sense of how the relevant information is encoded in our learned representation. A viable way to study dependencies between the input and latent variables is to analyse
the sensitivity of the latent variables with respect to flips of a particular spin $s_j$ while keeping all other spins fixed.
This information can be stored in the matrix 
\begin{equation}
\label{Ising1D_Sensitivity}
M_{ij} = \frac{\langle\left( f_{i}( s_{1}, \dots ,s_{j} ,\dots s_{N}) - f_{i}( s_{1}, \dots ,-s_{j} ,\dots s_{N})\right)^{2}\rangle}
{\frac{1}{N}\sum _{k=1}^{N}\langle\left( f_{i}( s_{1}, \dots ,s_{k} ,\dots s_{N}) - f_{i}( s_{1}, \dots ,-s_{k} ,\dots s_{N})\right)^{2}\rangle}\,,
\end{equation}
where the expectation values are to be computed for the complete (test) dataset. Heuristically, this matrix encodes the average sensitivity
of the components $f_{i}$ of the transformed representation with respect to flips of a particular spin $s_j$, normalised by 
the average sensitivity of $f_{i}$  to flips of any spin. For the actual duality transformation~\eqref{1DIsing_DualityTransformation},
the numerator takes precisely the values 0 or 4, leading to a staircase-like structure as depicted on the left hand side in Figure~\ref{fig:Ising1DN10n2_Sensitivity}. 

We trained 25 constrained autoencoders for the simple setting $N=10$ and $n=2$ and compared the transformation
behaviour of the learned variables to that of the true duality transformation  \eqref{1DIsing_DualityTransformation}.  
Interestingly, there exist many instances of networks with structurally similar dependencies as the proper duality transformation. These commonly include 
components $f_{i}$ depending strongly on neighbouring pairs of spins and a distinguished value $f_{N}$ which is highly sensitive
to one particular spin - the matrix $M_{ij}$ for one such 
example is presented on the right hand side in Figure~\ref{fig:Ising1DN10n2_Sensitivity}.

Notice that this basically represents the way the duality transformations \eqref{1DIsing_DualityTransformation} 
encode the information of the original system in that there exist $N-1$ terms $\sigma _{i}, i=1,\dots, N-1$ describing the
nearest-neighbour interactions  and one value $\sigma _{N}$ which does not interact with the external field and stores the
overall sign of the system.

\subsection{Distributional properties}
The next question we analysed is to which degree neural networks are capable of learning the relation between dual Ising models on the square lattice. A minimal requirement for this is that the duality map between the two systems
can be learned if samples from both data representations are provided explicitly.

Here we start with no one-to-one mapping between states of a system at temperature $T$ and 
those of a system at dual temperature $\widetilde{T}.$ Instead, we match features of the dual representation on the level of the probability 
distributions, i.e.~that the learned representation shares features with the target dual distribution. For this purpose, we consider the following architecture: States $s$ sampled from the temperature
$T$ are used as input for a deep convolutional network 
and mapped onto a lattice of the same shape whose entries are interpreted as probabilities of the the 
respective spins to take the value 1. 

Binary states are then sampled by utilising the Gumbel trick 
to preserve differentiability of the network. In the discussed setting, this can be realised by sampling for each
site $p_i$ of the lattice some value $\varepsilon _i \sim U(0,1)$ uniformly and map the input state $s$ to an output
state $f(s)$ with
\begin{equation}\label{GumbelTrick}
f_{i}(s)= 2\cdot\textrm{sig}\left[\gamma\left(\log (\varepsilon _i) - \log (1-\varepsilon _i) 
+ \log (p_i) - \log (1-p_i)\right)\right]-1\,,
\end{equation}
where $ \textrm{sig}$ denotes the sigmoid function $ \textrm{sig(x)}=\frac{1}{1+e^{-x}}$ and $\gamma$ 
is a scale parameter which can be used to force the output values closer to the extremal values 0 and 1\footnote{Some caution is needed when choosing $\gamma$  as high values can lead to vanishing or exploding gradients, 
resulting in poor training.} .

The output states $f(s)$ are then fed into a hard-coded layer to compute their total energy, and the loss
function is defined as the Kullback-Leibler divergence
\begin{equation}
D_{\textrm{KL}}(P_{f}\lVert P_{\sigma})=-\sum _{E} P_{f}(E) \log \left(\frac{P_{\sigma}(E)}{P_{f}(E)} \right)
\end{equation}
between the energy distributions $P_{f}(E) $  and $P_{\sigma}(E)$  of
states sampled from the network and the true dual temperature, respectively.
 
The network produces 
binary outputs as desired, with the energy distributions closely resembling those of the actual dual system.
This is depicted for two examples in Figure~\ref{fig:KWDuality_Unet_40x40}. 

\begin{figure}[t!]
\begin{center}
\includegraphics[width=0.85\textwidth]{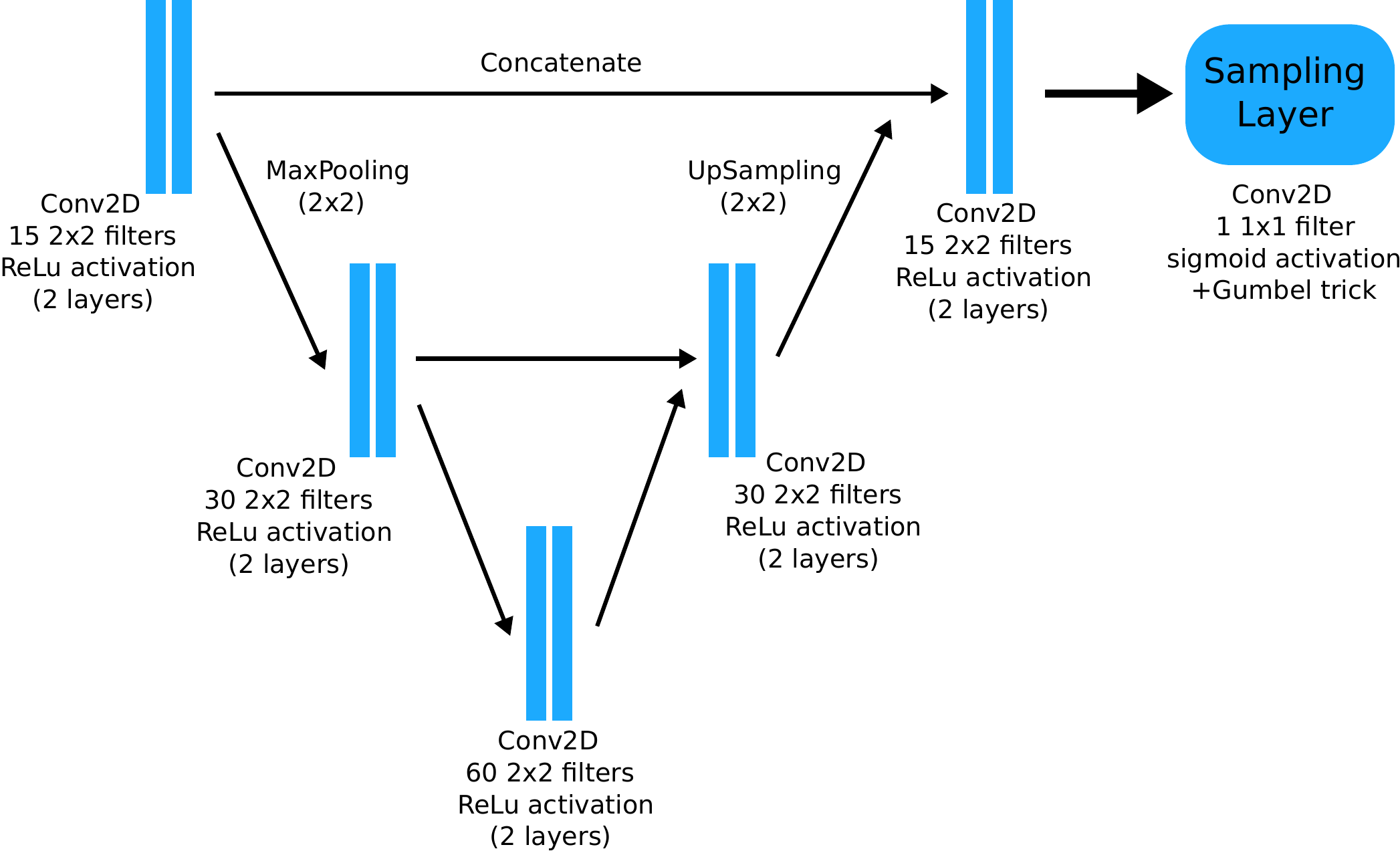}
\end{center}
\vspace*{-10pt}
\caption{Schematic illustration of a U-Net architecture. } 
\label{fig:U-Net_Schematic}
\end{figure}

\subsubsection*{Results}

We used a U-Net architecture as depicted in Figure~\ref{fig:U-Net_Schematic} with three levels consisting of two layers
of 15, 30 respectively 60 $2\times 2$ filters with ReLu activations. The scale parameter in \eqref{GumbelTrick} was set to 50.
Tests were conducted for a $40\times 40$ lattice at temperatures $T=0.25, 0.5, \dots , 2.25$ using standard Nesterov Adam
optimiser with initial learning rate 0.002 and learning rate decay. The dataset for each temperature was again split into 16000 training samples and 4000 test samples.
Training equilibrium was commonly reached within 50 epochs; no significant changes were noticed after 500 epochs. Tests were again performed for
10 random seeds per temperature and showed consistent overall performance, however, there were rare instances in which poor local minima 
required reinitialization of the network in particular when mapping to lower temperatures.

\begin{figure}[t]
\begin{center}
\includegraphics[width=0.4\textwidth]{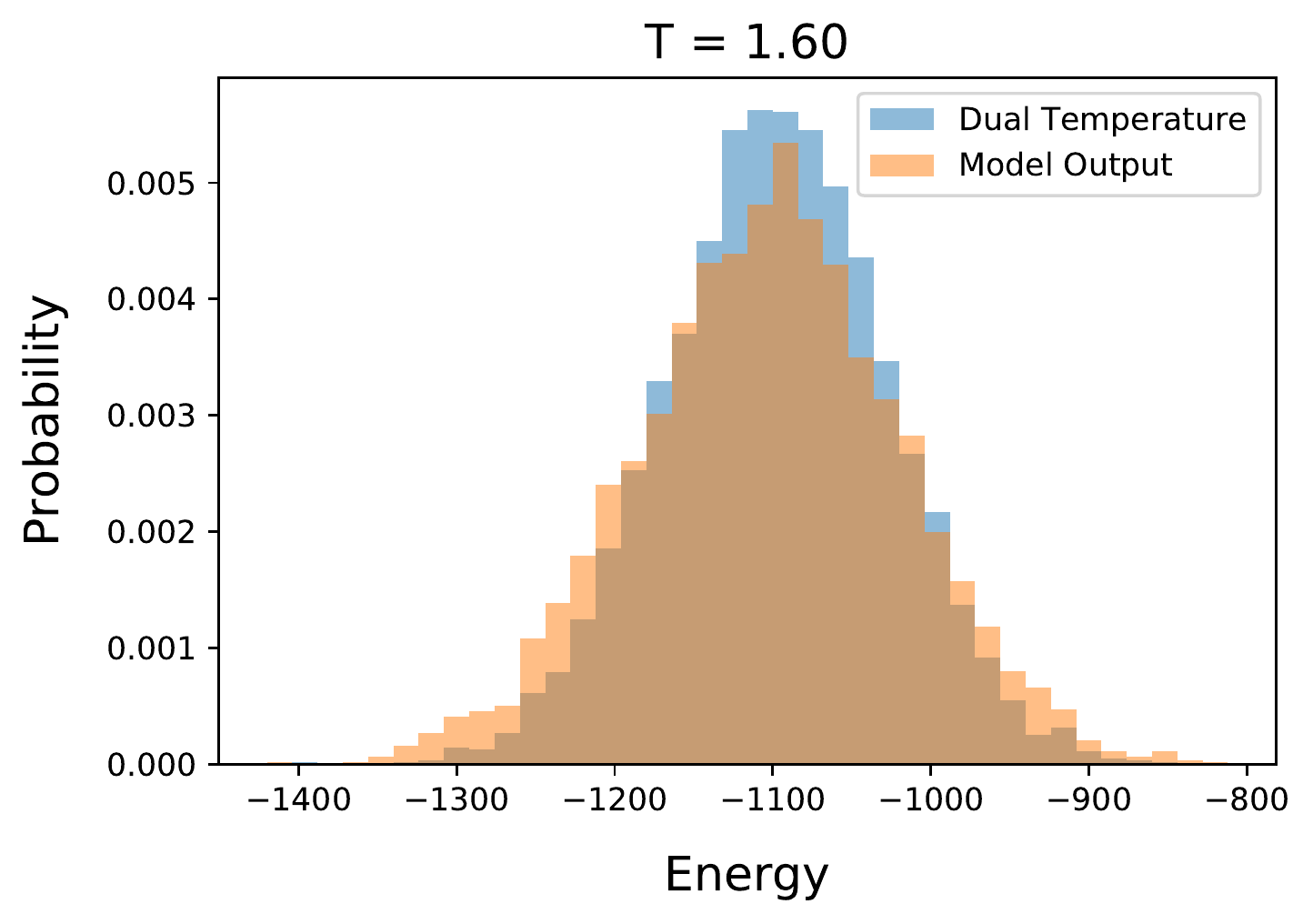}
\includegraphics[width=0.4\textwidth]{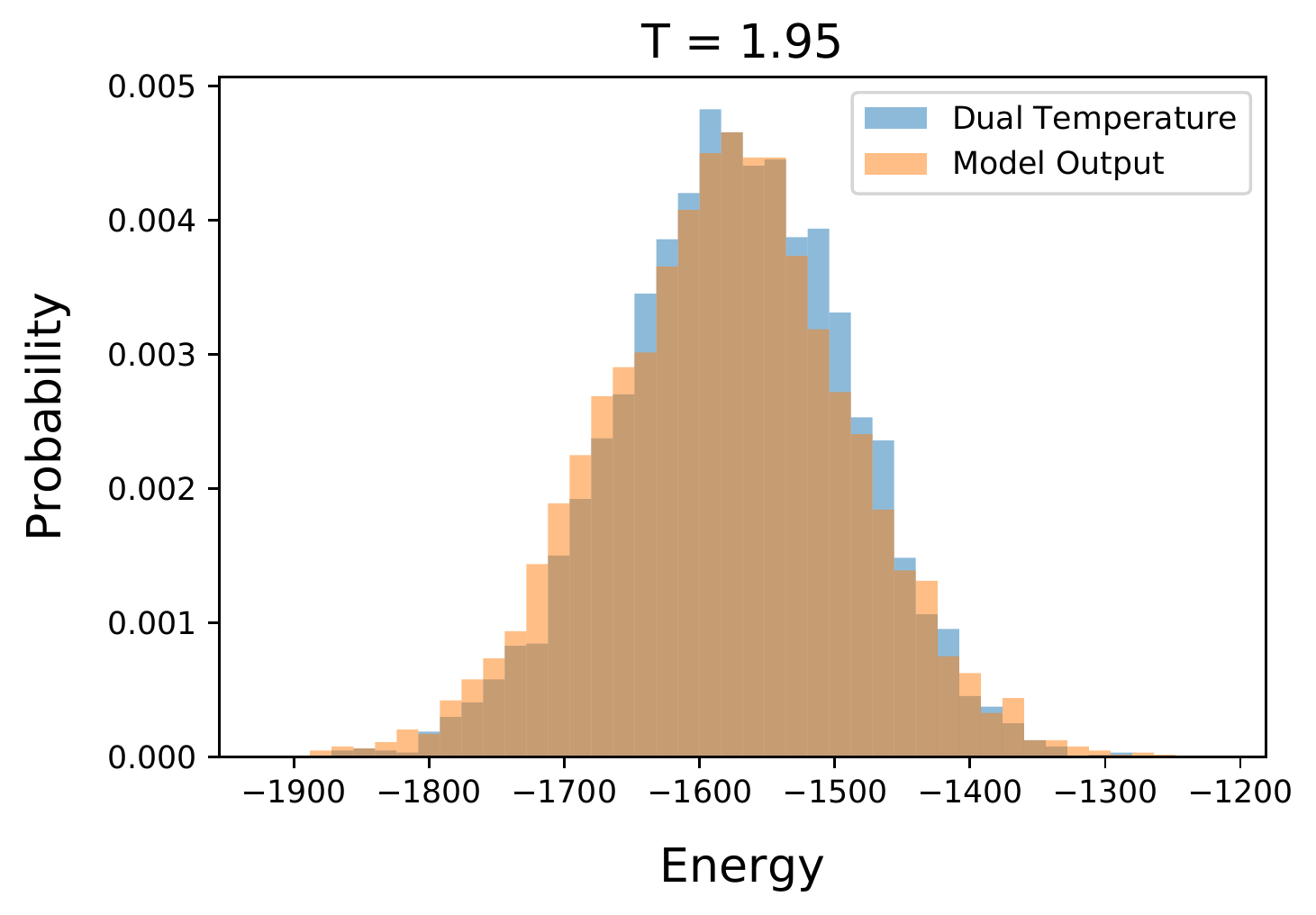}
\includegraphics[width=0.4\textwidth]{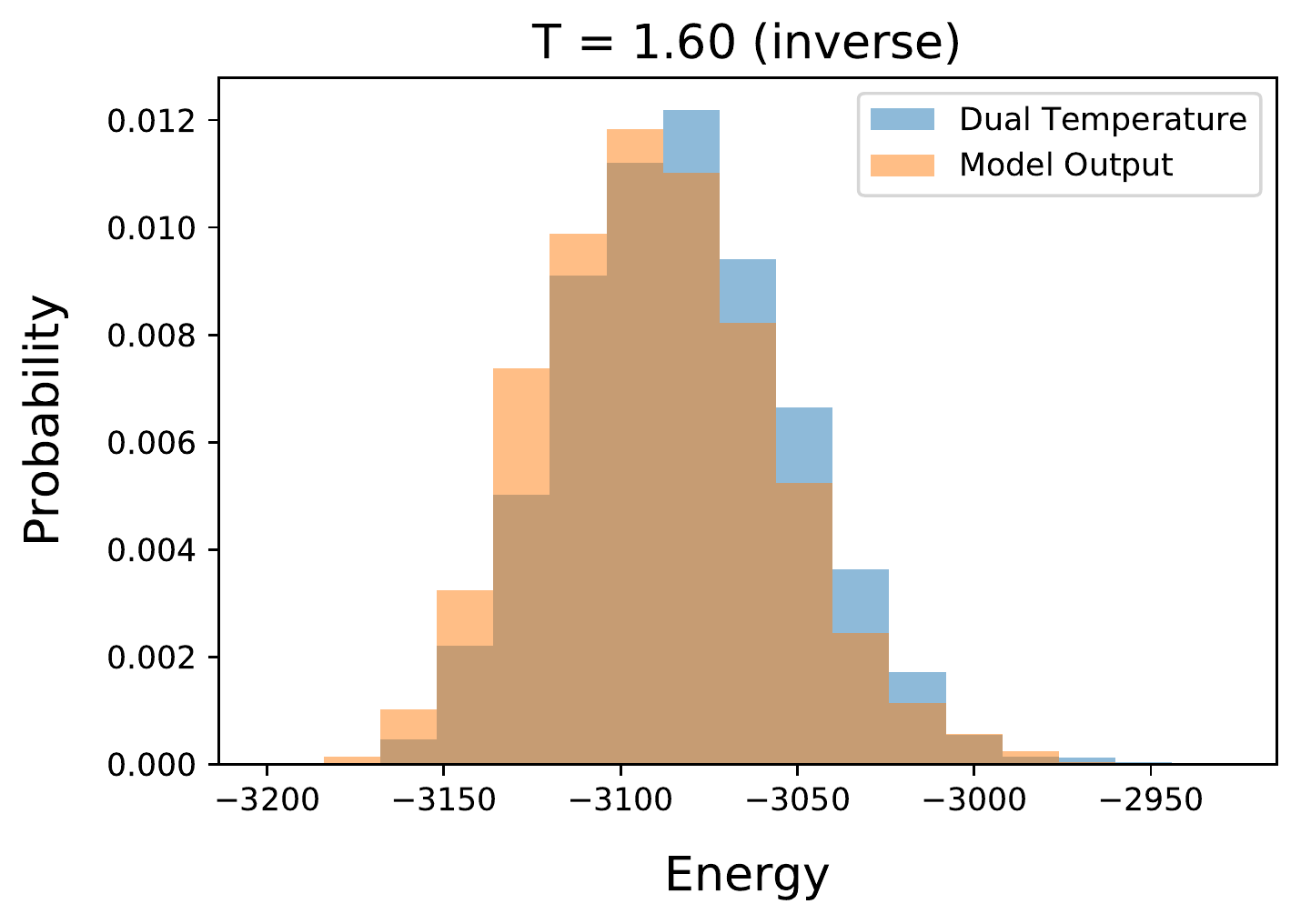}
\includegraphics[width=0.4\textwidth]{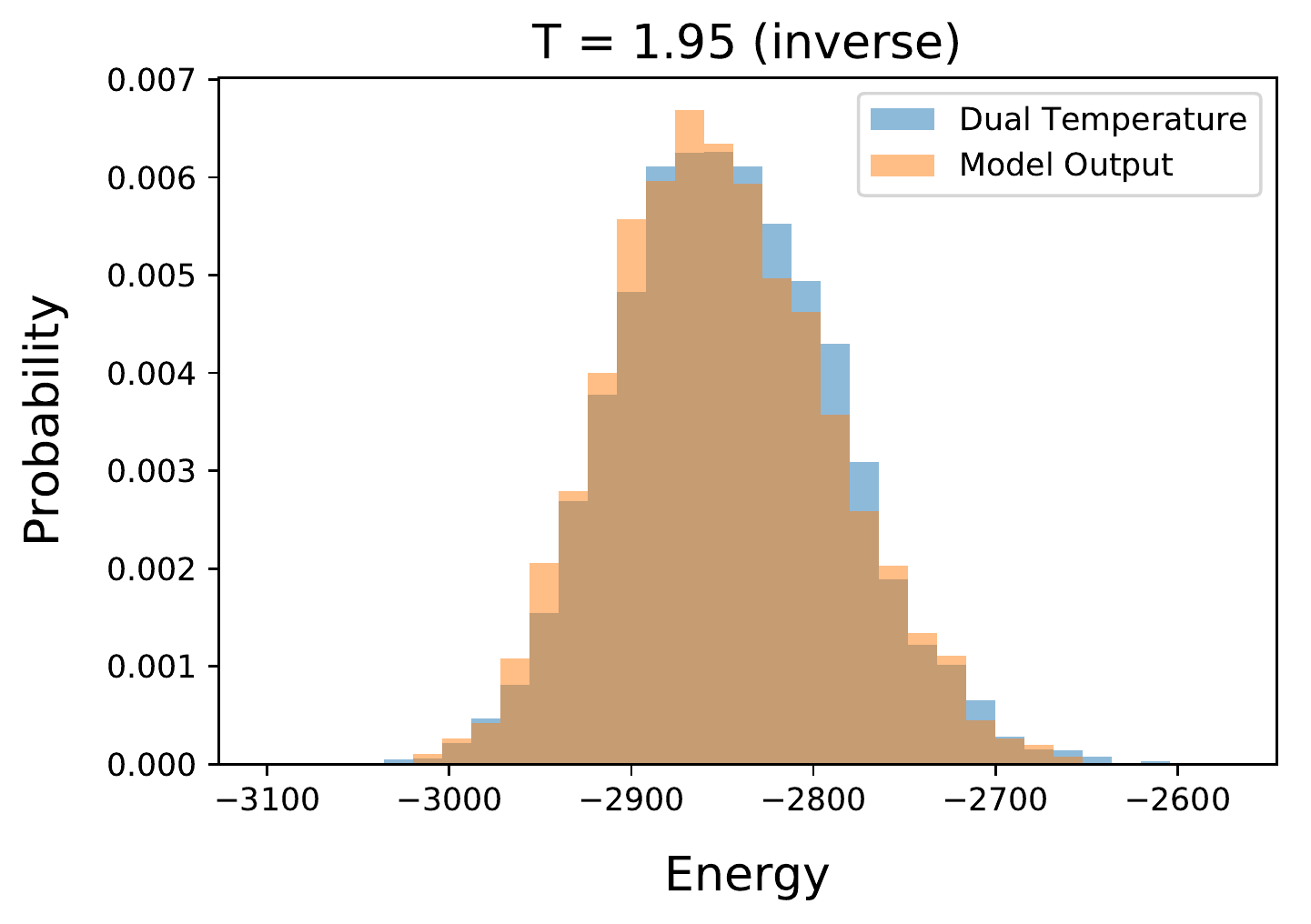}
\end{center}
\vspace*{-17pt}
\caption{Energy distributions of U-Net outputs and true dual temperatures. {\bf Top:} Mapping from low- to high-temperature regions. {\bf Bottom:} Mapping from  high- to low-temperature regions.}
\label{fig:KWDuality_Unet_40x40}
\end{figure}

The network produces 
binary outputs as desired, with the energy distributions closely resembling those of the actual dual system.
This is depicted for two examples in Figure~\ref{fig:KWDuality_Unet_40x40}. 

\begin{figure}
\begin{center}
\includegraphics[width=0.49\textwidth]{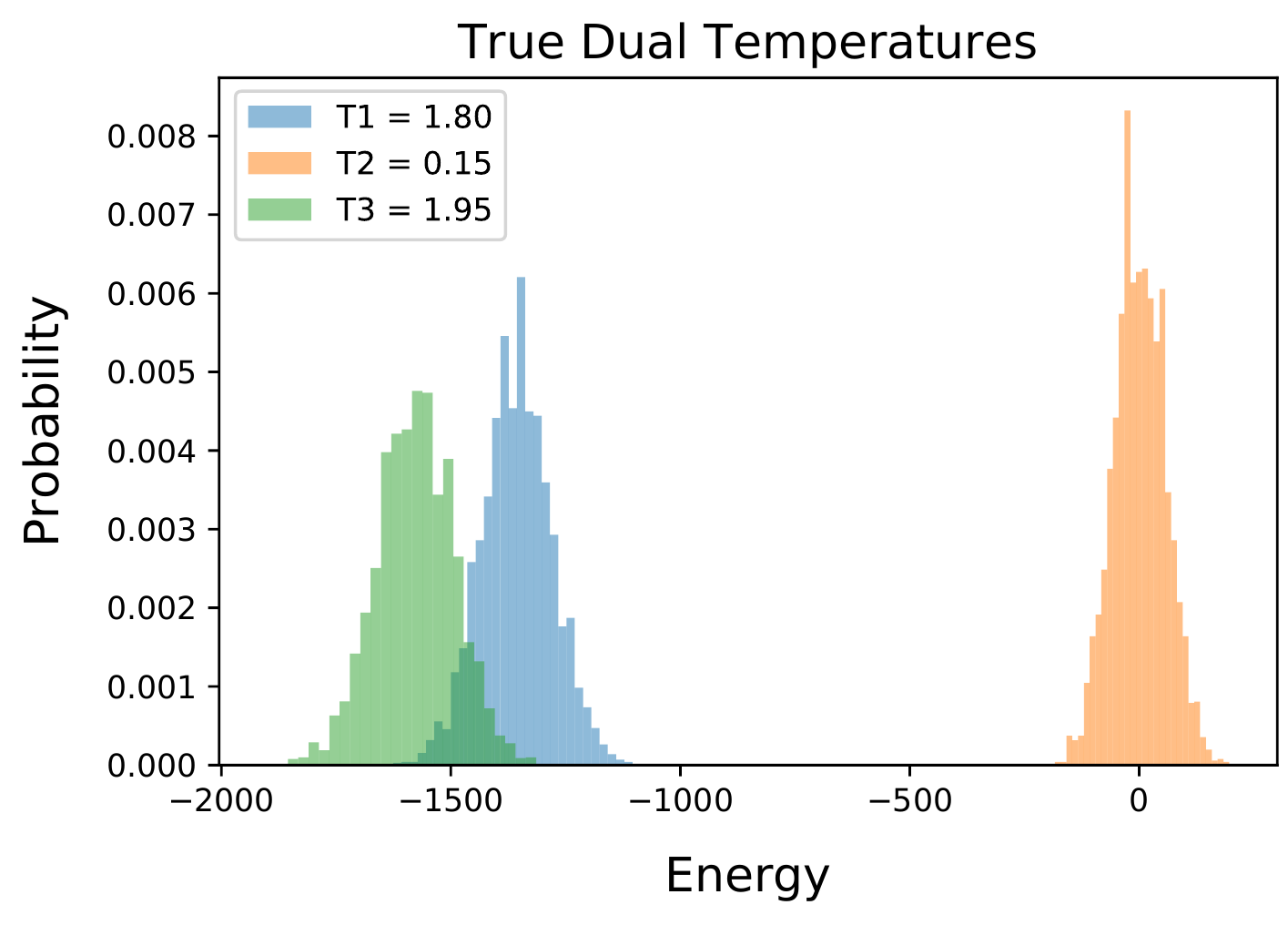}
\includegraphics[width=0.49\textwidth]{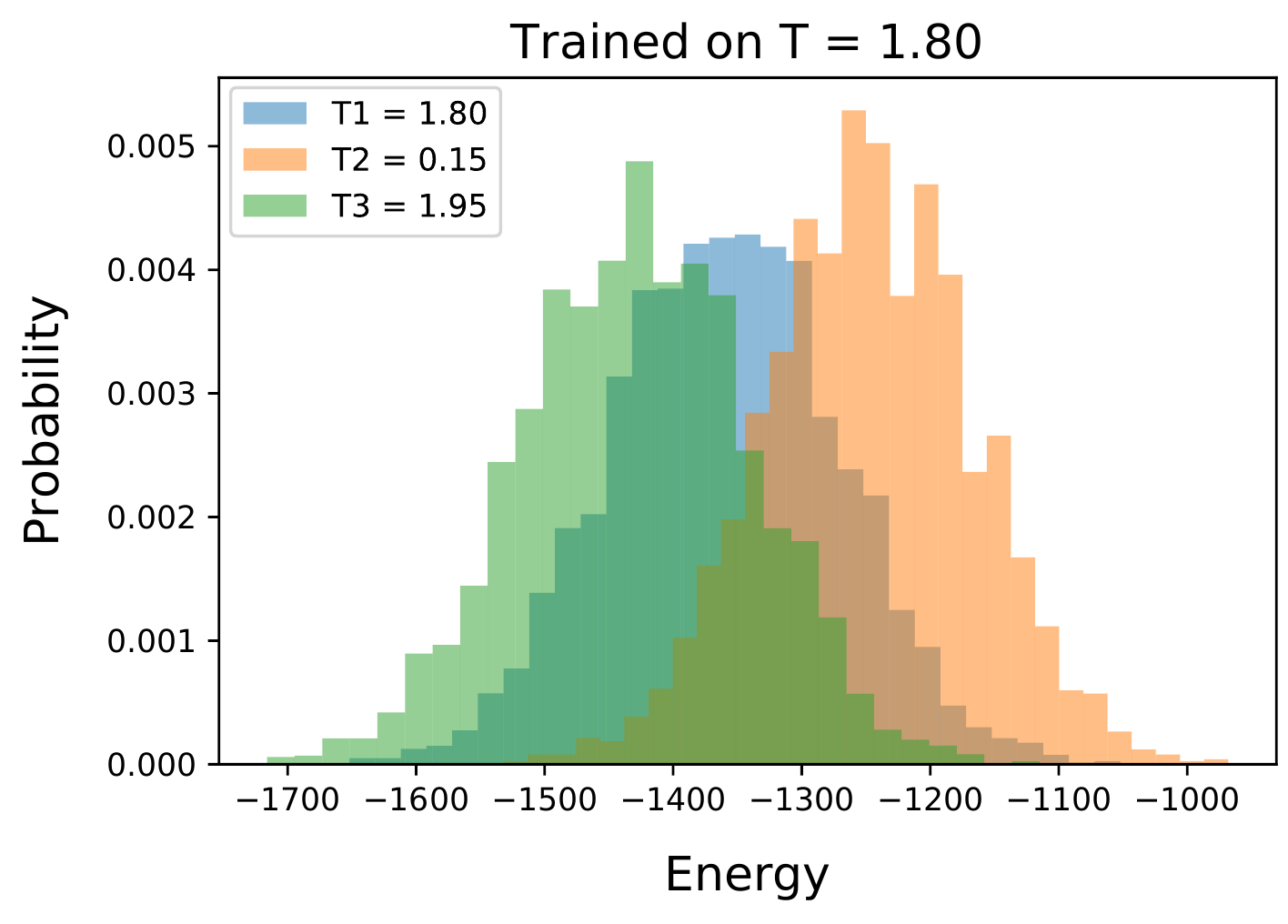}
\end{center}
\vspace*{-20pt}
\caption{
Output of U-networks trained on a single-temperature dataset for various temperatures. The ability
to distinguish between inputs depends strongly on the original temperature. Here we show results for $T=1.80$ where the network is able to distinguish between different inputs. 
\label{fig:Unet_CrossCheck}
}
\end{figure}

We next checked the output of U-nets trained on a single temperature for input states sampled from other temperatures. For networks trained on larger original
temperatures, the output energy distribution shows some resemblance of the true dual temperatures, albeit with wrong numerical values. This behaviour is shown in Figure~\ref{fig:Unet_CrossCheck} for temperature $T=1.80.$
 For lower training set temperatures,
the networks gradually lose their ability to distinguish between input states. 

When we trained the network with data from multiple temperatures, we have not (yet) found a significant improvement compared to Figure~\ref{fig:Unet_CrossCheck}.

Generally speaking, one can think of extending this method and incorporating more and more properties, i.e.~matching more and more correlators. This would lead to a more and more precise map which satisfies more and more properties of the respective dynamical system.

\section{Connection to Other Dualities in Physics}
\label{sec:connections}
We have seen in previous sections that dualities are a change in the basis which describes the system. Although we have already used this in the case of physical systems, such as the 2D Ising model (cf.~\ref{sec:2DIsing}), we would like to highlight how such a change in the basis appears analytically in physical systems and how it is connected to Fourier transformation. To do this we repeat the key steps from arguments presented for instance in~\cite{Polchinski:2014mva}.

To do this, one can consider electromagnetism in four dimensions without sources. The path integral is described by
\begin{equation}
\int {\cal D}A~e^{iS(A)/\hbar}~,\qquad S(A)=-\frac{1}{4g^2}\int d^4 x~(\partial^\mu A^\nu-\partial^\nu A^\mu)(\partial_\mu A_\nu-\partial_\nu A_\mu)
\end{equation}
This can be re-formulated as a path integral over the antisymmetric tensor field $F_{\mu\nu}$ subject to the constraint that the Bianchi identity $\partial_\mu \tilde{F}^{\mu\nu}=0$ is satisfied at each point $x$
\begin{equation}
\int {\cal D }F\prod_x\delta(\partial_\mu \tilde{F}^{\mu\nu}(x))e^{-\frac{i}{4\hbar g^2}\int d^4 x F_{\mu\nu}F^{\mu\nu}}~,
\end{equation}
where a potential Jacobian is ignored. By using an integral representation for the $\delta$ function and some integration by parts, this action can be rewritten as
\begin{equation}
\int {\cal D }F{\cal D}V~e^{-\frac{i}{\hbar}\int d^4 x \frac{1}{4g^2}F_{\mu\nu}F^{\mu\nu}-\frac{1}{4\pi}(\partial_\mu V_\nu-\partial_\nu V_\mu)\tilde{F}^{\mu\nu}}~.
\label{eq:full}
\end{equation}
In this formulation one can now also integrate out $F_{\mu\nu}$ as the integral is essentially Gaussian. This leads to
\begin{equation}
\int {\cal D}V e^{-\frac{i g^2}{16\pi^2}\int d^4 x (\partial_\mu V_\nu-\partial_\nu V_\mu)(\partial^\mu V^\nu-\partial^\nu V^\mu)}
\label{eq:piv}
\end{equation}
This path integral is now over a different field $V$ which was introduced merely as an auxiliary field. The relation between both representations can be seen from the equations of motion from the action involving both fields $A$ and $V:$
\begin{equation}
\tilde{F}_{\mu\nu}=-\frac{g^2}{2\pi}(\partial_\mu V_\nu-\partial_\nu V_\mu)\equiv -G_{\mu\nu}
\label{eq:locrelation}
\end{equation}
Electric and magnetic fields components are exchanged between these two descriptions and in addition the appearance of the coupling constant is inverted $g\to 1/g.$\footnote{Note that this becomes a real strong-weak duality once charged fields are introduced.} Despite the local relation~\eqref{eq:locrelation}, the map relating both representations is non-local as it involves the integration over space-time.

Note that the integration of a Gaussian from~\eqref{eq:full} to~\eqref{eq:piv} corresponds precisely to the transformation of a Gaussian from position space to momentum space in the Fourier transformation. This highlights the connection between Fourier transformation and mapping fields under duality.

This analysis for electromagnetism in four dimensions can be extended to the discussion of massive $p-$form fields in $D$ dimensions (cf.~\cite{Quevedo:1997jb} for a review). Again a relation between the variables in terms of Fourier transformation can be established.

\subsubsection*{Applications in Physics}
In the previous sections we have focused on the determination of classification tasks with the help of dual variables. In the context of physics, the use of dualities is generally speaking in the context of determining correlation functions more accurately. In turn this can be seen as properties of the data and hence can be connected with our classification tasks. To highlight the strength of these techniques we mention two major applications where the methods based on dualities outperform other techniques:
\begin{enumerate}
\item {\bf Hydrodynamic transport coefficients for quark gluon plasma:} In the context of holography, strongly coupled conformal field theories are related with weakly coupled gravitational systems\footnote{See~\cite{Hashimoto:2019bih} for the connection between holography and deep Boltzmann machines.} in one higher dimension. Field theory correlators can be calculated by performing the appropriate perturbation analysis in the gravitational system~\cite{Maldacena:1997re, Witten:1998qj, Aharony:1999ti}. One of the prime examples includes the calculation of the shear viscosity $\eta/s$ of ${\cal N}=4$ super Yang-Mills theory which effectively is a two-point correlation function of the stress energy tensor~\cite{Policastro:2001yc,Kovtun:2004de}. It has been argued that these calculations can be used to understand properties of the quark-gluon plasma and provide - at reasonably low calculational effort - quantitatively more accurate results than lattice predictions (cf.~\cite{CasalderreySolana:2011us} for a review and further interesting applications).
\item {\bf Yukawa couplings in the standard embedding for the heterotic string:} Here the duality in use is referred to as mirror symmetry, a generalisation of T-duality. In the heterotic standard embedding it facilitates the calculation of Yukawa couplings in the standard embedding. Concretely, in the dual frame the ${\bf 27}^3$ couplings are purely topological whereas in the original frame the couplings ($\overline{\bf 27}^3$) depend on the K\"ahler moduli. The topological couplings can be computed with standard methods in finding solutions to the Picard-Fuchs equations. Both couplings have to be identical due to mirror symmetry and utilising the mirror map between the dual moduli spaces allows a calculation of the K\"ahler moduli dependence in the $\overline{\bf 27}^3$ coupling. The direct calculation of these corrections requires counting of appropriate rational curves on the background Calabi-Yau manifold which is known as a hard problem in Mathematics. Using mirror symmetry this hard calculation can be avoided. For a physicist the Yukawa couplings in the original frame capture a tree-level part and non-perturbative corrections. It is these non-perturbative corrections which can be calculated using mirror symmetry.  For explicit constructions of these dualities and more details see for instance~\cite{Candelas,Candelas:1993dm,Hosono:1993qy,Hosono:1994av}. Note that the reduced calculational complexity required to calculate the Yukawa couplings in the dual frame was mentioned in~\cite{Halverson:2018cio}.
\end{enumerate}
Both examples highlight the capability of calculating far beyond the realm of standard perturbation theory. As a final comparison to showcase the connection of the dualities in the 1D Ising case, we discuss the connection with Seiberg duality. Here we identify a starting point for correlators which can serve as candidate replacements of metastability in the 1D Ising case.

\subsection{Seiberg duality}
Let us comment on the connection to the classical example of Seiberg duality in the context of SQCD~\cite{Seiberg:1994bz,Seiberg:1994pq,Intriligator:1995au}. Here two gauge theories share the same infrared physics but differ in the UV. These are referred to as the electric and magnetic phase. The electric phase ($3/2 N_c<N_f<3N_c$) is described by the field content presented in Table~\ref{tab:fieldcontentelectric} and the magnetic one in Table~\ref{tab:fieldcontentmagnetic}. The electric theory has no superpotential whereas the magnetic theory has a superpotential of the form $W=\tilde{M}q\tilde{q}$ where $\tilde{M}$ is related to the meson $M$ built out of quarks in the electric phase.
\begin{table}
\begin{center}
\begin{tabular}{c | c | c | c | c | c | c}
Field & $SU(N_c)$ & $SU(N_f)_{L}$ & $SU(N_f)_{R}$ & $U(1)_A$& $U(1)_B$& $U(1)_R$\\ \hline 
$Q$ &$\bf{N_{c}}$ & $\bf{N_{f}}$ & $1$ & $1$ & $1$ & $1-\frac{N_{c}}{N_{f}}$\\
$\tilde{Q}$ & $ \mathbf{\overline{N}_{c}}$ &$ 1$ &$ \mathbf{\overline{N}_{f}}$ &$ 1$ &$ -1$ &$ 1-\frac{N_{c}}{N_{f}}$\\
\end{tabular}
\end{center}
\vspace*{-10pt}
\caption{Field content of the electric phase.}\label{tab:fieldcontentelectric}
\end{table}
{

\subsubsection*{Electric Phase}
As a supersymmetric theory with zero
tree-level superpotential, the classical Lagrangian of the electric phase involves a D-term potential whose flat directions at vanishing value parameterise the moduli space of the theory. More precisely, the corresponding quark expectation values can be determined by imposing the D-flatness condition $D^{A}=0$ with
\begin{equation}
D^{A} = \sum _{i} {Q}_{i}^{\dagger}T_{i}^{A} Q_{i}+\tilde{Q}_{i}^{\dagger}T_{i}^{A} \tilde{Q}_{i}\,,
\end{equation}
where the $T^{A}$ denote the generators of the respective gauge group $SU(N_c)$. The classical moduli space is then defined as the space of quark vacuum expectation values modulo gauge equivalence. As argued in~\cite{Intriligator:1995au,Luty:1995sd}, this allows for an equivalent description in terms of expectation values of gauge-invariant polynomials in the fields subject to any classical relations. For the theories considered here, such combinations are given by the $2{N_f\choose N_c}$ baryon and $N_{f}^2$ meson operators

\begin{eqnarray}
\label{BaryonsMesons_Electric}
\nonumber B^{i_1\ldots i_{N_c}}&=& Q^{i_1}_{a_1}\cdots Q^{i_{N_c}}_{a_{N_c}}\epsilon^{a_1\ldots a_{N_c}}\,,\\
\tilde{B}_{i_1\ldots i_{N_c}}&=& \tilde{Q}_{i_1}^{a_1}\cdots \tilde{Q}_{i_{N_c}}^{a_{N_c}}\epsilon_{a_1\ldots a_{N_c}}\,,\\
\nonumber M^{i}_{j}&=&Q^{i}_{a}\tilde{Q}_{j}^{a}\,.
\end{eqnarray}
Due to the identity 
\begin{equation}
\epsilon _{a_{1}\dots a_{N_c}}\epsilon ^{b_{1}\dots b_{N_c}} = \delta _{a_1}{}^{[\underline{b_1}}\delta _{a_{N_c}}{}^{\underline{b_{N_c}}]}\,,
\end{equation}
these are subject to additional constraints 
\begin{equation}
 B^{i_1\ldots i_{N_c}}\tilde{B}_{j_1\ldots j_{N_c}}=M _{j_1}^{[\underline{i_1}}M _{j_{N_c}}^{\underline{i_{N_c}}]}\,,
\end{equation}
leaving a total of $2N_{f}N_{c}-(N_c^2-1)$ light D-flat directions (cf.~\cite{WechtLec}).  The physical interpretation of this is that the gauge group $SU(N_c)$ is completely broken, which is reflected in the number $N_c^2-1$ of broken generators~\cite{Intriligator:1995au}.

\subsubsection*{Magnetic Phase}

In the infrared, the above theory is dual to a magnetic description based on the gauge group $SU(\tilde{N}_c=N_f - N_c)$. The corresponding field content is listed in Table~\ref{tab:fieldcontentmagnetic}. Unlike the electric phase, the magnetic phase involves an additional superpotential 
\begin{equation}
\label{SeibergDuality_Superpotential}
W=\tilde{M}^i_j q_i \tilde{q}^j\,,
\end{equation}
where the magnetic meson $\tilde{M}$ defines a fundamental degree of freedom and is related to its electric counterpart defined in~\eqref{BaryonsMesons_Electric}  by a characteristic scale $\mu$,
\begin{equation}
\label{SeibergDuality:MagneticMeson}
\tilde{M}=\frac{1}{\mu}M\,.
\end{equation}
Often both mesons are identified and one uses the notation $M$ in either phase, which is indeed valid at the infrared fixed point.  The presence of the dimensionful parameter $\mu$ in \eqref{SeibergDuality:MagneticMeson} is only required to relate both meson operators in the ultraviolet limit: Here, the electric meson is a composite state with canonical dimension 2, picking up an anomalous dimension $3\frac{\tilde{N}_c}{N_f}$ during the renormalisation group flow to the infrared fixed point, while the latter defines a fundamental field of dimension one flowing to the same fixed point. It is therefore common to define a separate operator as in \eqref{SeibergDuality:MagneticMeson} to correctly describe the magnetic meson in the ultraviolet limit.
The characteristic scale $\mu$ also appears in the matching condition
\begin{equation}
\Lambda^{3N_c - N_f}\tilde{\Lambda}^{3\tilde{N}_c-N_f}=(-1)^{\tilde{N}_c}\mu ^{N_f}
\end{equation}
for the scales $\Lambda$ and $\tilde \Lambda$ of the electric and magnetic theory, respectively.  
From this, it can be seen that the duality relates different theories at strong and weak coupling, thus resembling the characteristic structure of a strong-weak duality.

\begin{table}[t]
\begin{center}
\begin{tabular}{c | c | c | c | c | c | c}
Field & $SU(\tilde{N}_c=N_f-N_c)$ & $SU(N_f)_{L}$ & $SU(N_f)_{R}$ & $U(1)_A$& $U(1)_B$& $U(1)_R$\\ \hline
$q$ & $\bf{\tilde{N}_c}$ & $\bf{\overline{N}_{f}}$ & $ 1$ & $ 1$ & $ \frac{N_{c}}{\tilde{N}_c}$ & $1-\frac{\tilde{N}_c}{N_{f}}$\\
$\tilde{q}$ & $\bf{\overline{\tilde{N}}_c}$ & $ $1 & $\bf{N_{f}} $ & $ 1$ &$ -\frac{N_{c}}{\tilde{N}_c}$ & $1-\frac{\tilde{N}_c}{N_{f}}$ \\
$\tilde{M}$ & 1 & $ \bf{N_{f}}$ & $ \bf{\overline{N}_{f}}$ & $ -2 $ & $ 0 $ & $2\frac{\tilde{N}_c}{N_{f}}$
\end{tabular}
\end{center}
\vspace*{-10pt}
\caption{Field content of the magnetic phase.}\label{tab:fieldcontentmagnetic}
\end{table}

Analogously to the electric phase, one can define $2{N_f\choose \tilde{N}_c}$ magnetic baryon operators as 
\begin{eqnarray}
\label{Baryons_Magnetic}
\nonumber b_{i_1\ldots i_{\tilde{N}_c}}&=& q_{i_1}^{a_1}\cdots q_{i_{\tilde{N}_c}}^{a_{\tilde{N}_c}}\epsilon_{a_1\ldots a_{\tilde{N}_c}}\,,\\
\tilde{b}^{i_1\ldots i_{\tilde{N}_c}}&=& \tilde{q}^{i_1}_{a_1}\cdots \tilde{q}^{i_{\tilde{N}_c}}_{a_{\tilde{N}_c}}\epsilon^{a_1\ldots a_{\tilde{N}_c}}\,,
\end{eqnarray}
which, due to the identity ${N_f\choose N_c}={N_f\choose N_f-N_c}$, carry the same number of degrees of freedom as their electrical counterparts.
Formally, further mesons could be defined by $\tilde{m}=q\tilde{q}$, however, these do not lead to new degrees of freedom in the moduli space due to additional equations of motion $\langle q\tilde{q}\rangle =0 $ arising from the presence of the superpotential \eqref{SeibergDuality_Superpotential}, thus avoiding inconsistency of the duality~\cite{WechtLec}. A more in-depth analysis of the moduli spaces as well as further consistency checks of the duality were performed (e.g. in~\cite{Intriligator:1995au}) and we would like to refer the interested reader to the original works for more details.

\subsubsection*{Application to Neural Networks}
At the infrared fixed point, there exists a direct relation between both types of  baryon operators,
\begin{eqnarray}
\nonumber B^{i_{1},\ldots i_{N_{c}}} & = \sqrt{-(-\mu)^{N_c - N_f}\Lambda^{3N_c - N_f}}\epsilon^{i_{1},\ldots i_{N_{c}}j_{1},\ldots j_{\tilde{N}_c}}  b_{j_1\ldots j_{\tilde{N}_c}}\,,\\
\widetilde{B}_{{i}_{1},\ldots{i}_{N_{c}}} & =\sqrt{-(-\mu)^{N_c - N_f}\Lambda^{3N_c - N_f}}\epsilon_{{i}_{1},\ldots{i}_{N_{c}}{j}_{1},\ldots{j}_{\tilde{N}_c}} \tilde{b}^{j_1\ldots j_{\tilde{N}_c}}\,.
\end{eqnarray}
As can be seen, the baryons in the electric and magnetic phase involve products of $N_c$ and $\tilde{N}_c=N_f - N_c$ quarks, respectively. 
This is similar to our discussion of the 1D Ising chain, in which determining the total energy required the computation 
of $n$-spin products in the original representation, while the dependency was linear in the dual frame and therefore significantly easier to learn for neural networks. As the degree $n$ of interactions was raised, the value of the total energy became increasingly sensitive to flips of single spins due to their involvement in an increasing number of $n$  local interaction terms (cf. Figure~\ref{fig:1DIsing_Dualities_n3_Metastability}), which eventually led to a complete deterioration of performance at very high $n$.

In the above setting, the baryon operators in \eqref{BaryonsMesons_Electric} and \eqref{Baryons_Magnetic} take the form of sums over products of $N_c$ or $\tilde{N}_c$ quarks, with each particular component appearing in $(N_c-1)!$ or $(\tilde{N}_c-1)!$ non-vanishing products (taking the role of the ``local interaction terms"). Similar to the 1D Ising chain, such dependencies are likely to be learned more easily in the phase for which the number of factors is lower. In the setting discussed here, there exists a range $3/2 N_c  < N_f < 2N_c$ for which $\tilde{N}_c<N_c$, implying that baryon relations might be easier to be accessed in the magnetic theory. Conversely,  the electric phase might be preferable in the region $2 N_c  < N_f < 3N_c$, where generically $\tilde{N}_c>N_c$. 

It is a natural question to explore whether this fact can be used to re-discover Seiberg-like dualities following the strategy successfully applied for the 1D Ising case in Section~\ref{sec:1DIsing}. As this analysis promises to be too lengthy for this proof of concept paper, we leave this issue for the future.

\section{Conclusion}
\label{sec:conclusions}

Dualities offer a more efficient way of calculating correlation functions in physics. In particular, in the context of strongly coupled regions they provide in several examples the best technique to calculate properties of these dynamical systems. We have presented several examples where this improved way of calculating correlation functions via dual representations can be related to improved classification tasks.

Such different and more efficient data descriptions are clearly desirable, but how can one get them without knowing about the explicit map between such representations. We have shown in this work how such beneficial representations can be obtained in an unsupervised fashion, i.e.~without telling the network about its existence. By reproducing several human-made dualities automatically we provide a proof of concept that machines can be programmed to find dualities. Clearly, further and more involved types of dualities need to be addressed with these kind of techniques, which then will enable the search for new dualities.

Undoubtedly our tasks are relatively simple and can be achieved for instance in the case of the 1D Ising and Fourier analysis by more sophisticated architectures. However, we want to stress that these settings serve as an important first step to address tasks which are not accessible with state-of-the-art techniques with the same strategies used here.

The dual representations obtained by our networks can be analysed and we have found a representation which is interpretable, e.g.~we could recognise a Fourier-like transformation or transformations similar to the duality transformation in the 1D Ising example. This is encouraging as the neural network provides us with the explicit map to this interpretable representation.

Where will further steps in this new field of exploring dualities between different descriptions of dynamical systems with the help of machine learning take us?

\section*{Acknowledgments}
We would like to thank Jim Halverson, Fernando Quevedo, and Fabian Ruehle for discussions.
SK thanks the Aspen Center for Physics, which is supported by National Science Foundation grant PHY-1607611, and the Simons Center for Geometry and Physics during the Neural Networks and the Data Science Revolution program for providing a very stimulating work environment to develop some of this work. Parts of these results have been presented already at the following conferences and workshops: String Phenomenology 2019, QTS 2019 (Montreal), Corfu Summer Institute, DLAP in Kyoto, 1st French-German Meeting in Physics, Mathematics and Artificial Intelligence Theory, and XAIENCE in Seoul.

\appendix
\section{Details on Discrete Fourier Transformation}
This appendix contains further details on the experimental setup used for our discussion of the Fourier transform in section
\label{app:fourier}
\subsection{Data}
The dataset is split into two categories ``pure noise" (0) and ``noise with signal" (1).
We consider a discretised space of size 1000 and generate $10^5$ signals $p_k$ in the Fourier domain
taking the form 
\begin{equation}
p_k= |p_k|e^{i \varphi _{k}}\,,
\end{equation}
where $|p_k|\sim\mathcal{N}(2,0.1)$, $\varphi _{k}\sim\mathcal{U}(0,2\pi)$ and $k$ uniformly 
sampled between 0 and 1000.  A signal
in position space is generated by computing the inverse Fourier transform, which
relates the position and Fourier domains via 
\begin{eqnarray}\label{DiscreteFT}
\nonumber p_k&=&\frac{1}{\sqrt{N}}\sum_{j=1}^N x_j e^{-2\pi i j k/N}~,\\
 x_k&=&\frac{1}{\sqrt{N}}\sum_{j=1}^N p_j e^{2\pi i j k/N}~,
\end{eqnarray}
with $N=1000$. We then generate noisy signals (``class 1") by adding Gaussian noise
following the distribution $x_{k\textrm{, noisy}}\sim \mathcal{N}(0,0.1)$ and pure 
noise $x_{k\textrm{, pure noise}}\sim \mathcal{N}(0,\sigma)$ with $\sigma$ chosen such
that the samples of both classes show the same mean quadratic deviation from 0.
The number of samples for both classes is set to $10^5$ , and we employ a 4:1 train-test
split. The data is formatted in such a way that each sample contains one channel representing
its real part and one its imaginary part.

In this task a signal in the position 
space takes the form of a sine-cosine wave spread all over the domain, whereas its information is concentrated
in one (complex) bin in the Fourier space (cf. Figure~\ref{fig:Fourier_example}). 

\subsection{Experiments}
All experiments were performed using Keras with TensorFlow backend. Training equilibrium in all settings was commonly reached after less than 50 epochs; the training process was run for 200 epochs to ensure that no further improvements occur after stopping. 
Training was performed with Nesterov Adam optimiser with learning rate
$2\cdot 10^{-3}$, batch size 128 and binary crossentropy as loss function. Data generation, preprocessing
and training of networks was performed for ten different random seeds to prevent
results from getting skewed due to outliers.
A summary of all results
related to this setting can be found in Table~\ref{tab:Results_Fourier} at the end of this appendix.

\subsubsection*{Simple Networks}

We first checked whether simple networks are able to distinguish the classes in position space.
We used one-dimensional convolutional neural networks with one convolutional layer consisting of 4 filters of size one with ReLu activation followed
by a linear layer with sigmoid activation. 
The accuracy on the position space commonly stagnated at values around 0.53, which is only slightly superior to pure guessing.
The poor performance could be traced back to both underfitting and overfitting, with the training set accuracy commonly remaining below 0.5800
for the entire training process. Slight improvements to test set accuracies around 0.54 could be made by including one or two additional convolutional layers, however, no notable difference in performance was observed for more complex architectures.

\subsubsection*{Adding Dense Layers}

We tested whether adding a dense layer
of size $2N$ at the beginning of the above architecture can improve the results. The model was tested for the three different settings
\begin{itemize}
\item{all weights randomly initialised and trainable,}
\item{weights of the convolutional-layer pretrained on Fourier domain and trainable,}
\item{weights of the convolutional-layer pretrained on Fourier domain and fixed.}
\end{itemize}
Due to the high number of parameters, we tested the performance for L1 and L2 regularisation with parameters
$10^{-5},10^{-4},10^{-3},10^{-2},10^{-1}$, dropout with rates $0.1,0.2,0.5$ and batch normalisation.
To cover a wider range of architectures, we furthermore varied the number of dense layers between 1 and 3.

Interestingly, none of the architectures showed any improvements beyond pure guessing, implying that it is difficult
for neural networks to find the Fourier transform (or similar mappings) by themselves, even if given ``hints" by initializing parts
of the weights to perform well in the momentum space domain. 

Except for architectures with very strong
regularization, the poor performance in most settings could be attributed exclusively
to severe overfitting. This problem might in principle be avoidable by increasing the amount of the training data to extremely large values.
This would, however, defeat the point of finding a ``useful" network with reasonable resources.

\subsubsection*{Feature Separation}

Tests for representations learned by the architecture described in section \ref{FeatureSeparation} were performed using the same simple network
architecture as employed for comparing the position and actual momentum space. Preprocessing and training modalities were the same as before; the 
feature separation network was trained separately for each of the ten test-runs. The feature separation network was trained on separately-generated noisy signals with varying noise levels of up to $\sigma = 0.075$ and showed similarly good performance in all instances. All performance values and plots presented in this paper are with respect to networks trained on noisy data with $\sigma = 0.075.$ We found an improved performance when we reduced noise levels in the data.

\begin{table}
%\begin{subtable}[c]{0.95\textwidth}
\centering
\begin{equation*}
\begin{split}
 \renewcommand{\arraystretch}{1.1}
 \arraycolsep12pt
 \begin{array}{l||c}
\textrm{Model} &  \textrm{val acc }  \\
 \hline\hline
 \textrm{Simple Network x-space}  & 0.5317
 \\
 \textrm{Simple Network p-space} & 0.9879 
\\
 \textrm{Simple Network learned representation (feature separation)} & 0.7717
 \\
 \\
\textrm{Simple Network (2 Conv-Layers) x-space}  & 0.5449
 \\
\textrm{Simple Network (3 Conv-Layers) x-space}  & 0.5438
 \\
 \\
\textrm{Simple Network + Dense x-space (fixed pretrained weights) }  &0.5005
 \\
\textrm{Simple Network + Dense x-space (free pretrained weights) }  & 0.5013
 \\
\textrm{Simple Network + Dense x-space (free random weights) }  & 0.5016
\\
\textrm{Simple Network + 2 Dense x-space (free random weights) }  & 0.5018
 \\
\textrm{Simple Network + 3 Dense x-space (free random weights) }  & 0.5025
 \\
 \\
\textrm{Simple Network + Dense x-space (free random weights)  + L1-Reg (1e-5)}  &  0.5013
 \\
\textrm{Simple Network + Dense x-space (free random weights)  + L1-Reg (1e-4)}  & 0.5014
 \\
\textrm{Simple Network + Dense x-space (free random weights)  + L1-Reg (1e-3)}  & 0.5011
 \\
\textrm{Simple Network + Dense x-space (free random weights)  + L1-Reg (1e-2)}  & 0.5015
 \\
\textrm{Simple Network + Dense x-space (free random weights)  + L1-Reg (1e-1)}  & 0.5008
 \\
 \\
\textrm{Simple Network + Dense x-space (free random weights)  + L2-Reg (1e-5)}  & 0.5010
 \\
\textrm{Simple Network + Dense x-space (free random weights)   + L2-Reg (1e-4)}  & 0.5019
 \\
\textrm{Simple Network + Dense x-space (free random weights)  + L2-Reg (1e-3)}  & 0.5010
 \\
\textrm{Simple Network + Dense x-space (free random weights)  + L2-Reg (1e-2)}  & 0.5018
 \\
\textrm{Simple Network + Dense x-space (free random weights)  + L2-Reg (1e-1)}  & 0.5015
 \\
\\
\textrm{Simple Network + Dense x-space (free random weights)  + Dropout (0.1)}  &  0.5017
 \\
\textrm{Simple Network + Dense x-space (free random weights)  +  Dropout (0.2)}  &   0.5015
 \\
\textrm{Simple Network + Dense x-space (free random weights)  +  Dropout  (0.5)}  &  0.5004
 \\ 
\\
\textrm{Simple Network + Dense x-space (free random weights)  +  BatchNorm}  &  0.5013
 \\  
 \end{array}
\end{split}
\end{equation*}
%\vspace{-25pt}
\caption{Mean best test set accuracies for signal detection in noisy data reached after 200 epochs.}
\label{tab:Results_Fourier}
%\end{subtable}

\end{table}

\section{Details on 1D Ising Model}
\label{app:1DIsing}
This appendix contains further details on the experimental setup used for our discussion of the 1D Ising model in sections~\ref{sec:1DIsing} and~\ref{sec:CAE}.

\subsection{True Dual Representation}
In section~\ref{sec:1DIsing} we focused on the application of simple neural networks to detect (meta-)stable states in the 1D Ising model with multi-spin interactions.

\subsubsection*{Classifying (Meta-)Stable States}

For our large-scale tests, we used a single-layer perceptron with 128 hidden neurons,
ReLu activation for the hidden layer and sigmoid activation for the output layer. 
To ensure comparability of results, we did not include any regularisation techniques or more
advanced components. Weights were initialised randomly following common practice;
training was  performed with standard Nesterov Adam optimiser and learning rate decay.

The full dataset contained all $2^{18}$ states for the 1D Ising chain
with $N=18$, and tests were performed for varying orders of interaction $n$. We 
split the data into states labeled as ``not (meta-)stable" (0) or ``(meta-)stable" (1) and normalised the training and test
sets to contain an equal number of samples for each class. Experiments were performed for training set sizes of 600, 3000 and 9500
which were chosen for better comparability of results due different total numbers of metastable states for different~$n$.

The training  showed a slight dependence on the initial conditions and was therefore performed ten times for each setting
and data representation. Python, NumPy and TensorFlow random seeds were fixed by hand and stored for each result. We found
that 500 epochs was a viable cutoff after which no relevant changes in the overall performance occurred. 
The average best test accuracies and losses achieved in 10 training runs are listed in Table~\ref{table:val_acc_1DSimpleNets_original}.

\subsubsection*{Modifications of Architecture}

In order to check whether the results obtained for the above setting generalise to a wider class of neural networks, we performed sample-wise tests for one or more of the following modifications in architecture:
\begin{itemize}
\item{Varying the number of hidden layers within the range $1,2,3,4,5$.}
\item{Varying the number of neurons per layer within the range $16, 32,\dots, 1024$.}
\item{Employing linear, sigmoid or ReLu activation functions.}
\item{Using L2 weight regularisation with penalty 0.01 and 0.1,  dropout with rates 0.2 and 0.5 or batch normalisation.}
\item{Including up to five Convolutional layers into the network.}
\end{itemize}

Almost none of the modifications lead to any significant change in the overall results. One exception was the introduction of convolutional networks, which were able to reach close-to-perfect performance at very low $n\leq 4$, but resorted to pure guessing at higher-order interactions.

\subsection{Autoencoder with Latent Loss}
In section~\ref{sec:CAE} the discussion of (meta-)stability classification was extended to the output of a constrained autoencoder as depicted in Figure~\ref{fig:TaskConstrainedAE}.

\subsubsection*{Training of Autoencoders}

The constrained autoencoders consisted of one hidden layer of 128 neurons in their encoder and decoder components and bottleneck of dimension 18 and 50. The latent output of the bottleneck part was additionally fed into a linear layer. Training was then performed using standard Nesterov Adam optimiser to simultaneously minimise component-wise binary crossentropy as reconstruction loss and mean squared error as regression loss.  

Weights were initialised randomly for both autoencoders. We found that this generally led to better performance in both losses compared against hard-coded layers or pretraining on the actual dual representation (the latter only possible for latent dimension 18).

Achieving good performance in both tasks required relatively large amounts of training data. The autoencoders were therefore trained on 80\% of the full dataset of $2^{18}$ states. In case of poor performance, underfitting was prevalent, and there were no cases of overfitting.  Using L1 penalties to force sparse activations or representations generally led to poor performance and therefore were not used for our tests.

\subsubsection*{Classifying (Meta-)Stable States}

We tested the performance in classifying (meta-)stable states using the same setting as before,
with the duality transformation~\eqref{1DIsing_DualityTransformation} replaced by the intermediate output
of the previously described constrained autoencoders. In order to prevent information of the metastability
test set from leaking into the training set of the autoencoder, the same
train-test split was employed for the metastability classification. Otherwise, training and testing modalities were identical
with those of the original and true dual representation.

Data preprocessing and training of all involved networks were repeated ten times for different Python, NumPy and TensorFlow random seeds to prevent
outliers from skewing the results. The average best test set accuracies reached after at most 500 epochs are stored in Table~\ref{table:val_acc_1DSimpleNets_learned}.
More specifically, we observed the following behaviour:

\begin{itemize}
\item{For very simple settings such as $n=4$, the classifiers performed almost equally well on the intermediate output as on the actual dual representation. As shown in Table~\ref{table:val_acc_1D_energetic}, similar performances can also be reached by using pure energy considerations, and these results should therefore be treated with caution.}
\item{For all higher-degree interactions with $n\geq 4$, both most classifiers clearly beat the benchmark performance on the original representation. However, networks trained on outputs with latent dimension 18 often fall short of outperforming the benchmarks set by purely energetic arguments (cf. again Table~\ref{table:val_acc_1D_energetic}) in particular at low training set sizes. This is  not the case for latent dimension 50, and such networks are able to distinguish well between both classes  in regions of high energetic overlap. Their performance is, however, not equally well to the true dual representation in these critical cases.}
\item The method transfers well to other values of $N$ and $n$ as long as the task of energy regression is sufficiently easy to solve for the considered class of networks, but breaks
down in very complex cases such as $N=100$ and $n=50$. 

\end{itemize}
In addition to the above tests,  a sanity check similar to the Fourier setting was performed by placing a non-pretrained encoding architecture with latent dimension 18 or 50 in front the simple network and training on the original representation (see Table~\ref{table:val_acc_1DSimpleNets_stacked}). The performance in this case was slightly worse than that of the simple networks alone (cf. Table~\ref{table:val_acc_1DSimpleNets_original}), showing that the layer pretrained for energy regression indeed leads to a benefit beyond a mere improvement of network capacity.

\begin{table}
\begin{footnotesize}
\begin{center}
 \begin{tabular}{c||ccccc}
 lat (18) &  n=4 &  n=5 &  n=8  & n=9  & n=12  \\
 \hline\hline
$6\cdot 10^2$ & 0.9047 &  0.8404 &  0.8629  &  0.8507 & 0.8747
 \\
$3\cdot 10^3 $& - &   0.8983 &   0.9039 &  0.9011 & 0.9165 
 \\
$9.5\cdot 10^3$ & - &   -  &   0.9405 &  0.9400  & 0.9751
 \end{tabular}\qquad %\\[0.1cm]
 \begin{tabular}{c||cccccc}
lat (50) &  n=4 &  n=5 &  n=8  & n=9  & n=12  \\
 \hline\hline
$ 6\cdot 10^2$ & 0.9026 &  0.8570&  0.8616  &  0.8715 & 0.8632
 \\
$3\cdot 10^3$ & - &   0.9058 &   0.9002  &  0.8991 & 0.9176
 \\
$9.5\cdot 10^3$ & - &   -  &   0.9410  &  0.9360  & 0.9745
 \end{tabular}
\end{center}
\end{footnotesize}
\vspace*{-10pt}
\caption{Detection of (meta-)stable states in the 1D Ising chain for different interactions and amounts of training data.
The listed numbers describe the average best test accuracy over 10 training runs of 500 epochs each when trained in the original representation
with a non-pretrained encoding architecture with latent dimension 18 {\bf (Left)} and 50 {\bf (Right)} placed in front of the simple network.
Missing values indicate that the number of required samples exceeds the total number of metastable states for the
considered setting.}
\label{table:val_acc_1DSimpleNets_stacked}
\end{table}

}

\clearpage

\bibliography{dualityrefs}
\bibliographystyle{utphys}

\end{document}